\documentclass{pasj01}
\usepackage{subfig}
\usepackage[flushleft]{threeparttable}
\usepackage{booktabs}
\usepackage{marginnote}

\usepackage{comment}

\draft 
\Received{$\langle$reception date$\rangle$}
\Accepted{$\langle$acception date$\rangle$}
\Published{$\langle$publication date$\rangle$}

\begin{document}

\title{Investigating the gas-to-dust ratio in the protoplanetary disk of HD~142527}
\author{Kang-Lou
	\textsc{Soon}\altaffilmark{1}}
\author{Munetake
	\textsc{Momose}\altaffilmark{1}}
\author{Takayuki
	\textsc{Muto}\altaffilmark{2}}
\author{Takashi
	\textsc{Tsukagoshi}\altaffilmark{3}}
\author{Akimasa
	\textsc{Kataoka}\altaffilmark{3}}
\author{Tomoyuki
	\textsc{Hanawa}\altaffilmark{4}}
\author{Misato
	\textsc{Fukagawa}\altaffilmark{3}}
\author{Kazuya
	\textsc{Saigo}\altaffilmark{3}}
\author{Hiroshi
	\textsc{Shibai}\altaffilmark{5}}

\altaffiltext{1}{
College of Science, Ibaraki University,
2-1-1 Bunkyo, Mito, Ibaraki 310-8512, Japan}
\email{munetake.momose.dr@vc.ibaraki.ac.jp}
\altaffiltext{2}{Division of Liberal Arts, Kogakuin University,
1-24-2 Nishi-Shinjyuku, Shinjyuku-ku, Tokyo 163-8677, Japan}
\altaffiltext{3}{National Astronomical Observatory Japan,
Osawa 2-21-1, Mitaka, Tokyo 181-8588, Japan}
\altaffiltext{4}{Center for Frontier Science, Chiba University,
1-33 Yayoi-cho, Inage-ku, Chiba, Chiba 263-8522, Japan}
\altaffiltext{5}{Department of Earth and Space Science, Graduate School of Science, Osaka University,
1-1, Machikaneyama-cho, Toyonaka, Osaka 560-0043, Japan}


\maketitle

\begin{abstract}
We present ALMA observations of the $98.5~\mathrm{GHz}$ dust continuum
and the $\mathrm{^{13}CO}~J = 1 - 0$ and $\mathrm{C^{18}O}~J = 1 - 0$ line emissions
of the protoplanetary disk associated with HD~142527. 
The $98.5~\mathrm{GHz}$ continuum shows a strong azimuthal-asymmetric distribution similar 
to that of the previously reported $336~\mathrm{GHz}$ continuum, with a peak emission in dust concentrated region in the north. 
The disk is optically thin in both the $98.5~\mathrm{GHz}$ dust continuum and the $\mathrm{C^{18}O}~J = 1 - 0$ emissions. 
We derive the distributions of gas and dust surface densities, $\Sigma_\mathrm{g}$ and $\Sigma_\mathrm{d}$, 
and the dust spectral opacity index, $\beta$, in the disk from ALMA Band 3 and Band 7 data.
In the analyses, we assume the local thermodynamic equilibrium
and the disk temperature to be equal to the peak brightness temperature of $\mathrm{^{13}CO}~J = 3 - 2$ with a continuum emission.
The gas-to-dust ratio, $\mathrm{G/D}$, varies azimuthally with a relation $\mathrm{G/D} \propto \Sigma_\mathrm{d}^{-0.53}$, 
and $\beta$ is derived to be $\approx 1$ and $\approx 1.7$ in the northern and southern regions of the disk, respectively. 
These results are consistent with the accumulation of larger dust grains in a higher pressure region. 
In addition, our results show that the peak $\Sigma_\mathrm{d}$ is located ahead of the peak $\Sigma_\mathrm{g}$. 
If the latter corresponds to a vortex of high gas pressure, the results indicate that the dust is trapped ahead of the vortex, as predicted by some theoretical studies.
\end{abstract}

\section{Introduction} \label{sec:introduction}
Dust particles in protoplanetary disks are the foundations of planet formation \citep{armitage2010}.
In the minimum-mass Solar Nebula model,
gas and dust particles are distributed smoothly in a radial direction,
with surface densities that
follow a piecewise power law \citep{weidenschilling1977b,hayashi1981}.
The high angular resolution observations by the Atacama Large Millimeter/submillimeter Array (ALMA), however,
revealed complicated morphologies in protoplanetary disks as traced by the dust continuum and molecular line emissions,
such as dust-depleted gaps, spiral-arms,
and crescent-like distributions
\citep{casassus2013a,vandermarel2013,isella2013,perez2014,
almapartnership2015,vandermarel2016,nomura2016,boehler2018,andrews2018,tsukagoshi2019}.
These observations also show that the spatial distributions of the gas and dust
are not necessarily similar, thus resulting in a gas-to-dust ratio
that spatially varies within the disks.

Various mechanisms in protoplanetary disks
can lead to a different evolution 
of the gas and dust and thus result in spatial variation of the
gas-to-dust ratio.
For example, dust particles can lose their angular momentum
due to gas-dust friction and
radially drift towards the central star \citep{weidenschilling1977a}.
The gas-dust friction may also result in the settling of larger grains
toward the disk midplane \citep{dominik2007,pinte2016}.
Dust filtration may also occur
at the edges of gaps in the dust that has been cleared by planets,
in which smaller particles migrate inward to the disk inner region
while larger particles are retained at the edges \citep{rice2006}.
Large-scale high pressure gas vortices can also trap dust particles
in the azimuthal direction, which may explain the asymmetric structure observed in some protoplanetary disks
\citep{barge1995,klahr1997,birnstiel2013,zhu2016,baruteau2016}.
Other mechanisms that can change the gas-to-dust ratio
are the growth and fragmentation of dust particles near the snowline \citep{zhang2015,okuzumi2016},
as well as secular gravitational instability \citep{takahashi2014,takahashi2016}.
In these cases the dust particles tend to accumulate in concentric rings around the star.
Gas may also be dispersed
from the disk by photoevaporation,
creating regions with a low gas-to-dust ratio within several astronomical units of the inner disk,
favorable for planet formation \citep{gorti2015}.
The existence of one or several planets 
can also dramatically alter the gas and dust disk structure  
\citep{dipierro2016,kanagawa2016,dong2017}.
While the dominant mechanisms that result in 
the distribution of the gas-to-dust ratio may differ from disk to disk,
the ratio may provide clues concerning the
processes that lead to the observed disk structures
and is crucial in understanding the back-reaction from dust to
gas if the ratio is low \citep{gonzalez2017,dipierro2018}.

HD~142527 is a binary system
consisting of two pre-main sequence stars: the primary star HD~142527A
and the secondary star HD~142527B.
The distance to HD~142527 derived by \citet{arun2019}
based on the \textit{Gaia} observations \citep{gaia2016a,gaia2016b}
is $157~\mathrm{pc}\pm1~\mathrm{pc}$.
The primary star is a Herbig $\mathrm{F6-F7IIIe}$ \citep{malfait1998,vandenancker1998}
star with a mass of approximately 
$2.4~M_\odot$ and age of $2.96~\mathrm{Myr}$ \citep{fukagawa2013,arun2019}.
The secondary star is a $\mathrm{M}$ dwarf with a mass of $0.13~M_\odot$, which
orbits around the primary star at an angular distance of approximately $0\farcs1$ \citep{biller2012,close2014,lacour2016}.
The binary system is surrounded
by an inner disk close to $30~\mathrm{au}$ 
and a massive outer disk whose gas content
extends to approximately $400~\mathrm{au}$ \citep{verhoeff2011,muto2015};
the inner disk is separated from the outer disk by a dust-depleted region, or gap, that extends to a radius of $150~\mathrm{au}$.
ALMA observations of HCO$^+$ hint at the existence of gas filaments 
across this gap, through which the material is funneled from the outer disk to the inner disk \citep{casassus2013a}.
The accretion rate is estimated to be
$10^{-7}~M_\odot~\mathrm{yr^{-1}}$ \citep{mendigutia2014}.
Near infrared images show that there are at least six spiral arms in the outer disk \citep{avenhaus2014}.
From the different spatial distribution of emission seen at the near- and mid-infrared wavelengths,
the outer disk is thought to be inclined to the line-of-sight,
with the northeastern half appearing to be the furthest
and the southwestern half the nearest \citep{fukagawa2006,fujiwara2006}.
The inclination angle of the outer disk and the position angle of the disk major axis
have been derived as $27^\circ$ and $161^\circ$, respectively,
from the kinematics traced by the $^{13}$CO $J=3-2$ line emission \citep{fukagawa2013}. 
On the other hand, the inner disk 
is modeled to be inclined at $70^\circ$ relative to the outer disk \citep{marino2015}.
The inner disk shadows the northern and southern regions of the outer disk from stellar irradiation
causing a drastic drop in the intensity of infrared wavelengths in the two regions \citep{avenhaus2014}.
At the submillimeter and longer wavelengths,
the dust continuum emission of the outer disk
shows a crescent structure in which the northern region
is significantly brighter than the southern region \citep{fukagawa2013,casassus2015}.
Simulations by \citet{price2018} show that the observed features of the disk (e.g., spiral arms, cavity, HCO$^+$ streamers)
may be explained by just considering the interaction between the disk and the binary system.
Their results also predict that the disk-binary interaction can create the asymmetric 
dust disk without invoking a gas vortex in the disk northern region (see also \cite{ragusa2017}).

In previous research, \citet{muto2015} and \citet{boehler2017} derived the 
gas and dust surface densities and the gas-to-dust ratio of the outer disk of HD~142527
by modeling ALMA observations at Band 7.
This research focused on the
northern and southern regions of the outer disk,
which correspond to the sectors where the dust continuum emission is brightest
and faintest at Band 7, respectively.
The gas-to-dust ratio was derived to be 
$\sim3$ and $\sim30$ in the northern and southern regions, respectively,
and the results indicate that dust is concentrated in the northern region.
However, discussions concerning the detailed spatial variations in the gas-to-dust ratio were
beyond the scope of these papers as only two regions were studied.
Herein, we extend on previous studies by
deriving the spatial distribution of the gas-to-dust ratio across the outer disk
of HD~142527.
We assume the local thermal equilibrium and derive the gas and dust surface densities by using 
ALMA observations of the 
$^{13}$CO and C$^{18}$O molecular line and dust continuum emissions,
at both Band 3 ($\nu \approx 100~\mathrm{GHz}$) and Band 7 ($\nu \approx 330~\mathrm{GHz}$).

This paper is organized as follows.
In Section \ref{sec:observation_data_reduction}
we introduce the ALMA Band 3 data 
as well as the previously published ALMA Band 7 data
for the HD~142527 system.
In Section \ref{sec:observational_results}, we present the calibrated images of the ALMA Band 3 data
and compare them to that from Band 7.
We describe the methods used to derive the gas and dust surface densities
in Section \ref{sec:analyses} and discuss the results in Section \ref{sec:discussions}.
Section \ref{sec:summary} provides a summary of our research.

\section{Observations and data reduction}\label{sec:observation_data_reduction}
We used the ALMA Cycle 2 Band 3\footnote{Project code: ADS/JAO.ALMA\#2013.1.00670S}
and Cycle 0 Band 7\footnote{Project code: ADS/JAO.ALMA\#2011.0.00318S} observational data
of HD~142527  
to investigate the distribution of the gas-to-dust ratio in the disk. 
The observational details are described in the following subsections.

\subsection{ALMA band 3 data}
Band 3 data were taken at seven execution blocks
carried out on the nights of the 4th, 5th, and 15th July 2015. 
Depending on the block, 
the bandpass calibrator used was either the J1427$-$4206 or J1617$-$5848,  
while the absolute flux calibrator was either Ceres, Pallas, Titan, or
J1427$-$4206\footnote{The source name for J1427$-$4206 was designated
as J1427$-$421 by mistake in the execution block of ADS/JAO.ALMA\#2013.1.00670S.
We use the correct source name in this paper.}.
The flux of the quasar J1427$-$4206 at $103.5~\mathrm{GHz}$ was monitored 
once in every $\approx 20$ days from the late June to the early August in 2015,
and judging from these results, the flux variation during the observation period should be
less than $10\%$.
For all the execution blocks J1604$-$4228 was used as the phase calibrator.
We used CASA pipeline version 4.3.1 to perform the data reduction and calibration,
and during the data reduction two blocks
were completely discarded for lacking calibration information.
After flagging aberrant data, the total on-source integration time was $2.94$ hours
and the number of $12~\mathrm{m}$ antennas involved were $37$ to $40$,
thereby forming a range of baselines 
between $25.05~\mathrm{m}$ and $1566.19~\mathrm{m}$.

The ALMA correlator was configured to store linear XX and YY polarizations
in four separate spectral windows. 
Two spectral windows were optimally configured for the continuum observation
with frequencies centered at $97.50~\mathrm{GHz}$ and $99.50~\mathrm{GHz}$,
and both had an effective bandwidth of $1.875~\mathrm{GHz}$.
The other two windows, each having 3840 channels, were centered 
at $109.78~\mathrm{GHz}$ and $110.19~\mathrm{GHz}$ to target
the $J=1-0$ line emission of C$^{18}$O and $^{13}$CO
with a spectral resolution of $15.259~\mathrm{kHz}$ ($\Delta v \approx 0.04~\mathrm{km~s^{-1}}$).

We used CASA version 5.1.0 to image the calibrated visibility and
combined the two wide band spectral windows to obtain a continuum centered at 
$98.5~\mathrm{GHz}$ with a total bandwidth of $4~\mathrm{GHz}$.
We then applied the multi-scale CLEAN algorithm with Briggs weighting (robust parameter $0.5$)
and deconvolution scale parameters of
$0$ (corresponding to a point source), $1$, and $2$ times the average beam size.
These parameters are constant throughout the imaging process.
In order to improve the signal-to-noise ratio, we performed self-calibration to the continuum image
as follows. 
First, we solved the gain phase of the initial CLEAN model for the continuum image
starting from a time interval equal to the total time duration of each scan (which is between 
$2~\mathrm{minutes}$ and $7~\mathrm{minutes}$) of the target,
followed by shorter time intervals in the order of $240~\mathrm{s}$, 
$120~\mathrm{s}$, and $60~\mathrm{s}$.
After every phase calibration,
we performed CLEAN to the continuum image to obtain a new model 
that could be used in the succeeding phase calibration.
Once the phase calibration was completed, we solved the gain amplitude of the last
phase-calibrated model at a time interval equal to the time duration of each scan.
Lastly, we applied the phase-calibrated and gain-calibrated models
to the visibility data of the $98.5~\mathrm{GHz}$ continuum and performed  final CLEAN  
to create an image of the continuum.
With self-calibration, the resulting noise of the continuum image is lowered by a factor of approximately five, and
a root mean square noise level of $\sigma = 9.6~\mu\mathrm{Jy~beam^{-1}}$ is reached.
The synthesized beam in full width at half maximum (FWHM) of the
final image is $\timeform{0.54''}\times\timeform{0.44''}$, oriented along $\mathrm{P.A.}=  78.1^\circ$.

Before imaging the $J=1-0$ line emission
of $^{13}$CO and C$^{18}$O, we applied the phase and amplitude solutions
that were derived from the
self-calibration to the visibility data of the continuum
to these CO visibility data.
We used Briggs weighting and multiscale deconvolution, similar to the continuum imaging.
We then smoothed the frequency channels to an
equivalent velocity resolution of $0.30~\mathrm{km~s^{-1}}$.
The velocity resolution is set at a slightly smaller value than 
the observed velocity dispersion,
which is approximately $0.4~\mathrm{km~s^{-1}}$
as shown in figures 3 and 4 of \citet{muto2015}
as well as in figure \ref{fig:moment_13C16O_030kms}(d) in this paper,
to reveal the emission with the best sensitivity.
Finally, we applied a CASA task \textit{imsmooth} to
smooth the image cubes of $^{13}$CO and C$^{18}$O
so that they had the same angular resolution
in the $98.5~\mathrm{GHz}$ continuum image,
i.e., $\timeform{0.54''}\times\timeform{0.44''}$ ($\mathrm{P.A.}=  78.1^\circ$).
The noise level is $\sigma = 2.0~\mathrm{mJy~beam^{-1}}$.

We also created a spectral cube of $\Delta v = 0.12~\mathrm{km~s^{-1}}$
for both the $^{13}$CO and C$^{18}$O line emissions
to derive their peak brightness temperature.
The beam size was also smoothed to match that of the $98.5~\mathrm{GHz}$ continuum image.
The noise level is $\sigma = 2.9~\mathrm{mJy~beam^{-1}}$.

\subsection{ALMA band 7 data}
The observational setup and calibration process of the Band 7 data
are described in detail by \citet{fukagawa2013} and \citet{muto2015}.
In this study, we use the calibrated images of the
$336~\mathrm{GHz}$ continuum
and the $^{13}$CO $J=3-2$ and C$^{18}$O $J=3-2$ lines
produced by \citet{muto2015}. The velocity resolution of the line images 
is $0.12~\mathrm{km~s^{-1}}$. 
The synthesized beams of the Band 7 images are smaller than that from Band 3,
so we applied \textit{imsmooth} to the Band 7 images 
to obtain a spatial resolution identical to that of the Band 3 images, i.e.,
a synthesized beam of 
$\timeform{0.54''}\times\timeform{0.44''}$ ($\mathrm{P.A.}=  78.1^\circ$)
The resultant noise rms was $150~\mu\mathrm{Jy~beam^{-1}}$ for the continuum,
and $5.9~\mathrm{mJy~beam^{-1}}$ and $7.5~\mathrm{mJy~beam^{-1}}$
for the $^{13}$CO and C$^{18}$O image cubes, respectively.

\section{Results}\label{sec:observational_results}
\subsection{$98.5~\mathrm{GHz}$ continuum emission}\label{sec:98.5_continuum}
\begin{figure*}[ht!]\centering
\begin{tabular}{ccc}
\includegraphics[width=0.3\textwidth]{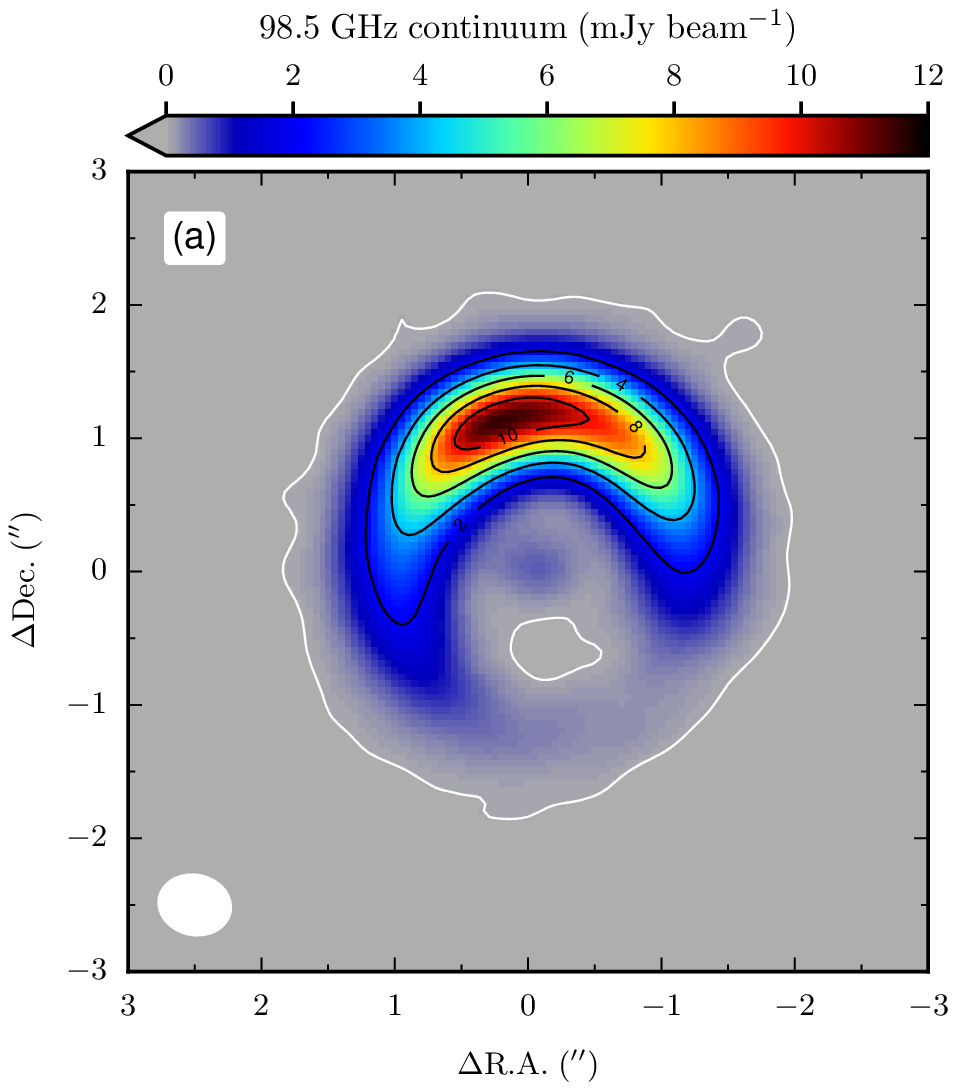} &
\includegraphics[width=0.3\textwidth]{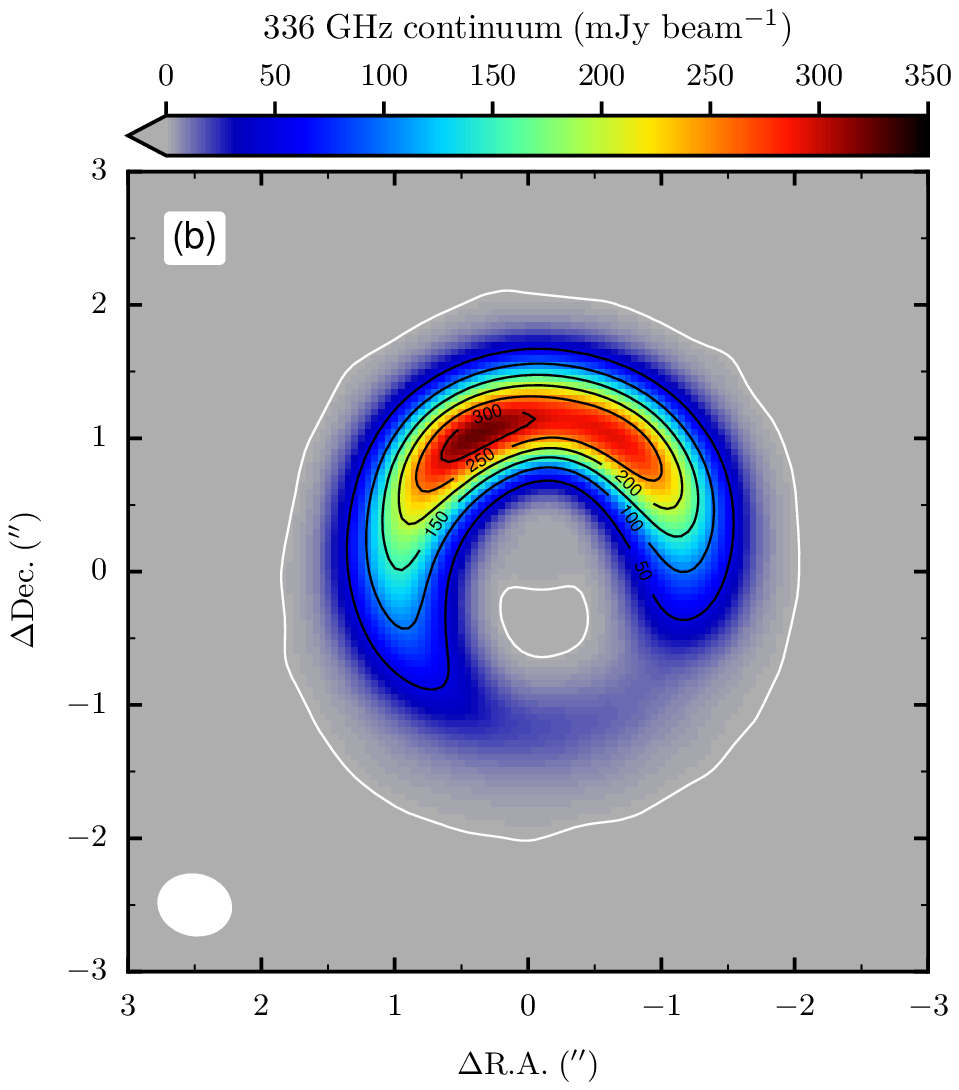} &
\includegraphics[width=0.3\textwidth]{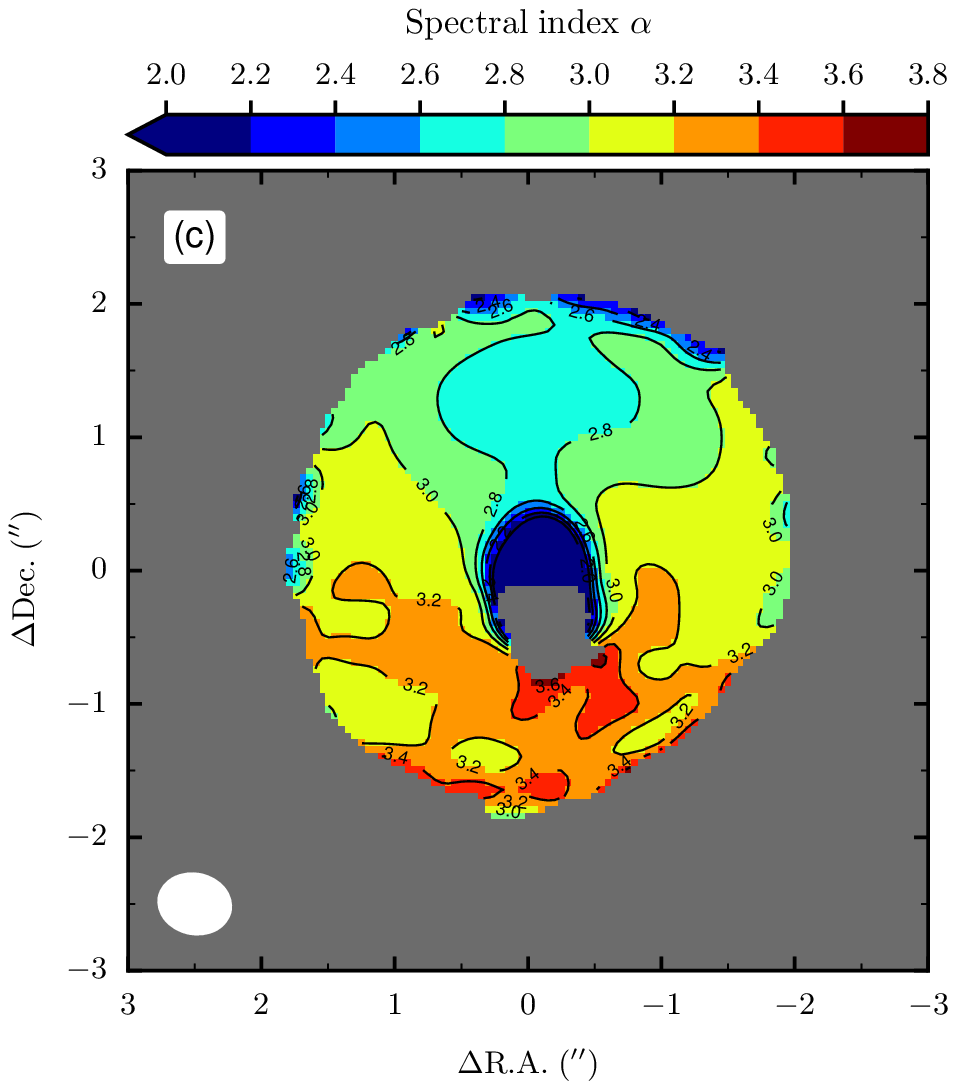} \\
\includegraphics[width=0.3\textwidth]{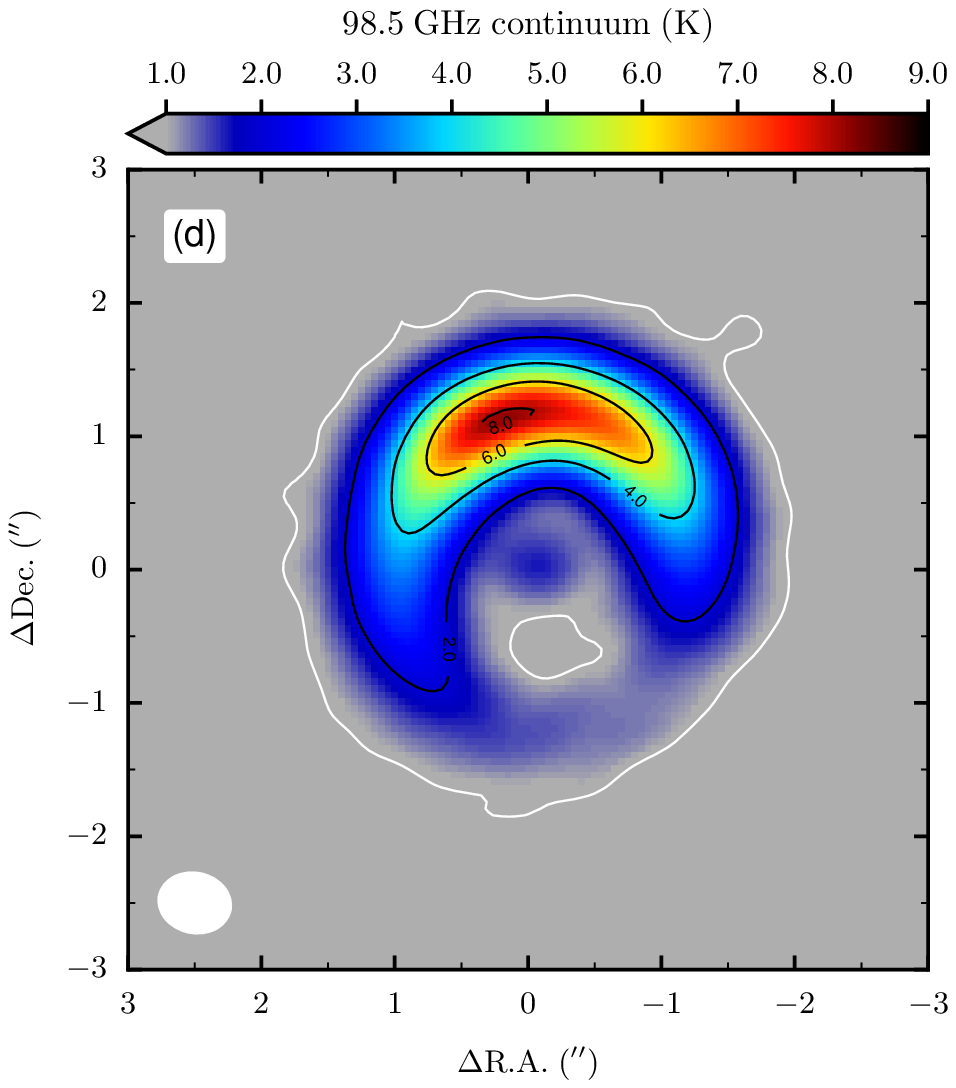} &
\includegraphics[width=0.3\textwidth]{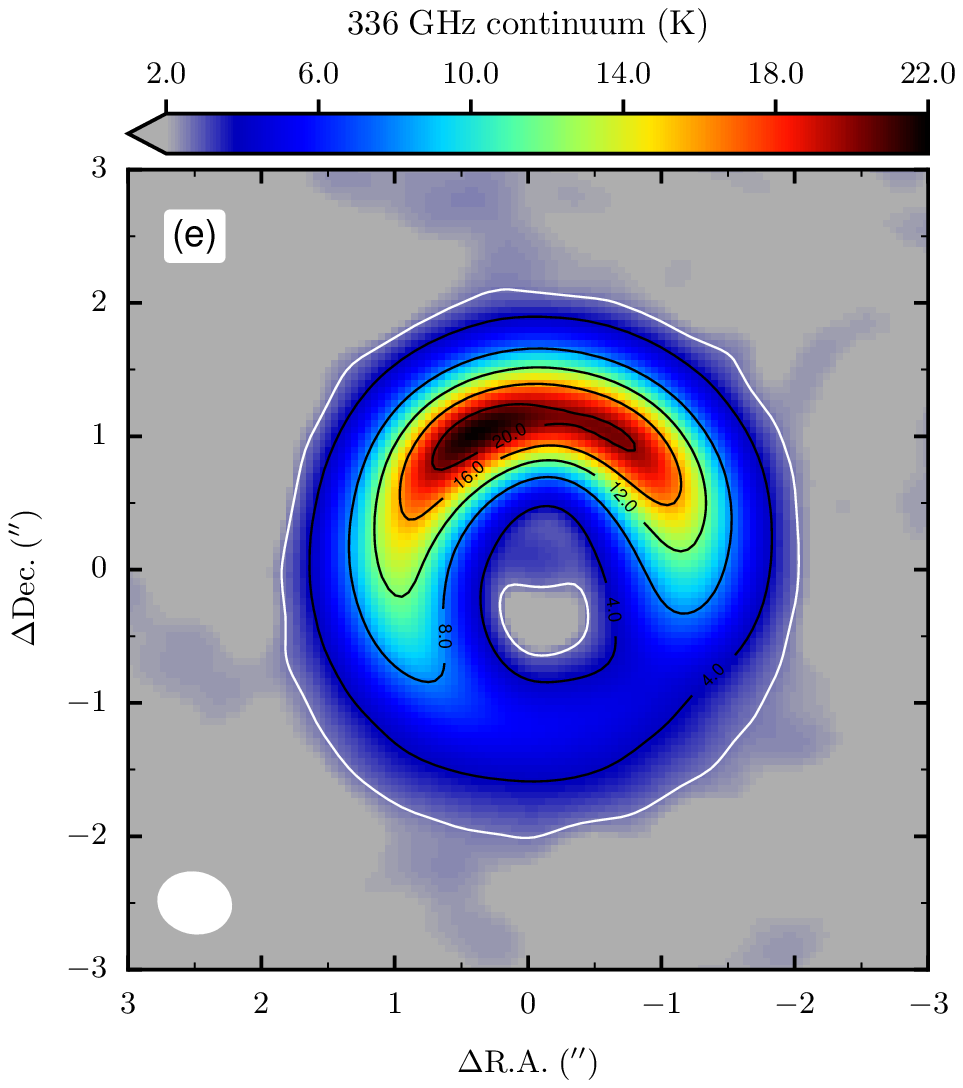} &
\end{tabular}
\caption{
Images of the $98.5~\mathrm{GHz}$ (left panels) and $336~\mathrm{GHz}$ (middle panels) dust continuum emission;
panels (a) and (b) show the emission in units of $\mathrm{mJy~beam^{-1}}$,
while panels (d) and (e) in units of $\mathrm{Kelvin}$.
North is up and east is to the right. 
The white contours denote the $5\sigma$ level, which is
$48~\mu\mathrm{Jy~beam^{-1}}$ or $0.91~\mathrm{K}$ for the $98.5~\mathrm{GHz}$ images
and $750~\mu\mathrm{Jy~beam^{-1}}$ or $2.63~\mathrm{K}$ for the $336~\mathrm{GHz}$ images.
The spectral index $\alpha$ derived from these dust continuum emission is shown in panel (c).
The white ellipse at the bottom left corner indicates the synthesized beam
($\timeform{0.54''}\times\timeform{0.44''}$, $\mathrm{P.A.}=  78.1^\circ$).
}\label{fig:continuum} \end{figure*}
\begin{figure}[ht!]\centering
\includegraphics[width=0.45\textwidth]{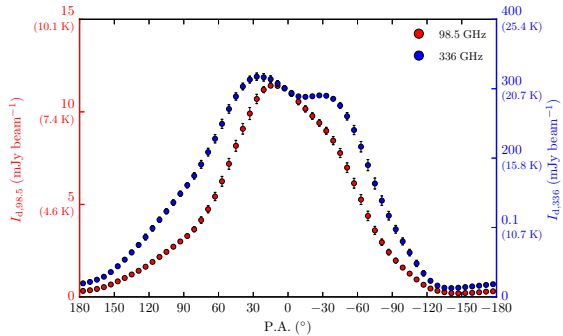}
\caption{
Dust continuum emission at $98.5~\mathrm{GHz}$ (red plots, left axis) and at $336~\mathrm{GHz}$ (blue plots, right axis) on the ridge
plotted as a function of $\mathrm{P.A.}$.
The values in the parentheses denote the equivalent brightness temperature derived from the beam size of $0\farcs54 \times 0\farcs44$.
The error bars denote the standard deviation of the averaged intensity.
} \label{fig:PA_peakIv} \end{figure}
\begin{table*}[!ht]\centering
\caption{Maximum and minimum of the $98.5~\mathrm{GHz}$ and the $336~\mathrm{GHz}$ continuum emission along the ridges.}
\begin{threeparttable}\centering
\begin{tabular}[t]{lcccccc}\toprule
                  && \multicolumn{2}{c}{$98.5~\mathrm{GHz}$} && \multicolumn{2}{c}{$336~\mathrm{GHz}$}   \\
\cmidrule{3-4}\cmidrule{6-7}
                  && Max &Min &&Max&Min \\
\midrule
Coordinates $(r, \mathrm{P.A.})$\tnote{$^\dagger$}                       && $(1\farcs1,15^{\circ})$   & $(1\farcs4,213^{\circ})$ && $(1\farcs0,27^{\circ})$    & $(1\farcs2,219^{\circ})$    \\
$F_\mathrm{d}$ $\mathrm{(mJy~beam^{-1})}$\tnote{$^\ddagger$}  && $1.14\times10$      & $1.96\times10^{-1}$      && $3.17\times10^{2}$       & $1.27\times10^{1}$       \\
$T_\mathrm{B}$ $\mathrm{(K)}$\tnote{$^\ddagger$}                     && $8.1$                           & $1.2$                           && $21.5$                          & $4.8$                            \\
\cmidrule{1-7}
$\mathrm{Contrast} \equiv \mathrm{Max/Min}$	                   && \multicolumn{2}{c}{$58.2$}                                          && \multicolumn{2}{c}{$25.0$}                \\                
\bottomrule
\end{tabular}
\begin{tablenotes}\footnotesize
\item[$^\dagger$] The coordinates $(r,\mathrm{P.A.})$ indicate the center of an area of radial size $0\farcs1$ and angular size $6^\circ$.  
\item[$^\ddagger$] The standard deviations (not shown) of flux density $F_\mathrm{d}$ and brightness temperature $T_\mathrm{B}$
                               in the area centered at $(r,\mathrm{P.A.})$ are less than $1\%$ of their mean values.
\end{tablenotes} \end{threeparttable} \label{tab:continuum_ridge} \end{table*}

The dust continuum emissions of the disk surrounding HD~142527 at $98.5~\mathrm{GHz}$ ($\lambda = 3.0~\mathrm{mm}$) and 
$336~\mathrm{GHz}$ ($\lambda = 0.89~\mathrm{mm}$) are 
shown in figure \ref{fig:continuum}.
The bottom panels in figure \ref{fig:continuum}
present the dust continuum emission 
in terms of brightness temperature $T_\mathrm{B}$ (in units of $\mathrm{Kelvin}$),
which was converted from the flux density $F_\mathrm{d}$ (in units of $\mathrm{mJy~beam^{-1}}$) 
using the inverse of the Planck function
\begin{equation}
\label{eq:inverse_planck}
T_\mathrm{B} = \frac{h\nu}{k} \left[ \ln\left( \frac{2h\nu^3}{c^2 F_\mathrm{d}/ \Omega} +1  \right) \right]^{-1},
\end{equation}
where $c$, $h$, and $k$ denote
the speed of light, the Planck constant, and the Boltzmann constant, respectively.
The solid beam angle $\Omega$ is defined as 
\begin{equation}
\Omega = \frac{\pi \theta_\mathrm{maj} \theta_\mathrm{min}}{4\ln(2)}, \nonumber
\end{equation}
where $\theta_\mathrm{maj}$ and $\theta_\mathrm{min}$ are the FHWM of the beam major
and minor axes, respectively.
The signal-to-noise ratio of the peak continuum flux at $98.5~\mathrm{GHz}$ is $240$.
The spatially integrated flux at $98.5~\mathrm{GHz}$,
considering only emission above $5\sigma$ level, is $72.1~\mathrm{mJy}$.
On the other hand, the integrated flux at $86.3~\mathrm{GHz}$ (equivalent wavelength $3.476~\mathrm{mm}$) 
obtained with the Australia Telescope Compact Array (ATCA) at a coarser beam of 
$16\farcs2 \times 2\farcs9$ is $47.1~\mathrm{mJy} \pm 6.5~\mathrm{mJy}$, 
and the spectral index between $800~\mu\mathrm{m}$ and $3.476~\mathrm{mm}$
is derived to be $\approx 3$ \citep{verhoeff2011}.
With this spectral index, the integrated flux at $98.5~\mathrm{GHz}$ corresponds
to $48.5~\mathrm{mJy}$ at $86.3~\mathrm{GHz}$, indicating that the ALMA observation recovers all the flux.

Both $98.5~\mathrm{GHz}$ and $336~\mathrm{GHz}$
continuum images reveal an asymmetric ring-like structure in their emission.
The FWHM radial width of the continuum emission from the outer disk is approximately $100~\mathrm{au}$,
which means that the observation beam (linear scale $\approx 76~\mathrm{au}$)
marginally resolves the disk in the radial direction.
The $98.5~\mathrm{GHz}$ continuum emission
shows a radial extent of approximately $2~\mathrm{arcsec}$ above the $5\sigma$ level,
and it shares a similar distribution with that observed at $336~\mathrm{GHz}$, i.e.,
the outer disk exhibits a crescent-like structure as a result of 
the concentration of dust in the northern region.
Hereafter, we use the word \textit{ridge} to refer to the line that connects the radial peak of a physical 
quantity in every $\mathrm{P.A.}$ direction (as seen from the central star) on the outer disk.
To search for the maximum and minimum values along the ridge,
the averaged values 
from an area of radial size $0\farcs1$ and angular size $6^\circ$
with coordinates indicated by $(r,\mathrm{P.A.})$ will be used.
The maximum and minimum values of the continuum emission along the ridge
are listed in Table \ref{tab:continuum_ridge}.
The contrast of the $98.5~\mathrm{GHz}$ emission along its ridge
is approximately $58$
and is higher than that at $336~\mathrm{GHz}$, which is $25$ (see also \cite{fukagawa2013}).
This difference is because of the lower optical depth at $98.5~\mathrm{GHz}$;
the optical depth at $336~\mathrm{GHz}$ in the northern region of the disk 
is $\gtrsim 1$, hence its emission is saturated \citep{casassus2015}.
The lower optical depth at $98.5~\mathrm{GHz}$
also accounts for the lower peak $T_\mathrm{B}$,
i.e., $8.1~\mathrm{K}$, compared to
$21.5~\mathrm{K}$ at $336~\mathrm{GHz}$.

The spectral index $\alpha$ is defined as 
\begin{equation}
\label{eq:beta}
\alpha \equiv \left.  \log \left[ \frac{ F_\mathrm{d,336}}{F_\mathrm{d,98.5}} \right] \right/
              \log \left[   \frac{336~\mathrm{GHz}}{98.5~\mathrm{GHz} }     \right],
\end{equation}  
and is shown in figure \ref{fig:continuum}(c).
The index varies smoothly in the azimuthal direction;
in
the northern region $\alpha \approx 2.8$,
while in the southern region $\alpha \approx 3.4$.
When the Rayleigh-Jeans approximation is valid,
the flux density is proportional to
$\nu^{2+\beta}$,
where $\beta$ is the dust opacity spectral index.
The smaller value of $\alpha$ might indicate a smaller $\beta$ \citep{beckwith1990,beckwith1991,miyake1993},
but the spectral slope also gets flatter as the optical depth at $336~\mathrm{GHz}$ gets higher.
We will derive $\beta$ and optical depth from the dust continuum images at $98~\mathrm{GHz}$
and $336~\mathrm{GHz}$ in Section \ref{sec:analyses_dust}. 

Figure \ref{fig:PA_peakIv} shows the flux density of the continuum emission 
on the ridge as a function of $\mathrm{P.A.}$.
A dip in intensity at $\mathrm{P.A.} \sim 0^\circ$ is seen at $336~\mathrm{GHz}$,
rendering the emission morphology to be a double-peaked structure;
however, there is no dip at the same $\mathrm{P.A.}$ direction at $98.5~\mathrm{GHz}$.
The $336~\mathrm{GHz}$ emission
is so opaque that it
would be more sensitive to dust temperature
than dust surface density;
the dip may therefore reflect a temperature drop in the disk atmosphere \citep{casassus2015}.
In fact, the intensity at infrared wavelengths is also lower
at $\mathrm{P.A.} \sim 0^\circ$
(as well as at $\mathrm{P.A.} \sim 160^\circ$, see \cite{avenhaus2014})
and is thought to be the result of a shadow from 
a warped inner disk \citep{marino2015}.
On the other hand, the $98.5~\mathrm{GHz}$ emission
is optically thin and should also depend on the dust surface density;
here, the shadow is not apparent.

\subsection{$^{13}$CO $J=1-0$ line emission} \label{sec:13C16O_observations}
\begin{figure*}\centering
\begin{tabular}{cc}
\includegraphics[width=0.3\textwidth]{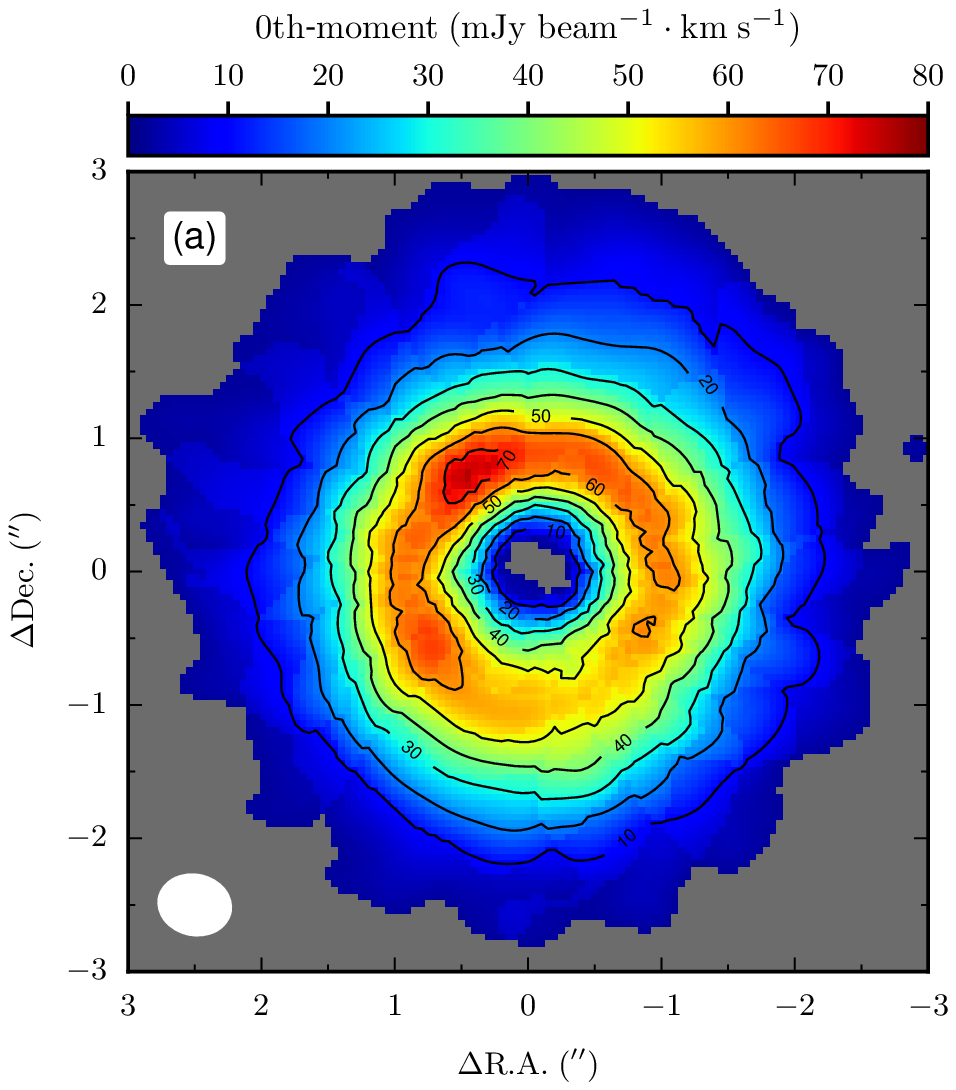} &
\includegraphics[width=0.3\textwidth]{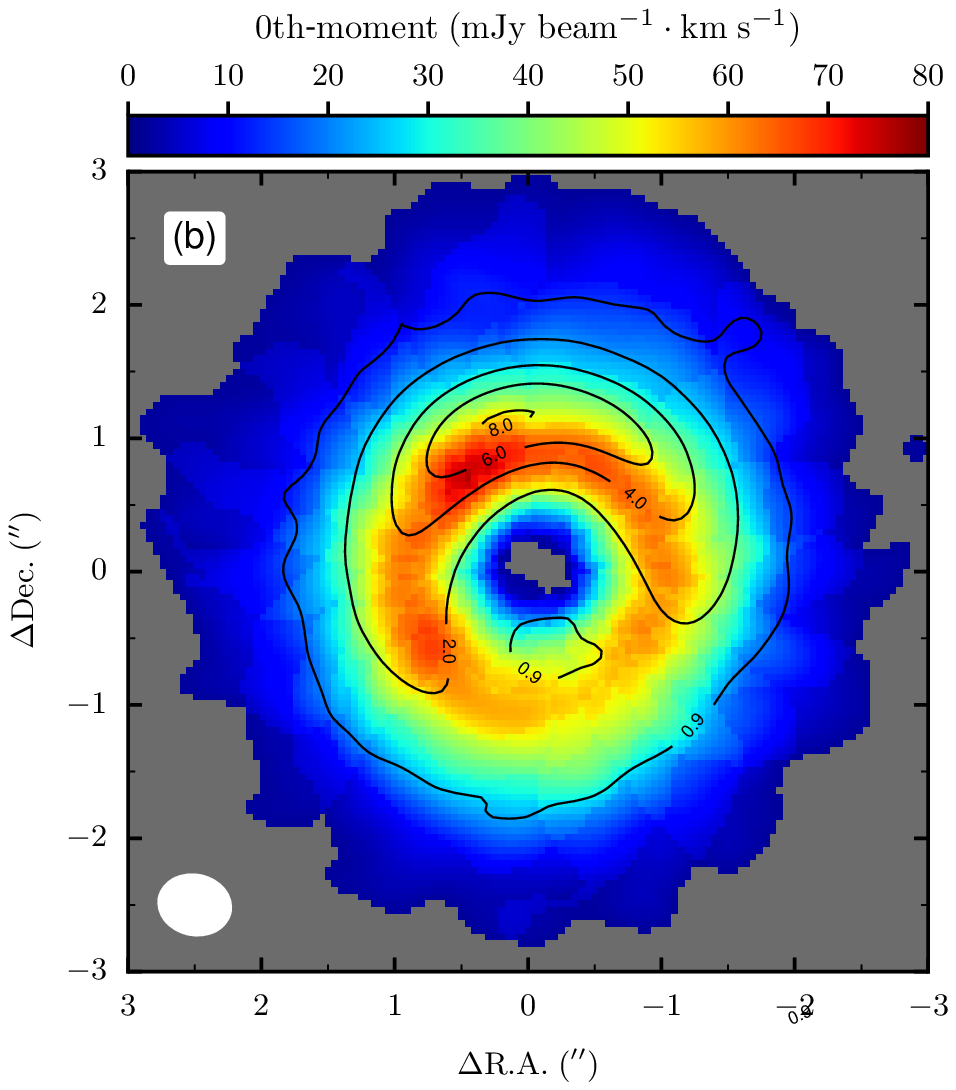} \\
\includegraphics[width=0.3\textwidth]{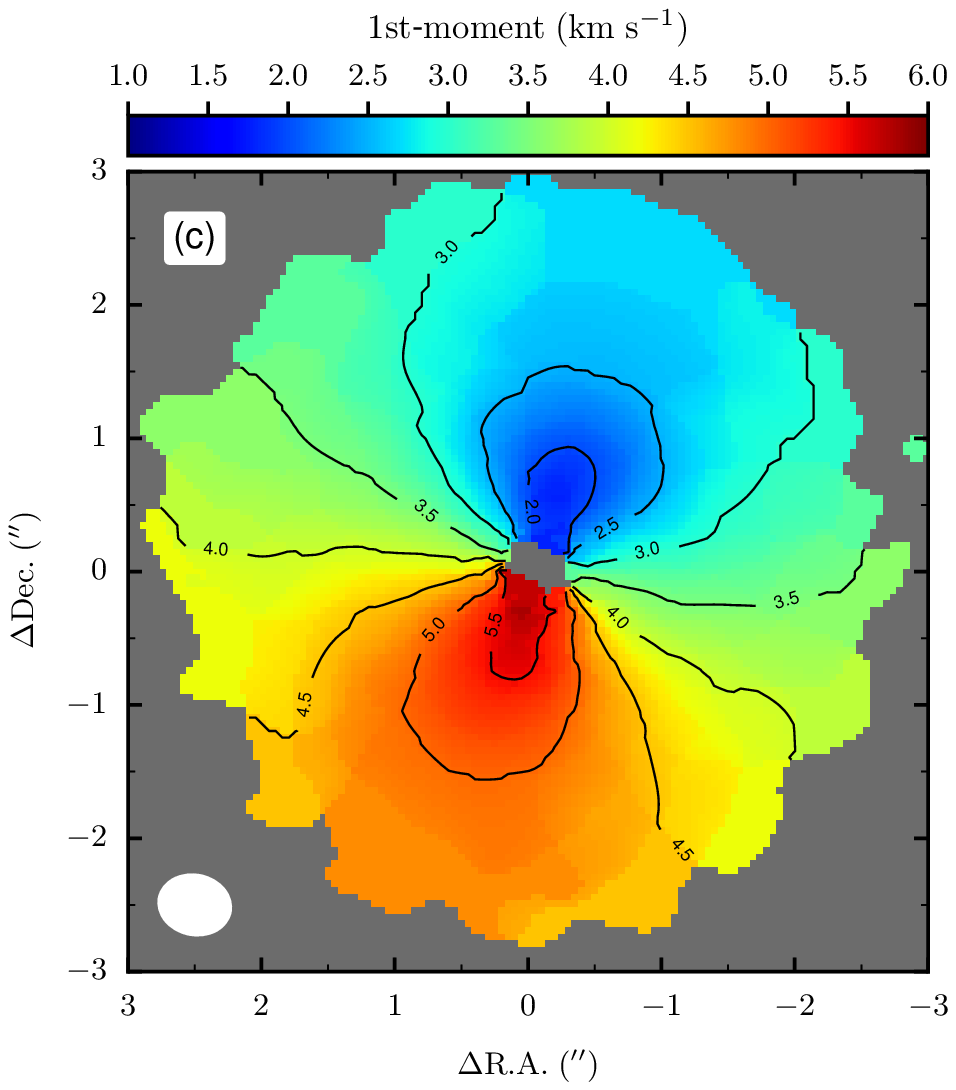} &
\includegraphics[width=0.3\textwidth]{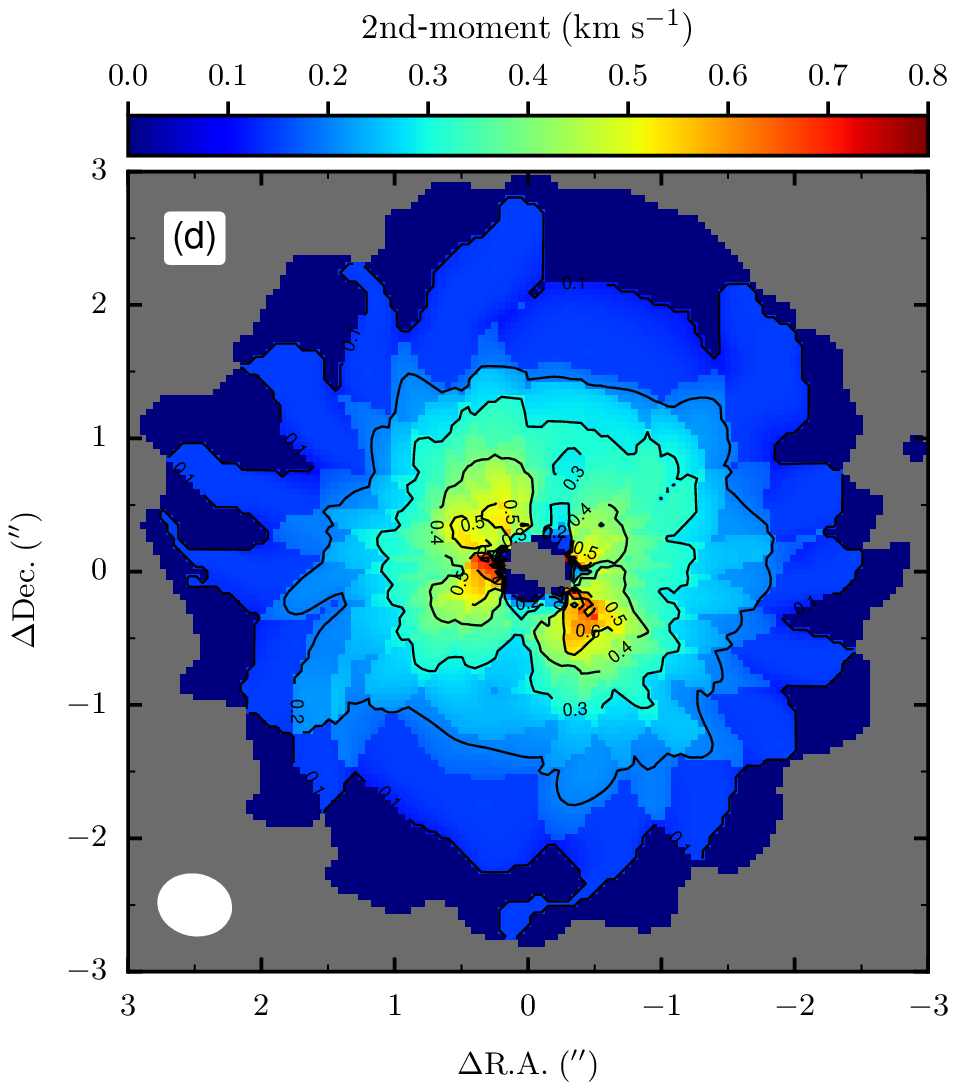}
\end{tabular}
\caption{
Moment maps of the $^{13}$CO $J=1-0$ line emission
created from the image cube of spectral resolution $\Delta v = 0.30~\mathrm{km~s^{-1}}$. 
The white ellipse in the left bottom corner indicates the synthesized beam,
which is identical to that of the $98.5~\mathrm{GHz}$ continuum image
($\timeform{0.54''}\times\timeform{0.44''}$, $\mathrm{P.A.}=  78.1^\circ$).
The moment maps are created from velocity channels in the range of $v_\mathrm{lsr} = 1.2~\mathrm{km~s^{-1}} - 6.3~\mathrm{km~s^{-1}}$
(see Appendix \ref{sec:channel_map}) after applying a Keplerian mask in the channels and
then clipping emissions lower than $3.5\sigma$ ($\sigma = 2.0~\mathrm{mJy~beam^{-1}}$).
In panel (b), the $98.5~\mathrm{GHz}$ continuum emission shown in figure \ref{fig:continuum}(d)
are superimposed as contours on the 0th-moment image.
}\label{fig:moment_13C16O_030kms}\end{figure*}
\begin{figure*}[h!]\centering
\begin{tabular}{cc}
\includegraphics[width=0.33\textwidth]{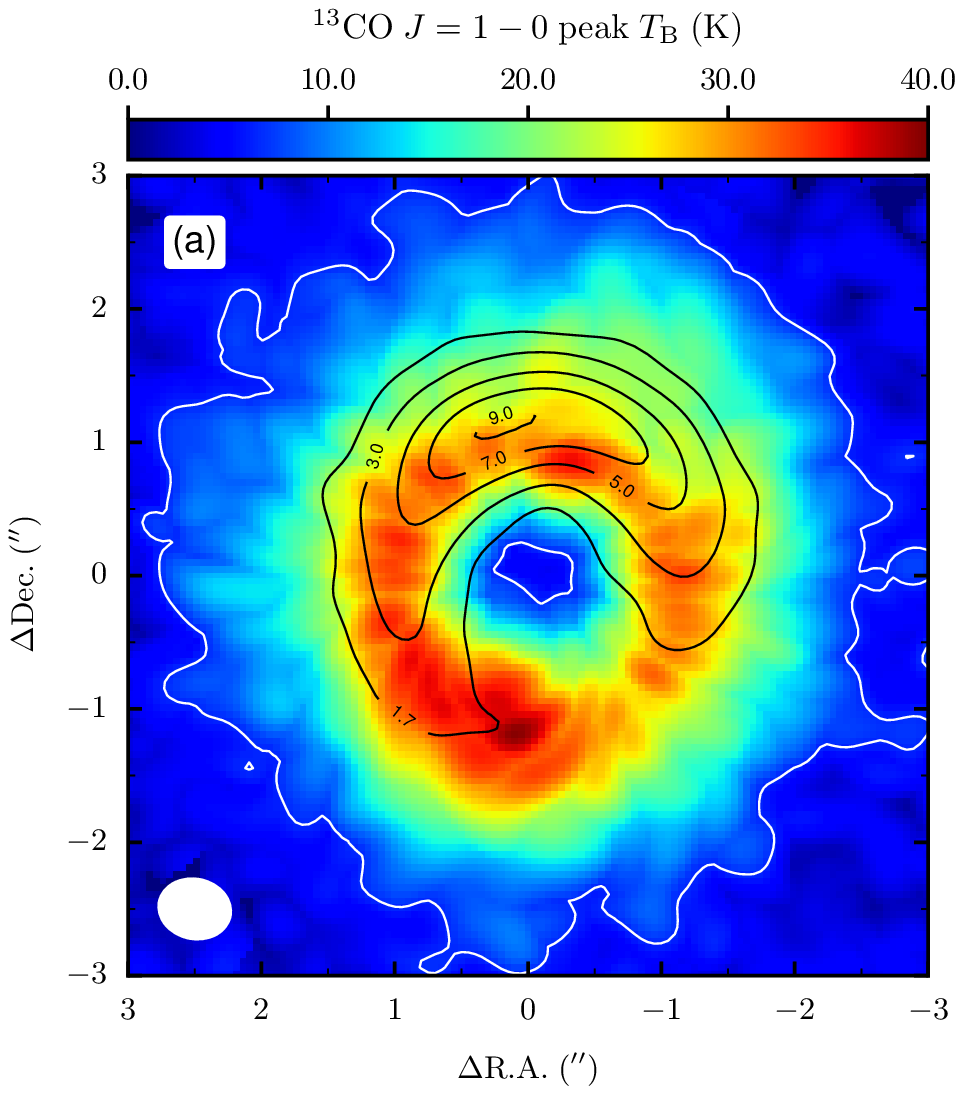} &
\includegraphics[width=0.33\textwidth]{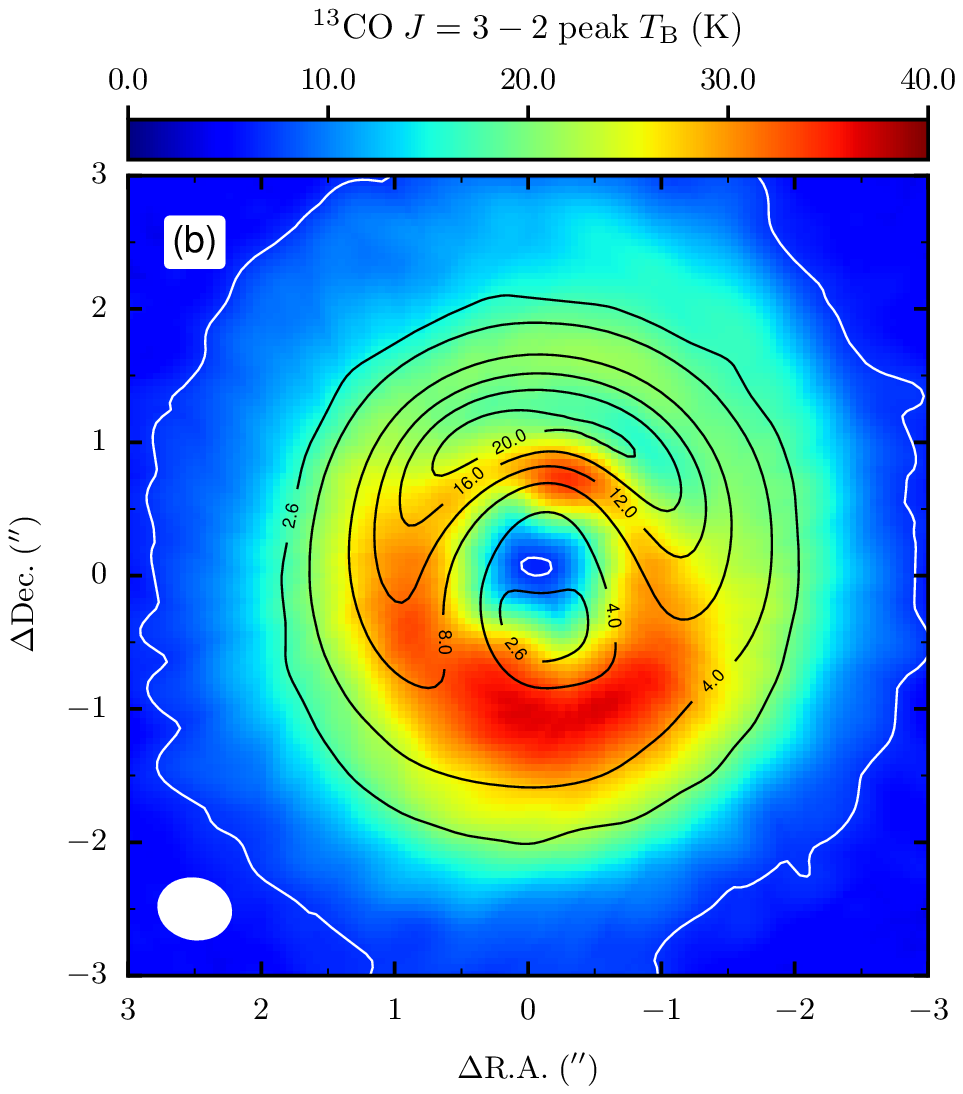} \\
\includegraphics[width=0.33\textwidth]{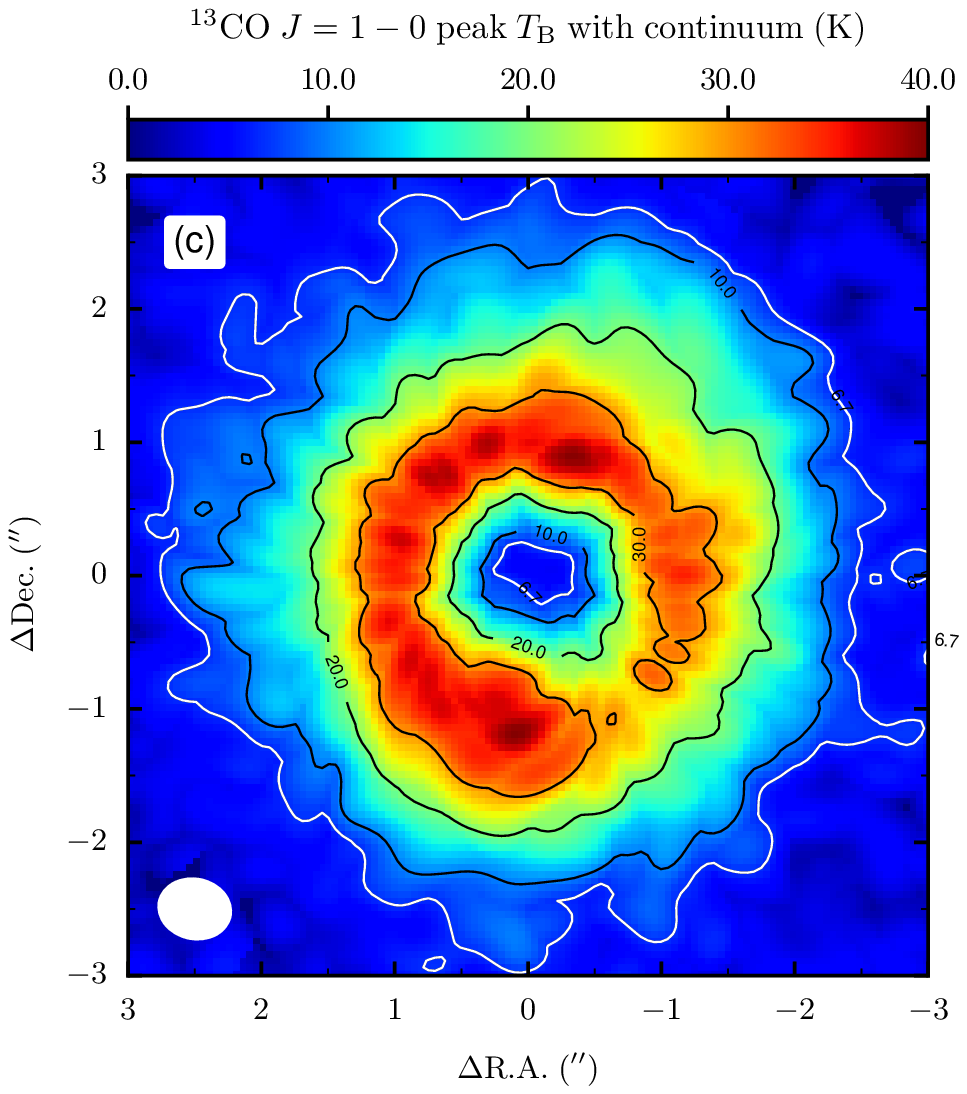} &
\includegraphics[width=0.33\textwidth]{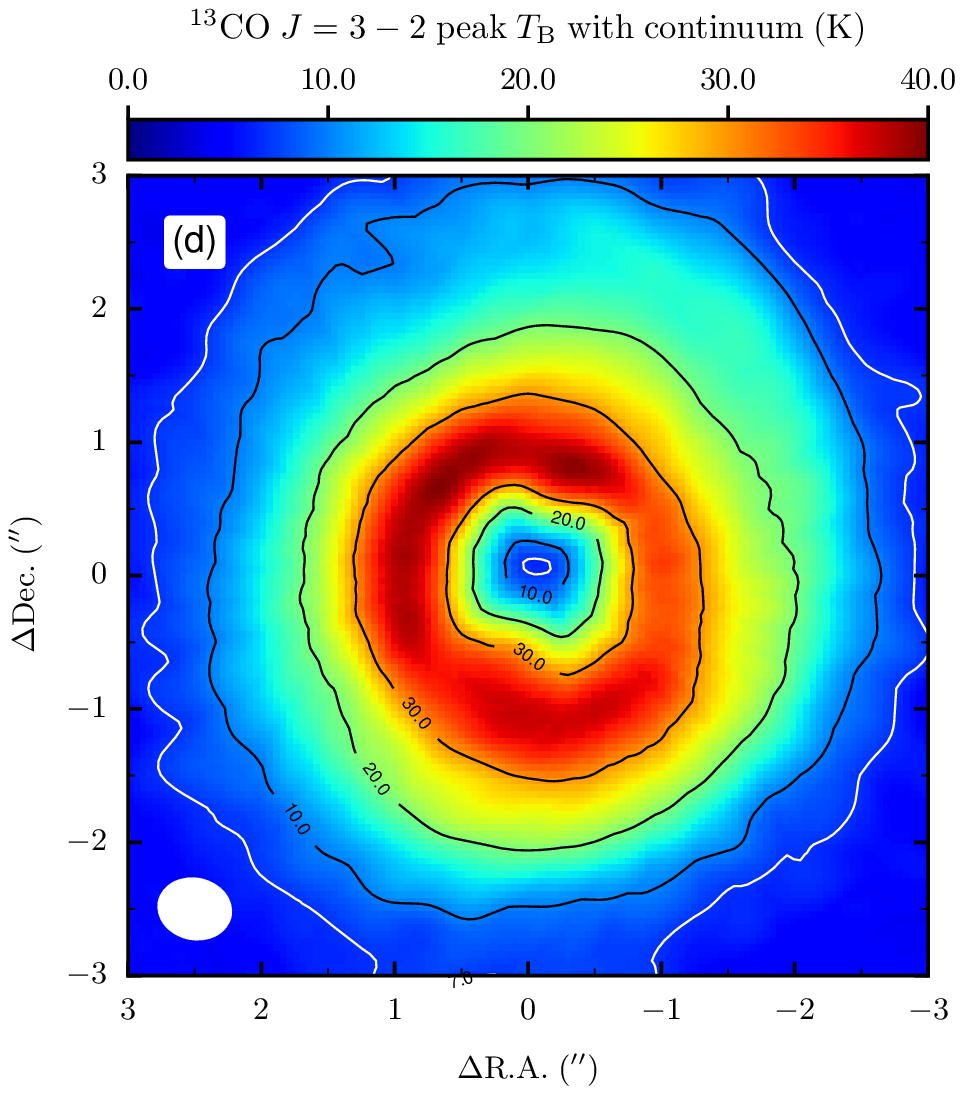}
\end{tabular}
\caption{
Peak surface brightness of $^{13}$CO;
the left and right panels show the peak $T_\mathrm{B}$ of the $J=1-0$ and $J=3-2$ line emission, respectively.
The maps in the top row are the peak $T_\mathrm{B}$ after subtracting the continuum.
The black contours in panel (a) denote the narrow-band continuum emission (see text for details),
while those in panel (b) denote the $336~\mathrm{GHz}$ continuum emission as shown in figure \ref{fig:continuum}(d).
The maps in the bottom row are the peak $T_\mathrm{B}$
with the continuum emission. In panels (a) and (c), the white contours
denote the $3.5\sigma$ level, which are $6.56~\mathrm{K}$ and $6.74~\mathrm{K}$, respectively.
In panels (b) and (d), the white contours denote the $5\sigma$ level,
which are $6.91~\mathrm{K}$ and $6.97~\mathrm{K}$, respectively.
}\label{fig:13co_peakTB}\end{figure*}

Figure \ref{fig:moment_13C16O_030kms}
displays the moment maps of the $^{13}$CO $J=1-0$ line emission.
The moment maps are created by applying a Keplerian masking and a noise masking to the channel maps
(see Appendix \ref{sec:channel_map}).

Compared to the $98.5~\mathrm{GHz}$ continuum emission,
the $^{13}$CO 0th-moment is observed to have a wider radial extent of approximately $3~\mathrm{arcsec}$
and is more axisymmetric.
The contrast between the north and the south is approximately $1.4$;
the ridge in the 0th-moment map has a maximum at $(0\farcs8, 33^\circ)$
where the integrated intensity is 
$F_\mathrm{int} = 71.1~\mathrm{mJy~beam^{-1}~km~s^{-1}}$
and a minimum at $(1\farcs0, 213^\circ)$
where $F_\mathrm{int} =49.5~\mathrm{mJy~beam^{-1}~km~s^{-1}}$.
In the northern region,
the ridge of the
0th-moment is located inwards of that associated with the continuum emission,
which is owing to the higher optical depth of the $^{13}$CO line.

Figures \ref{fig:13co_peakTB}(a) and (c) show
the peak $T_\mathrm{B}$ of the $^{13}$CO $J=1-0$ line spectrum;
as mentioned in Section \ref{sec:observation_data_reduction},
these maps are created from the image cube of the velocity resolution of $0.12~\mathrm{km~s^{-1}}$.
Figure \ref{fig:13co_peakTB}(a) shows the peak $T_\mathrm{B}$ of the line emission after subtracting 
the continuum level;
the continuum level is estimated from the line free channels of the spectral window containing
the $^{13}$CO line emission,
which is centered at $110.2~\mathrm{GHz}$ and is denoted by black contours in the maps.
This continuum emission has a total bandwidth of $54~\mathrm{MHz}$.
On the other hand, figure \ref{fig:13co_peakTB}(c) shows the 
peak $T_\mathrm{B}$ of $^{13}$CO including the $110.2~\mathrm{GHz}$ continuum emission.
The continuum-subtracted peak $T_\mathrm{B}$ in figure \ref{fig:13co_peakTB}(a)
shows a dip in the northern region where the 
dust continuum emission is brightest.
This is similar to the $J=3-2$ line emission of
$^{12}$CO, $^{13}$CO (figure \ref{fig:13co_peakTB}b), and C$^{18}$O (figure \ref{fig:c18o_peakTB}b)
\citep{fukagawa2013,perez2015,boehler2017},
as well as that of HCN $J=4-3$ and CS $J=7-6$ \citep{vanderplas2014}.
As discussed by \citet{weaver2018},
the dip is due to a high optical depth of dust continuum
and/or molecular line emission.
The peak $T_\mathrm{B}$ of $^{13}$CO with the continuum emission included,
shown in figure \ref{fig:13co_peakTB}(d),
has a higher brightness temperature in the north
and there is no dip observed at the peak of the continuum emission.
In each $\mathrm{P.A.}$ direction,
the peak $T_\mathrm{B}$ of $^{13}$CO $J=1-0$ with the continuum included (figure \ref{fig:13co_peakTB}c) 
is in the range of $T_\mathrm{B} = (26 - 40)~\mathrm{K}$;
it is similar to
that of the $^{13}$CO $J=3-2$ (figure \ref{fig:13co_peakTB}d),
implying that both lines are optically thick at the center of the line.
An exception is seen in the region of $\mathrm{P.A.} = 200^\circ -240^\circ$
where the peak $T_\mathrm{B}$
of $^{13}$CO $J=1-0$ with continuum emission drops to $T_\mathrm{B} \approx 26~\mathrm{K}$,
compared to that of $^{13}$CO $J=3-2$, which is $T_\mathrm{B} \approx 36~\mathrm{K}$,
suggesting that the $J=1-0$ line emission in this region is optically thin.

\subsection{C$^{18}$O $J=1-0$ line emission} \label{sec:12C18O_observations}
\begin{figure*}\centering
\begin{tabular}{cc}
\includegraphics[width=0.33\textwidth]{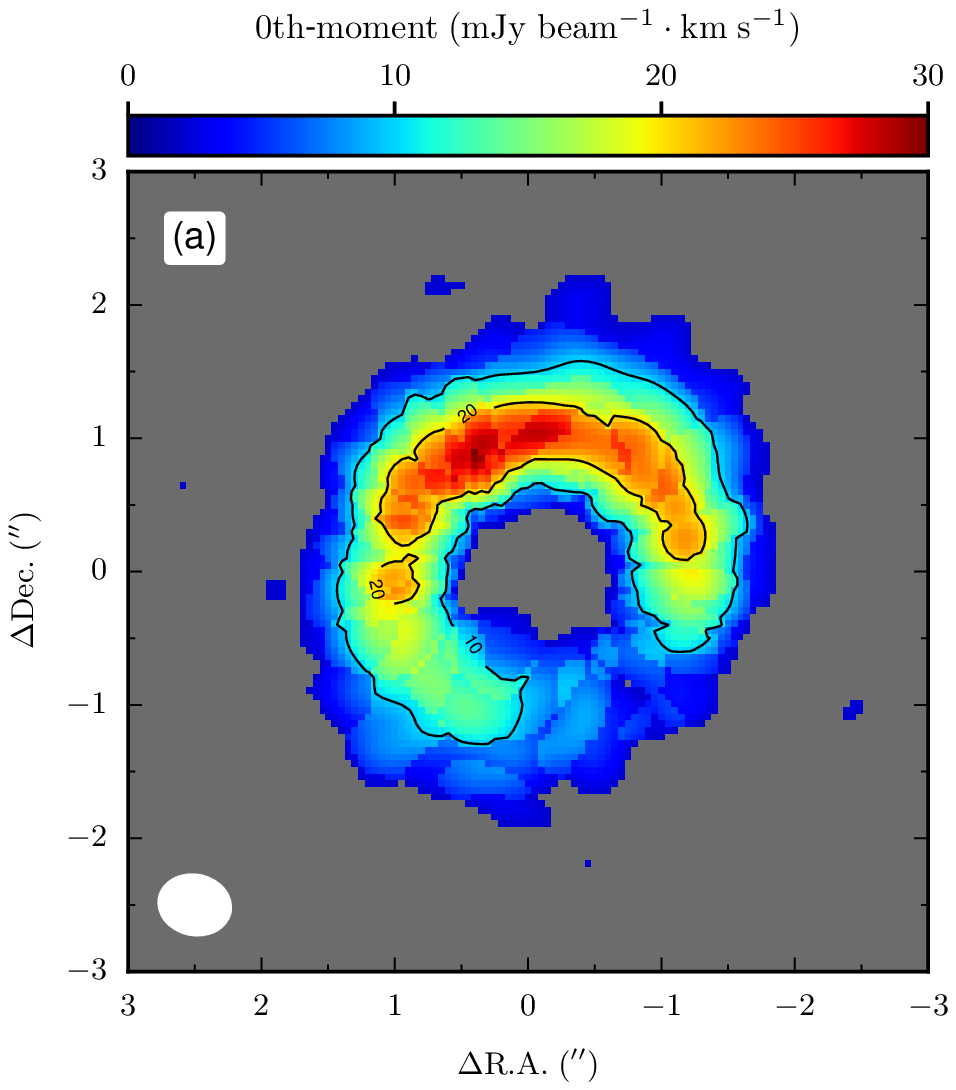} &
\includegraphics[width=0.33\textwidth]{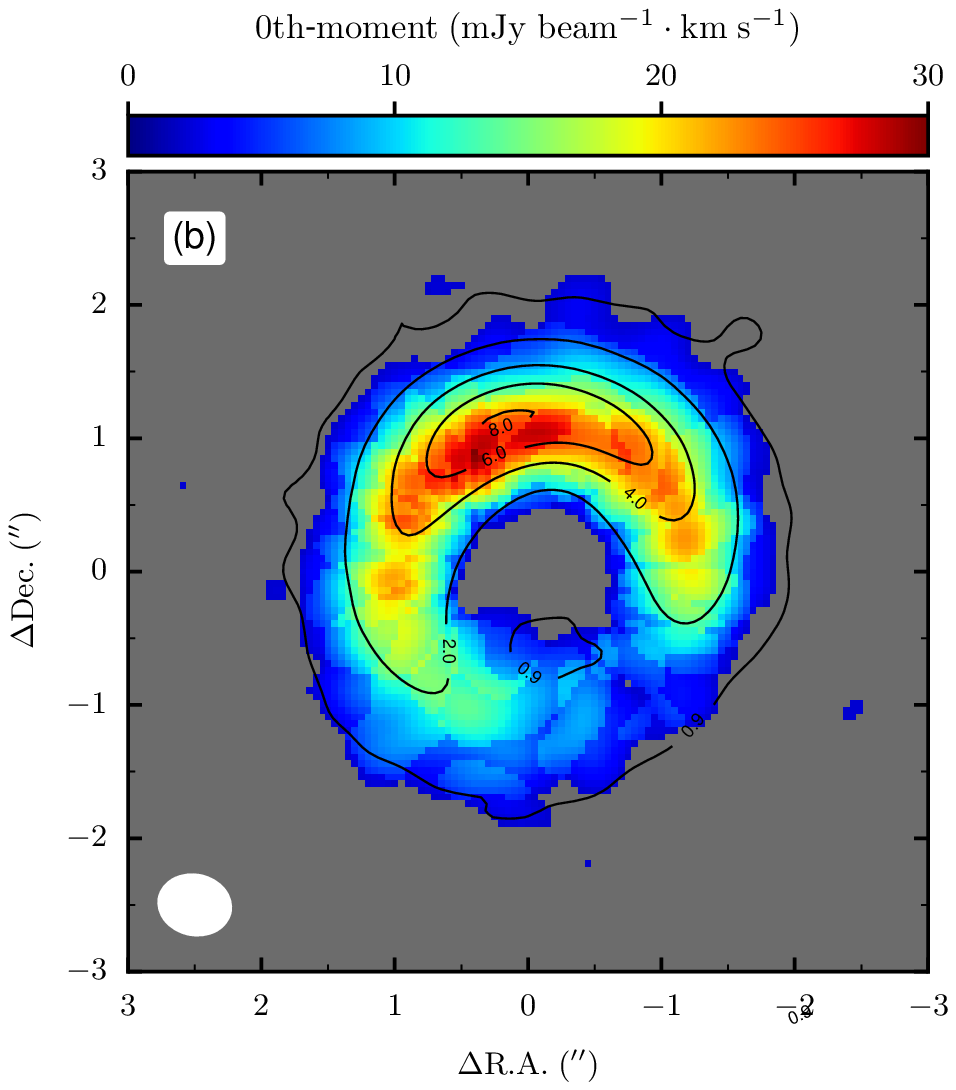} \\
\includegraphics[width=0.33\textwidth]{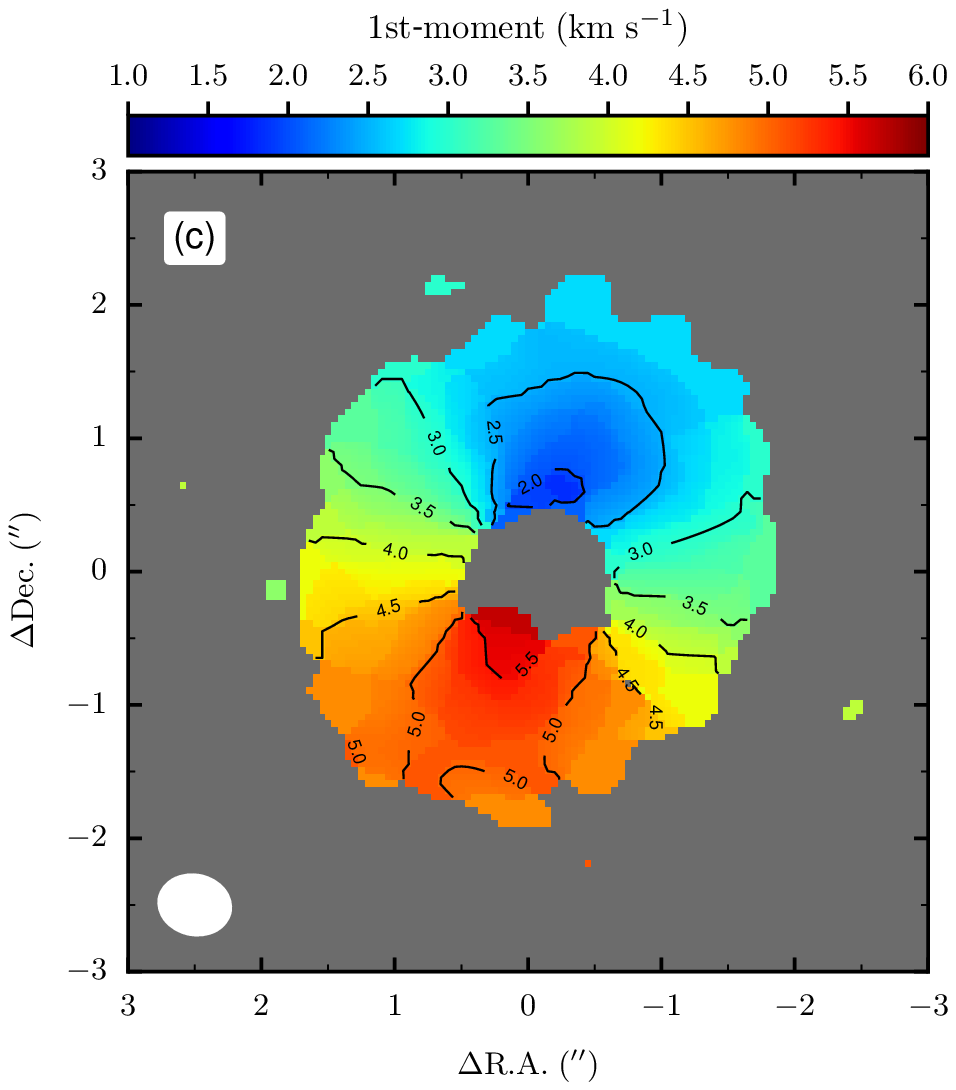} &
\includegraphics[width=0.33\textwidth]{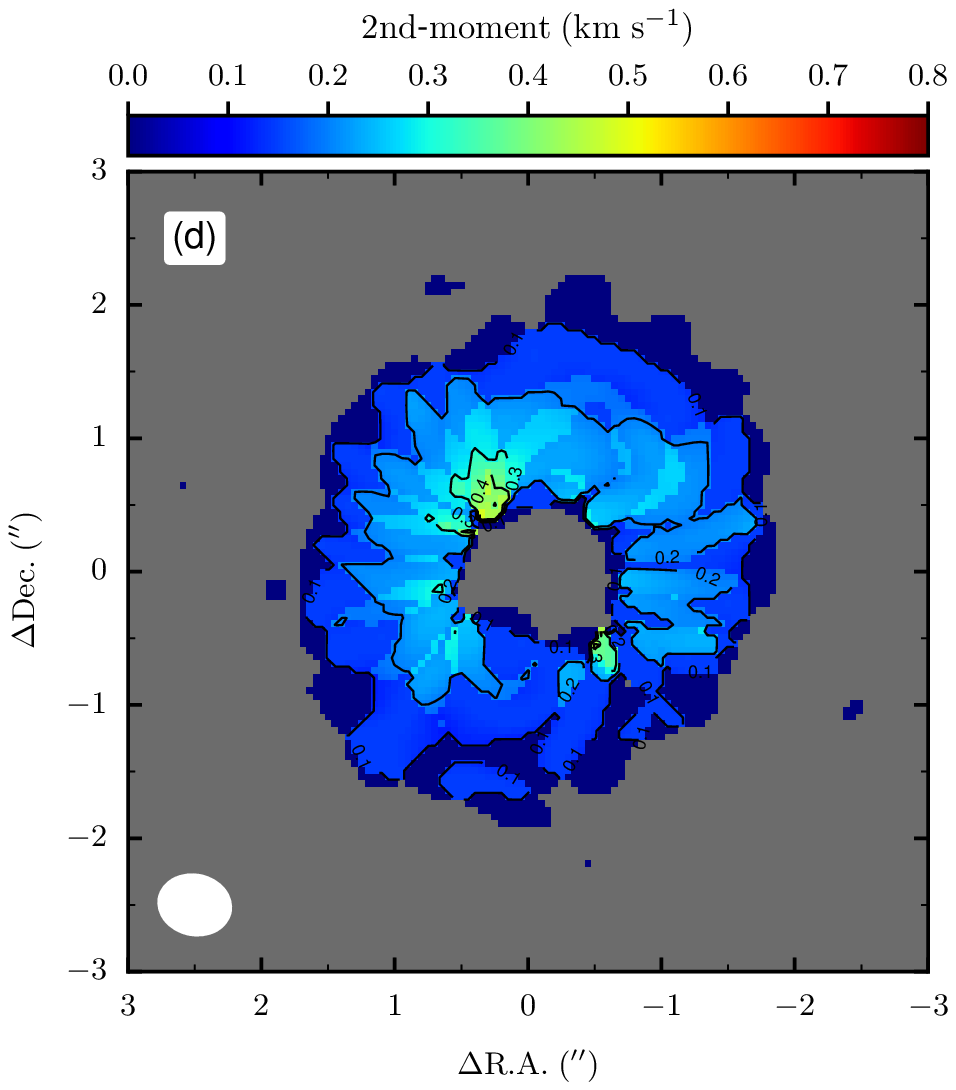} 
\end{tabular}
\caption{
Similar to figure \ref{fig:moment_13C16O_030kms}, for the C$^{18}$O $J=1-0$ line emission.
The spectral resolution is $\Delta v = 0.30~\mathrm{km~s^{-1}}$.
The synthesized beam has the same size and shape as that of the $98.5~\mathrm{GHz}$ continuum image
($\timeform{0.54''}\times\timeform{0.44''}$, $\mathrm{P.A.}=  78.1^\circ$)
The moment maps are created from velocity channels in the range of $v_\mathrm{lsr} = (1.5 - 5.7)~\mathrm{km~s^{-1}}$
(see Appendix \ref{sec:channel_map}),
after applying a Keplerian mask in the channels and
then clipping emissions lower than
$3.5\sigma$ ($\sigma = 2.9~\mathrm{mJy~beam^{-1}}$).
}\label{fig:moment_12C18O_030kms}\end{figure*}
\begin{figure*}[h!]\centering
\begin{tabular}{cc}
\includegraphics[width=0.33\textwidth]{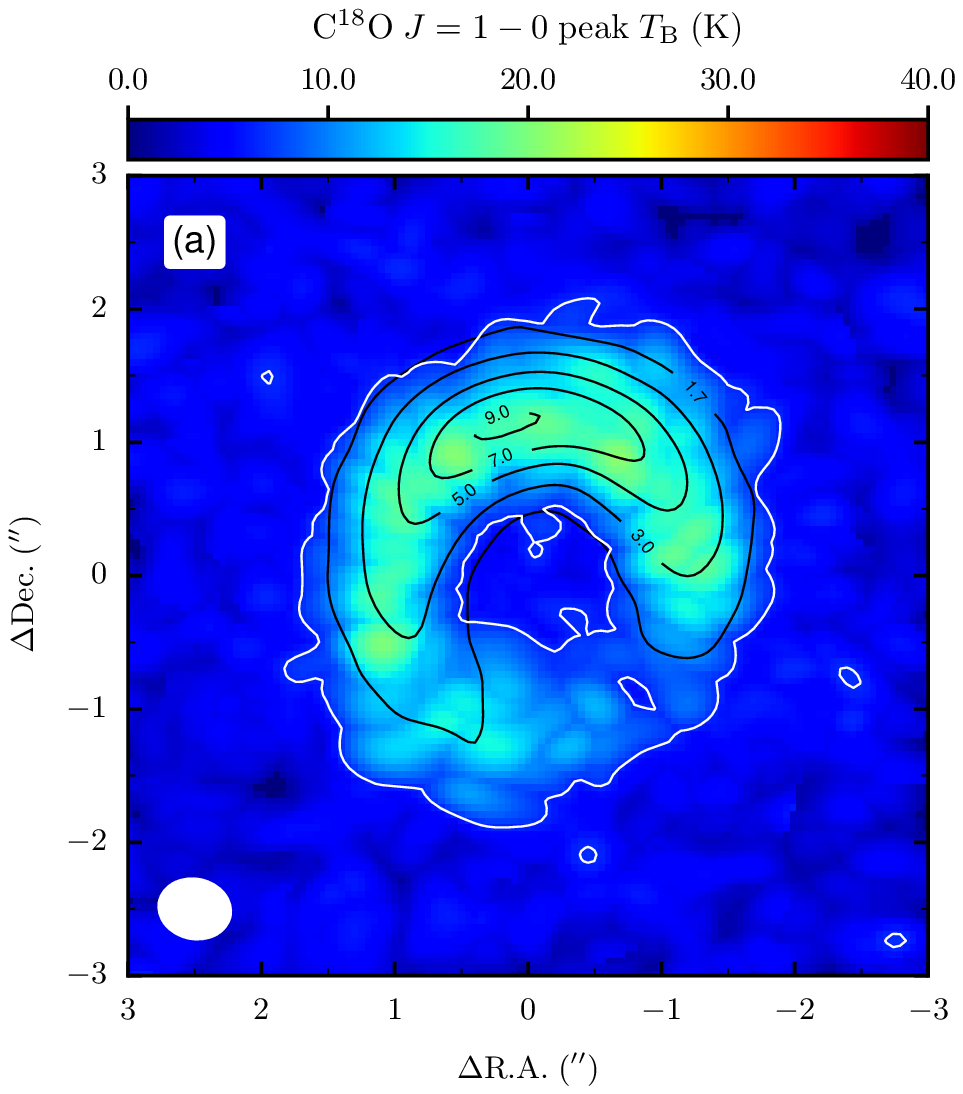} &
\includegraphics[width=0.33\textwidth]{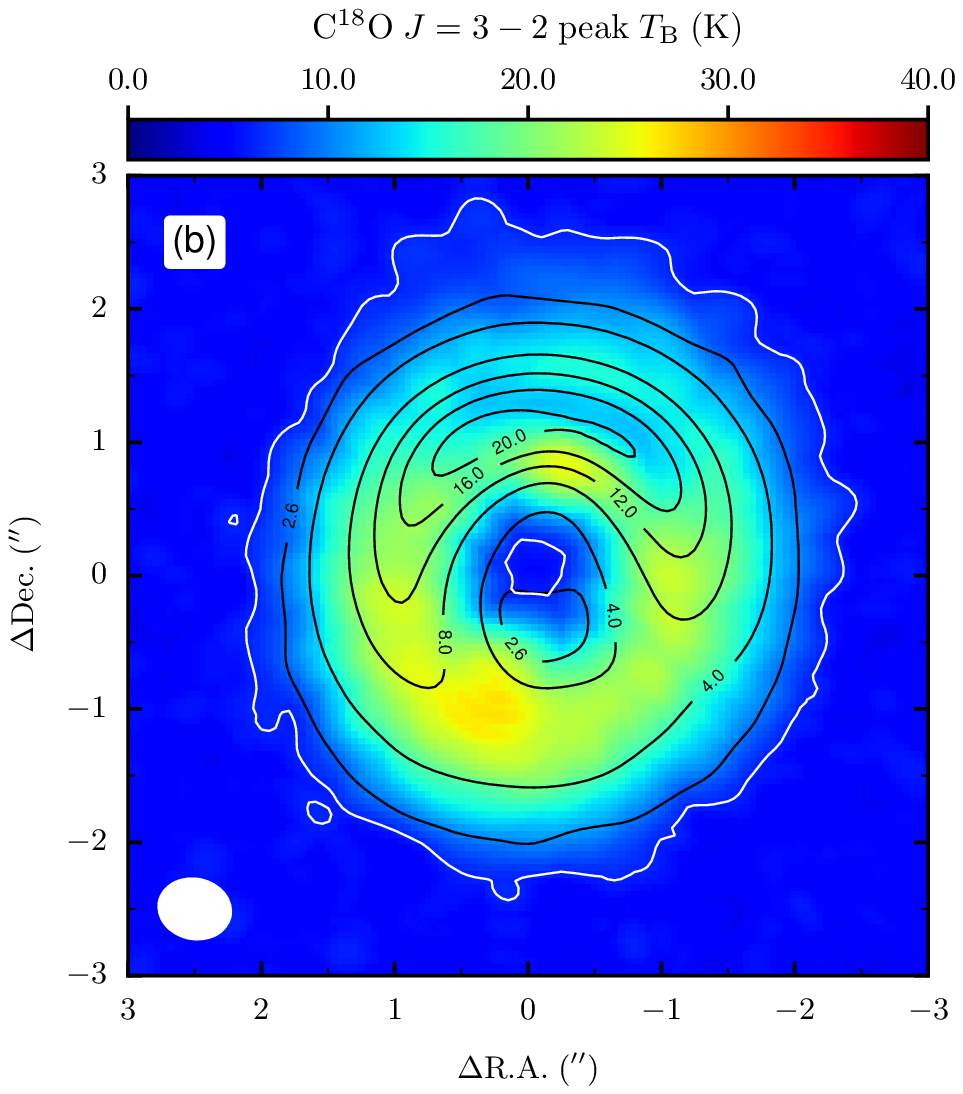} \\
\includegraphics[width=0.33\textwidth]{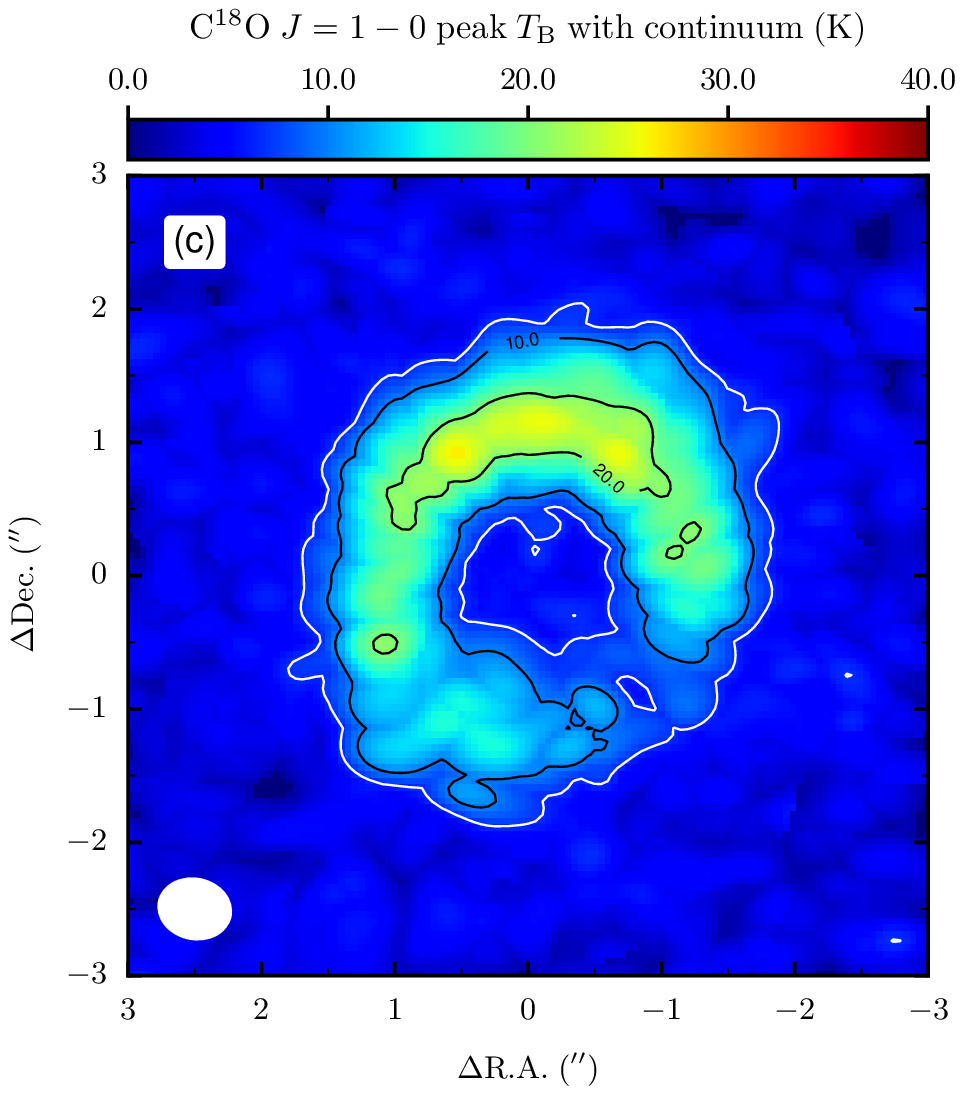} &
\includegraphics[width=0.33\textwidth]{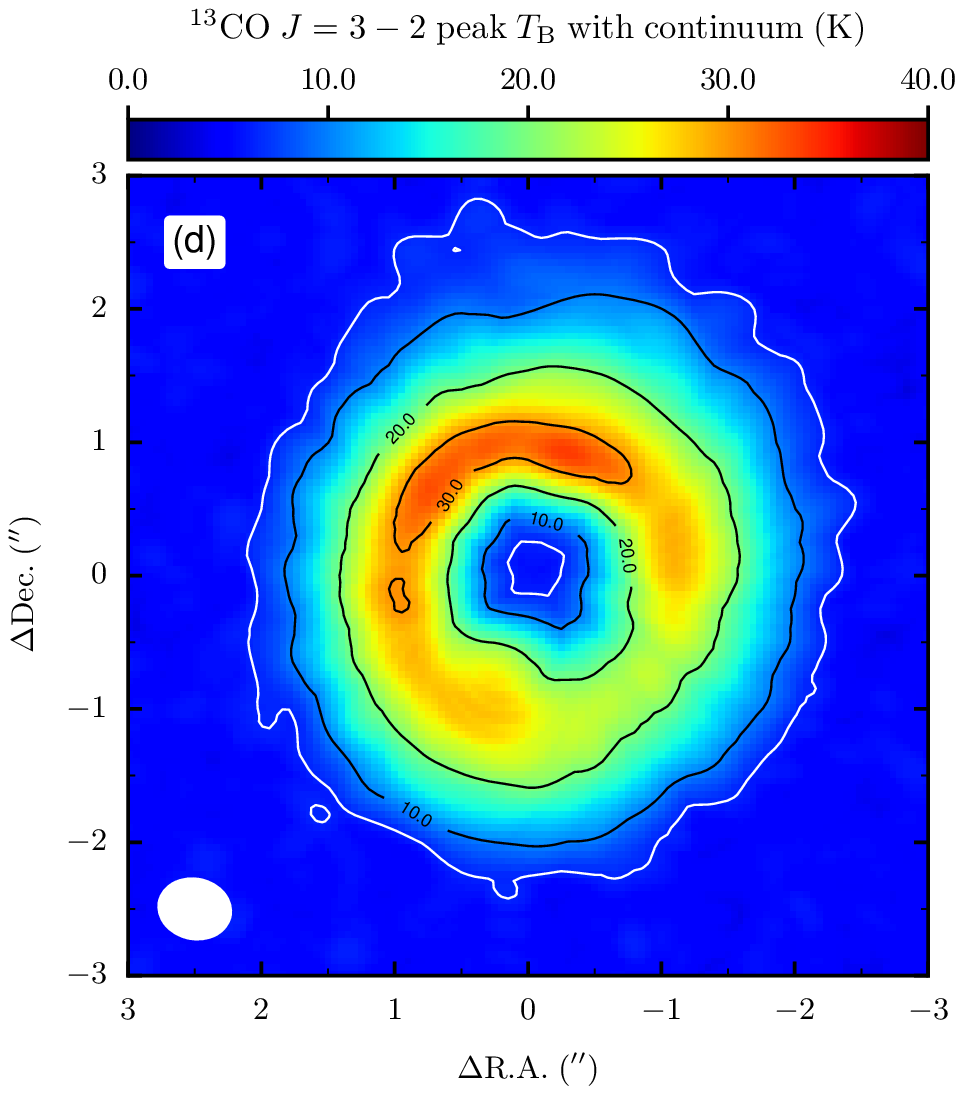}
\end{tabular}
\caption{
Similar to figure \ref{fig:13co_peakTB},
for the C$^{18}$O $J=1-0$ (left panels) and $J=3-2$ (right panels) line emission.
The black contours in panel (a) denote 
the narrow-band continuum emission centered at the rest frequencies of the $J=1-0$ line emission (see text for details),
while those in panel (b) denote the $336~\mathrm{GHz}$ continuum emission
as shown in figure \ref{fig:continuum}(d).
In panels (a) and (c), the white contours
denote the $3.5\sigma$ level,
which are $6.53~\mathrm{K}$ and $6.70~\mathrm{K}$, respectively.
In panels (b) and (d), the white contours
denote the $5\sigma$ level,
which are $6.32~\mathrm{K}$ and $6.38~\mathrm{K}$, respectively.
}\label{fig:c18o_peakTB}\end{figure*}

Figure \ref{fig:moment_12C18O_030kms} displays the moment maps of the 
C$^{18}$O $J=1-0$ line emission created with the same Keplerian masking and noise masking
as those of $^{13}$CO $J=1-0$ shown in figure \ref{fig:moment_13C16O_030kms}.
Compared to that of $^{13}$CO,
the C$^{18}$O is confined in a narrower radial extent of $r \approx 0\farcs5 - 2\farcs0$.
This is because of the weaker emission of C$^{18}$O $J=1-0$.
In the southwestern region of $\mathrm{P.A.} = 180^\circ - 240^\circ$, 
the emission is much weaker and is below $10\sigma$.

The peak of the 0th-moment in figure \ref{fig:moment_12C18O_030kms}(a)
is located at $(0\farcs9, 27^\circ)$,
close to the continuum peak 
with $F_\mathrm{int} = 28.7~\mathrm{mJy~beam^{-1}~km~s^{-1}}$.
The ridge of 0th-moment is lowest at
$(0\farcs9, 225^\circ)$ where
$F_\mathrm{int} = 6.95~\mathrm{mJy~beam^{-1}~km~s^{-1}}$.
The contrast of the ridge is thus $4.13$.

Figure \ref{fig:c18o_peakTB}(a) shows
the peak surface brightness of the C$^{18}$O $J=1-0$,
while figure \ref{fig:c18o_peakTB}(c) shows
the peak $T_\mathrm{B}$ with the continuum emission.
The continuum emission in figure \ref{fig:c18o_peakTB}(a) is centered at $109.8~\mathrm{GHz}$ with a bandwidth of $54~\mathrm{MHz}$,
created using the same method as that shown in figure \ref{fig:13co_peakTB}(a).
The surface brightness of C$^{18}$O $J=1-0$ with continuum emission
is significantly weaker than that of the $^{13}$CO $J=1-0$ and $J=3-2$,
as well as that of the C$^{18}$O $J=3-2$ shown in figures \ref{fig:c18o_peakTB}(b) and (d),
indicating that the C$^{18}$O $J=1-0$ is optically thin.
The peak $T_\mathrm{B}$ is $\approx 25~\mathrm{K}$ in the north
while it is $\approx 10~\mathrm{K}$ in the south,
and its distribution 
is similar to that of the continuum emission.

\section{Analyses} \label{sec:analyses}
In this section, we derived the 
gas and dust surface density for the disk 
of HD~142527 under the assumptions of local thermodynamic equilibrium (LTE).
We do not consider dust sedimentation at the midplane and
assume that the gas and dust are well-mixed in the disk.
We assume the physical temperatures of the gas and dust,
$T_\mathrm{d}$ and $T_\mathrm{g}$, to be identical.
These temperatures are taken from
the peak $T_\mathrm{B}$ of $^{13}$CO $J=3-2$ including the
continuum emission, as shown in figure \ref{fig:13co_peakTB}(d);
this temperature is referred to as the disk temperature in the following.
Since the $^{13}$CO $J=3-2$ emission in the disk inner region is optically thin
and does not reflect the physical temperature,
in the following analyses we mask out 
the inner region in which the brightness temperature is less than $30~\mathrm{K}$.
Due to the disk inclination,
the western region of the disk ($\mathrm{P.A.} = 161^\circ - 341^\circ$)
is closest to us while the eastern region is the furthest from us.
The disk temperature at the far side is higher by about $3~\mathrm{K}$,
since the surface of the disk that is irradiated by the central star is exposed to us.
The disk is assumed to be isothermal in a vertical direction.
A model where $T_\mathrm{d}$ is assumed to be $80\%$ of $T_\mathrm{g}$
is discussed in Appendix \ref{sec:two_layer_disk}.

\subsection{Derivation of dust surface density} \label{sec:analyses_dust}
\begin{figure*}\centering
\begin{tabular}{ccc}
\includegraphics[width=0.3\textwidth]{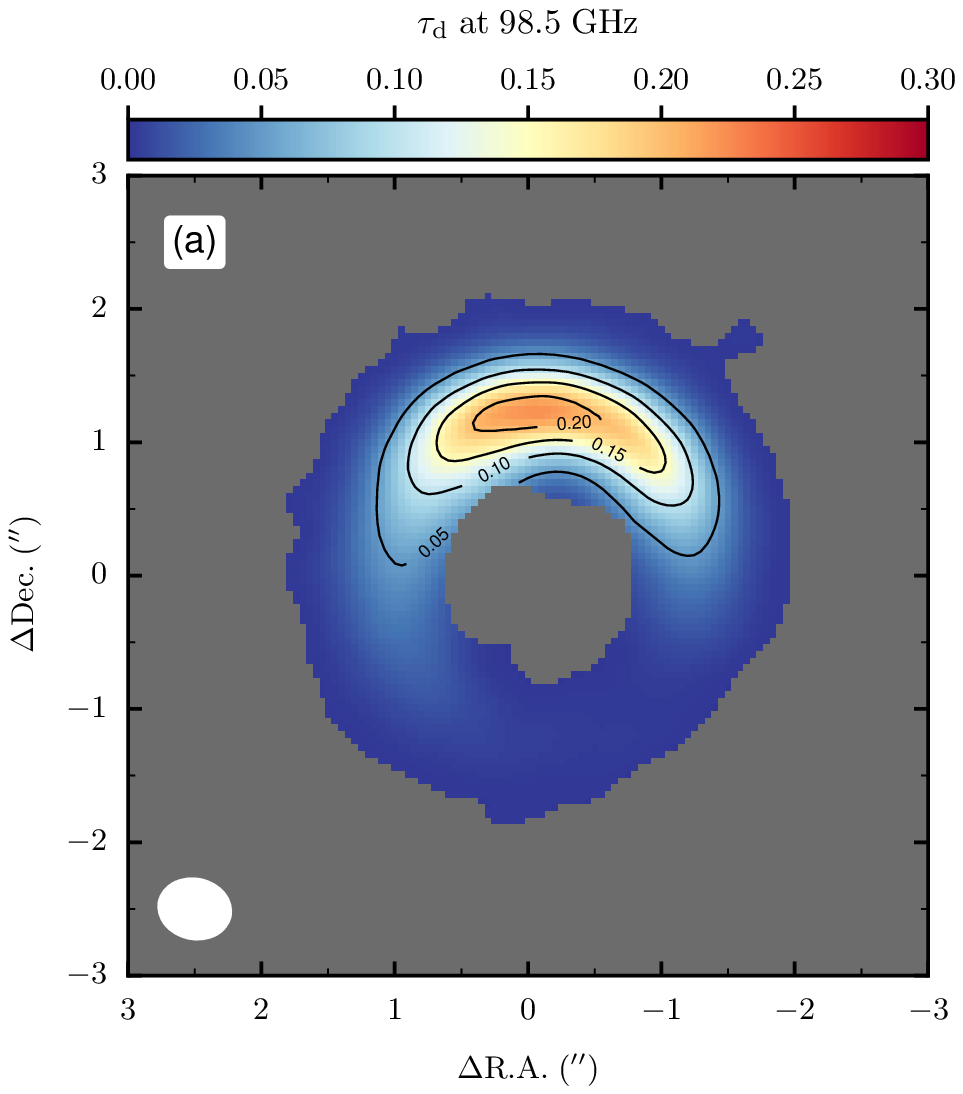} &
\includegraphics[width=0.3\textwidth]{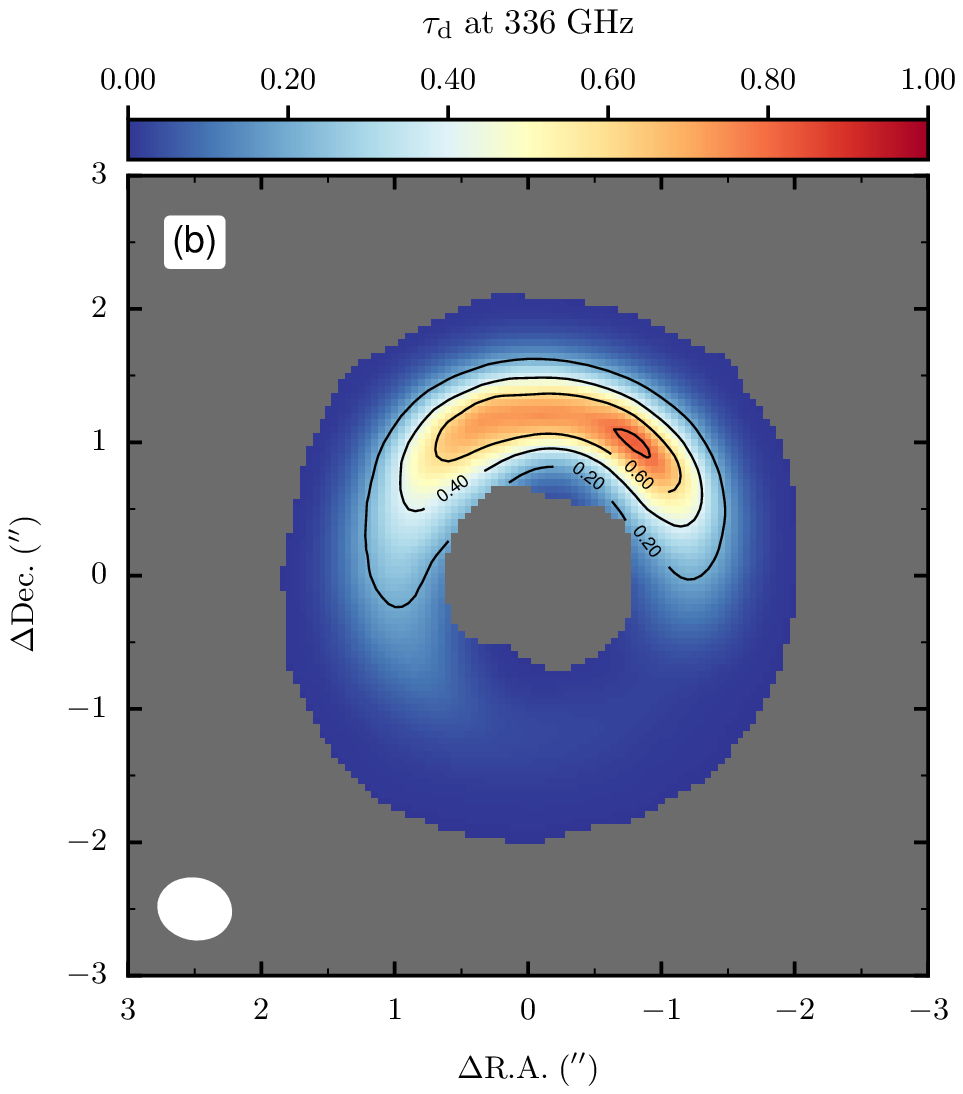} &
\includegraphics[width=0.3\textwidth]{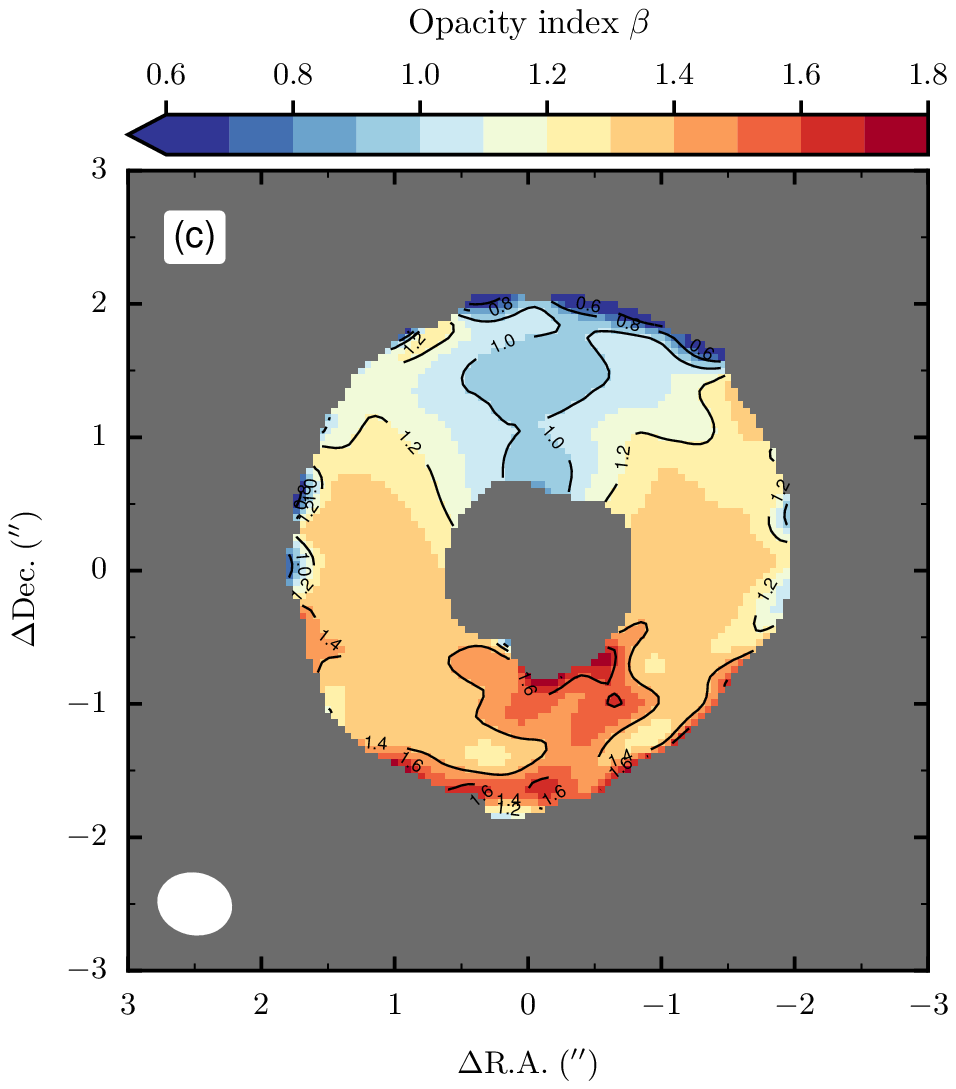}
\end{tabular}
\caption{
Optical depth $\tau_\mathrm{d}$ of the (a) $98.5~\mathrm{GHz}$ and (b) $336~\mathrm{GHz}$ continuum emission.
Dust spectral opacity index $\beta$ derived from Equation (\ref{eq:beta}) is shown in panel (c).
}\label{fig:continuum_tau}\end{figure*}

The dust surface density is derived from $\Sigma_\mathrm{d} = \tau_\mathrm{d}/\kappa_\mathrm{d}$,
where $\tau_\mathrm{d}$ is the optical depth of the dust continuum emission
and $\kappa_\mathrm{d}$ is the dust opacity.
We first calculate
$\tau_\mathrm{d}$ at $98.5~\mathrm{GHz}$ and $336~\mathrm{GHz}$ 
from the radiative transfer equation
\begin{equation}
\label{eq:dust_rte}
I_\mathrm{d} = \left[  B_\nu(T_\mathrm{d}) - B_\nu(T_\mathrm{bg}) \right] \left[  1 - \exp(-\tau_\mathrm{d})    \right] ,
\end{equation}
where $I_\mathrm{d}$ denotes the intensity of the continuum emission,
$B_\nu$ is the Planck function, $T_\mathrm{bg}= \mathrm{2.7~\mathrm{K}}$ 
the temperature of the cosmic background radiation.
The results are shown in figures \ref{fig:continuum_tau}(a) and (b).
The peak $\tau_\mathrm{d}$ at $98.5~\mathrm{GHz}$ 
is $0.24$ and is co-spatial to its peak continuum emission.
On the other hand, peak $\tau_\mathrm{d}$
at $336~\mathrm{GHz}$ is $0.82$, and is located at the
western component of the double-peaked structure seen at the $336~\mathrm{GHz}$ emission;
this is due to the lower disk temperature at the near side.
The dust opacity is highly uncertain and
depends on the particle compositions, structures, as well as size distributions \citep{miyake1993,draine2006,kataoka2014,birnstiel2018};
in the analyses, we adopt the canonical dust opacity (per dust mass) $\kappa_\mathrm{d}$
described by \citet{beckwith1990},
\begin{equation}
\label{eq:dust_opacity}
\kappa_\mathrm{d} = 10 \left(  \frac{\nu}{10^{12}~\mathrm{Hz}}    \right) ^\beta \quad \mathrm{cm^2~g^{-1}},
\end{equation}
where $\beta$ is the dust opacity spectral index and is calculated from the $\tau_\mathrm{d}$ distributions via
\begin{equation}
\label{eq:beta}
\beta =\left.  \log \left[ \frac{ \tau_\mathrm{d,336}}{\tau_\mathrm{d,98.5} } \right] \right/
              \log \left[   \frac{336~\mathrm{GHz}}{98.5~\mathrm{GHz} }     \right]
\end{equation}
and shown in figure \ref{fig:continuum_tau}(c).
The opacity index varies throughout the disk;
in the southern region, $\beta$
is close to the interstellar value of $1.7$ \citep{planck2014},
while in the northern region, it is $\approx1$.
The $\beta$ in the north can be interpreted as the consequence 
of the growth of dust grains
and is qualitatively consistent with the modeling results 
based on polarization observations \citep{ohashi2018}.
Dust scattering can strongly depend on 
the dust compositions and structures \citep{tazaki2016,tazaki2018},
but we have ignored it in this study.
This is because the observed intensity does not depend strongly 
on the dust scattering
if the scattering opacity is comparable to
the absorption opacity,
and the modeling of the dust continuum map shows that 
this is the case in the lopsided disk around HD~142527 \citep{soon2017,boehler2017}.

\begin{figure*}\centering
\begin{tabular}{cc}
\includegraphics[width=0.3\textwidth]{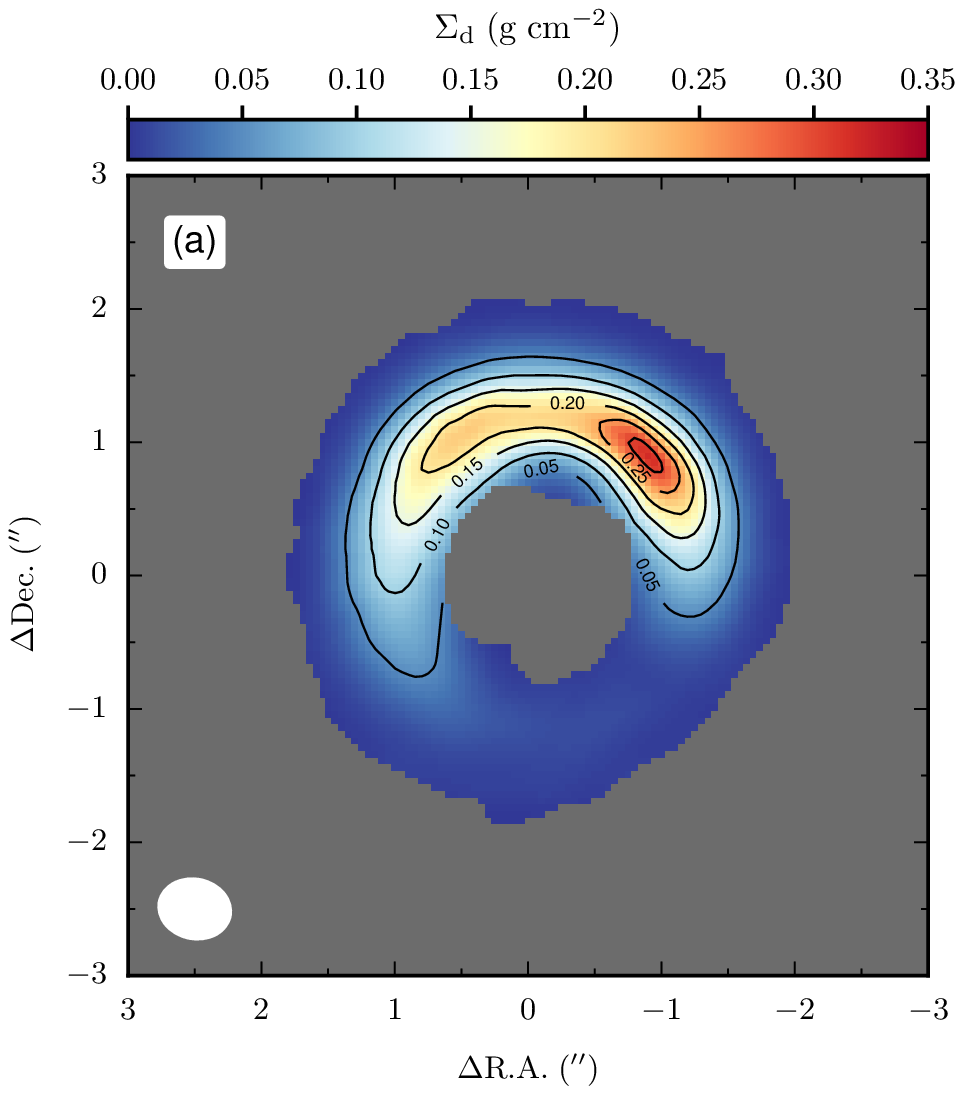} &
\includegraphics[width=0.3\textwidth]{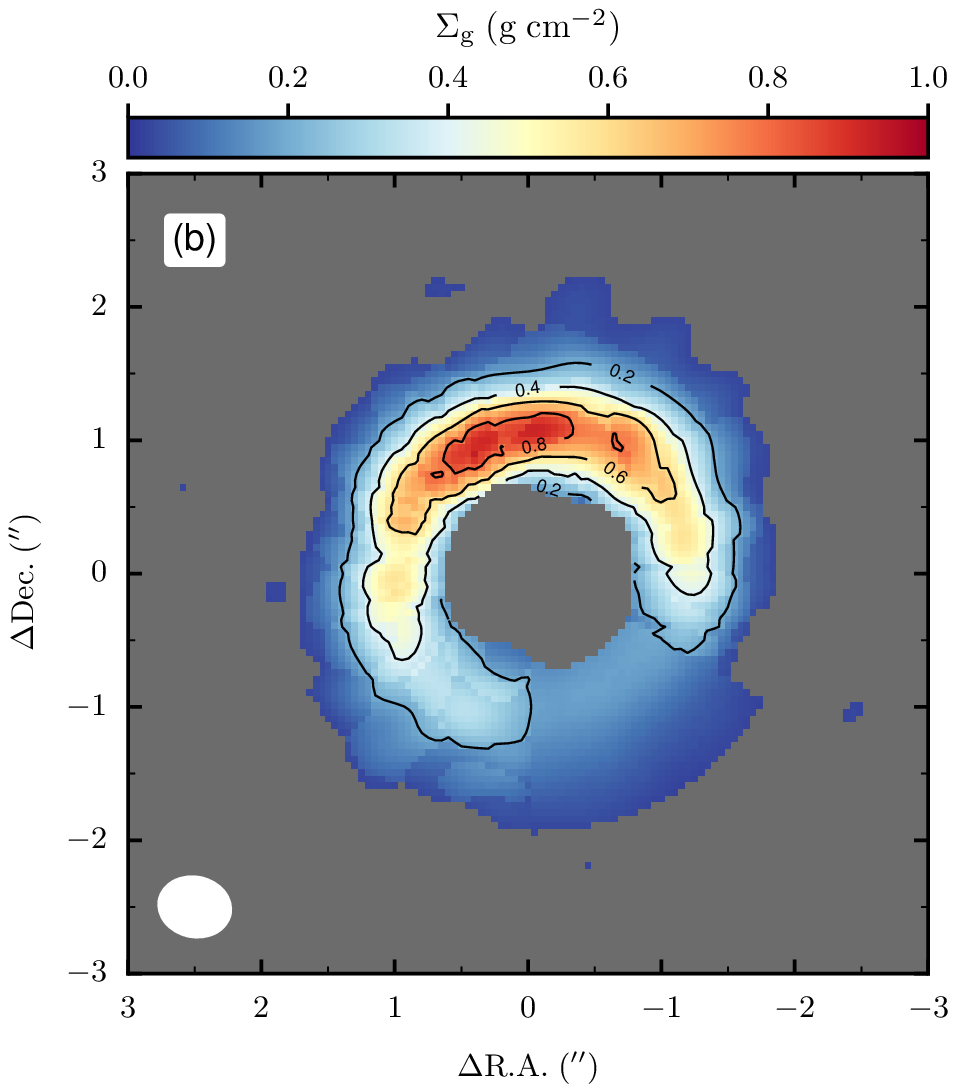} \\
\includegraphics[width=0.3\textwidth]{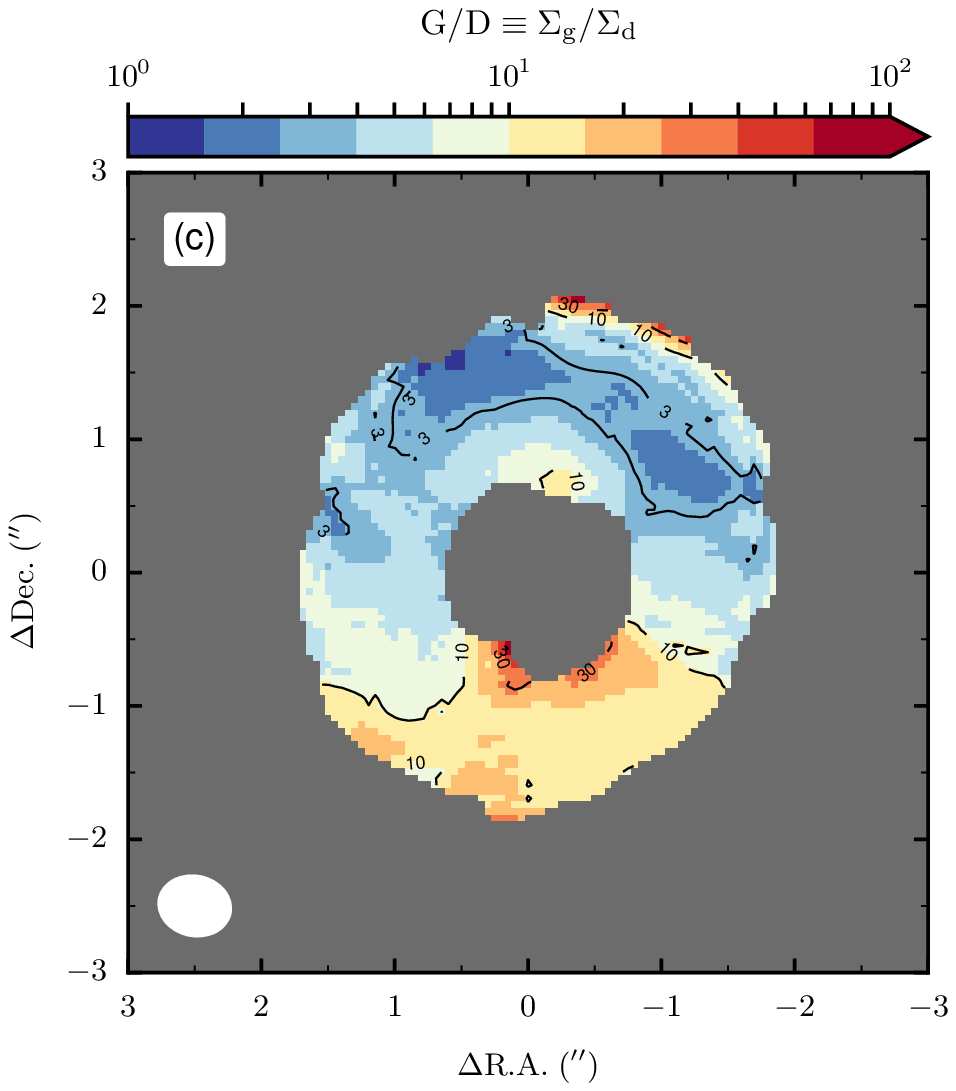} &
\includegraphics[width=0.3\textwidth]{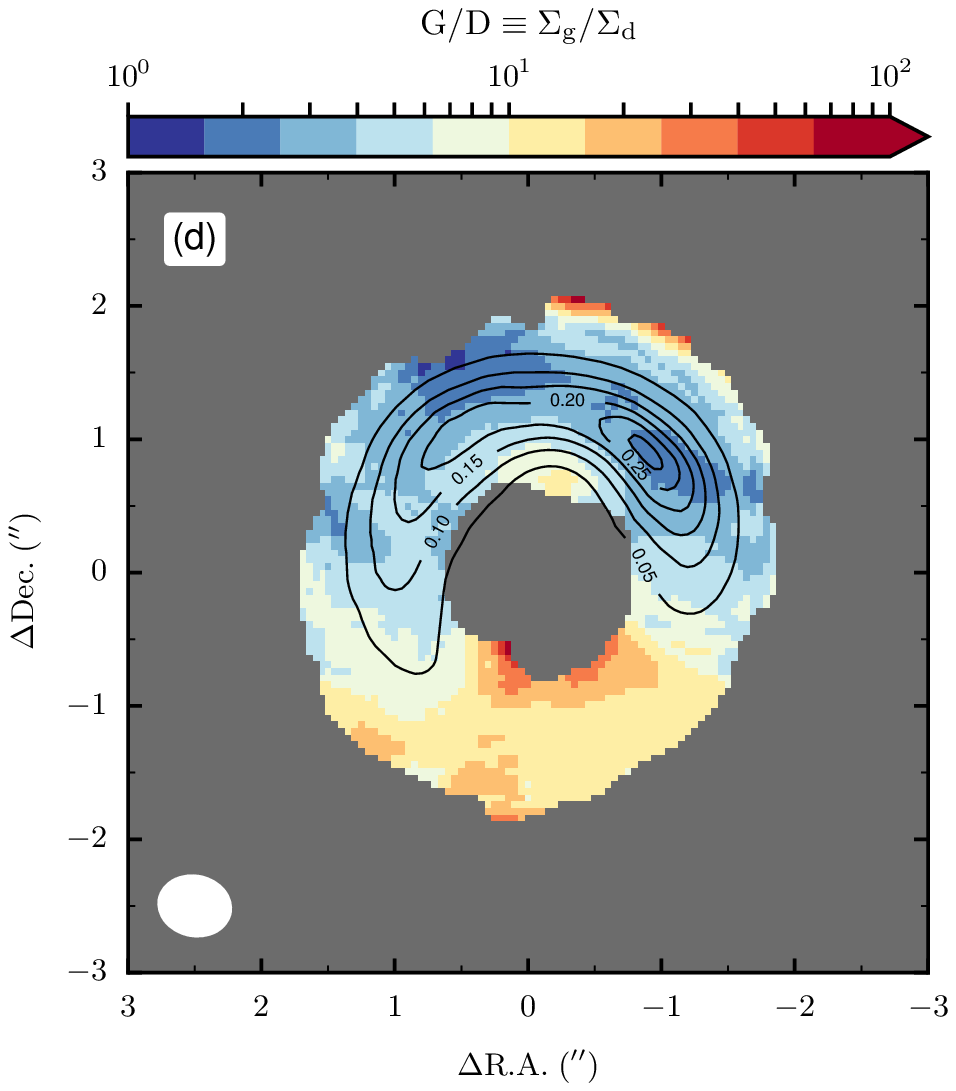}
\end{tabular}
\caption{
Derived results of (a) dust surface density $\Sigma_\mathrm{d}$,
(b) gas surface density $\Sigma_\mathrm{g}$, and
(c) gas-to-dust ratio $\mathrm{G/D}\equiv \Sigma_\mathrm{g}/\Sigma_\mathrm{d}$.
In panel (d) the contours of $\Sigma_\mathrm{d}$ in panel (a) are superimposed
on the $\mathrm{G/D}$ image.
}\label{fig:sigma}\end{figure*}

Figure \ref{fig:sigma}(a) shows the derived dust surface density $\Sigma_\mathrm{d}$.
Along the ridge of $\Sigma_\mathrm{d}$,
the maximum is located at $(1\farcs2, 315^\circ)$
where $\Sigma_\mathrm{d} = 3.08 \times 10^{-1}~\mathrm{g~cm^{-2}}$
and the minimum is located at $(1\farcs2, 225^\circ)$ where $\Sigma_\mathrm{d} = 9.14 \times 10^{-3}~\mathrm{g~cm^{-2}}$.
The derived spatial location of the peak $\Sigma_\mathrm{d}$ does not correspond to
that of the peak continuum emission at $98.5~\mathrm{GHz}$ which is optically thin;
this is mainly due to the lower temperature in the near side of the disk
as well as the $\beta$ distribution.
Similarly, because of the temperature and $\beta$ distributions,
the derived $\Sigma_\mathrm{d}$ ridge contrast of $33$ is
lower than that of the $98.5~\mathrm{GHz}$ continuum emission, which is $58$.

\citet{boehler2017} successfully reproduce
the dust continuum observations at $342~\mathrm{GHz}$ in $\mathrm{P.A.} = 16^\circ - 26^\circ$ (north $\mathrm{P.A.}$ sector)
and $\mathrm{P.A.} = 216^\circ - 226^\circ$ (south $\mathrm{P.A.}$ sector)
using a disk model,
in which the peak dust surface densities are $0.65~\mathrm{g~cm^{-2}}$ in the north sector 
and $0.012~\mathrm{g~cm^{-2}}$ in the south sector
(see Figure 13 and Table 1 of their paper).
Their modeling is based on the ALMA observations with a beam of $0\farcs26 \times 0\farcs21$.
By convolving the models of dust surface density
derived by \citet{boehler2017} with the beam size of our ALMA observations (i.e., $0\farcs54 \times 0\farcs44$, $\mathrm{P.A.} = 78.1^\circ$),
the peak dust surface densities in the north and south $\mathrm{P.A.}$ sectors 
are derived to be $\approx 0.4~\mathrm{g~cm^{-2}}$ 
and $\approx 0.006~\mathrm{g~cm^{-2}}$, respectively.
After correcting for the difference in dust opacity\footnote{
\citet{boehler2017} use a constant dust absorption opacity of $2.9~\mathrm{g~cm^{-2}}$.
The dust opacity $\kappa_\mathrm{d}$ used in this study
are $\approx 3.4~\mathrm{g~cm^{-2}}$ and $\approx 2.0~\mathrm{g~cm^{-2}}$
in the north and south $\mathrm{P.A.}$ sectors, respectively.}, 
these correspond to $\approx 0.3~\mathrm{g~cm^{-2}}$ 
and $\approx 0.009~\mathrm{g~cm^{-2}}$.
While our results in the south $\mathrm{P.A.}$ sector agrees with
\citet{boehler2017}, 
our results for the north $\mathrm{P.A.}$ sector, i.e., $\Sigma_\mathrm{d} \approx 0.2~\mathrm{g~cm^{-2}}$,
is approximately $70\%$ of that derived by \citet{boehler2017}.
This inconsistency
may be due to the high optical depth of the dust continuum in the northern regions;
the inherent uncertainty in the estimate on $\Sigma_\mathrm{d}$ is large \citep{soon2017}.
Furthermore, omitting the dust scattering may underestimate $\Sigma_\mathrm{d}$,
and hence the $\Sigma_\mathrm{d}$ ridge contrast
(\cite{soon2017,birnstiel2018}).

\subsection{Derivation of gas surface density} \label{sec:analyses_gas}
We derived the disk H$_2$ gas surface density $\Sigma_\mathrm{g}$
from the $J=1-0$ and $J=3-2$ line emission of C$^{18}$O
by assuming the interstellar abundance $\chi(\mathrm{C^{18}O/H_2}) = 1.79 \times 10^{-7}$ \citep{wilson1999};
the validity of the abundance is discussed in Section \ref{sec:validity_results}.
The $J=3-2$ emission is used for the derivation of $\Sigma_\mathrm{g}$ in the region of $\mathrm{P.A.} = 180^\circ - 240^\circ$,
where the $J=1-0$ emission is weak.

The optical depth of the molecular line $\tau_\mathrm{g}$ in the velocity channels 
is calculated from the radiative transfer equation,
which is similar to Equation (\ref{eq:dust_rte}) but takes the form
\begin{equation}
\label{eq:gas_rte}
I_\mathrm{g} = \left[  B_\nu(T_\mathrm{g}) - B_\nu(T_\mathrm{bg}) \right] \left[  1 - \exp(-\tau_\mathrm{g})    \right] \exp(-\tau_\mathrm{d}),
\end{equation}
i.e., the factor $\exp(-\tau_\mathrm{d})$ is included to
account for the line emission absorbed by the dust.
Here,
$\tau_\mathrm{d}$ is the optical depth of the narrow-band continuum emission 
at $109.8~\mathrm{GHz}$, which is shown as contours in figure \ref{fig:c18o_peakTB}(a),
in the case of the C$^{18}$O $J=1-0$ line emission.
When the above equation is applied to the C$^{18}$O $J=3-2$ line emission,
on the other hand, $\tau_\mathrm{d}$ refers to the $336~\mathrm{GHz}$ continuum emission,
i.e., the continuum image in figure \ref{fig:continuum}(b).
The optical depth $\tau_\mathrm{g}$ is related to the total (particle number) column density $N_\mathrm{tot}$ of the CO isotopologue 
by
\begin{eqnarray}
\label{eq:co_surface_density}
N_\mathrm{tot}  = &&\frac{3h}{8 \pi^{3} \mu^2 J_u}      \left ( \frac{kT_\mathrm{ex}}{hB_0}  + \frac{1}{3}  \right)  \exp\left ( \frac{E_{J_u}}{kT_\mathrm{ex}}  \right )   \nonumber \\
			    &&\quad \times \left [ \exp \left (  \frac{h \nu}{kT_\mathrm{ex}}   \right ) -1     \right ]  ^{-1}  \sum \left[ \tau_\mathrm{g}(v) \Delta v  \right]   \quad \mathrm{cm^{-2}},
\end{eqnarray}
where
$\mu$, $J_u$, and $B_0$, are 
the dipole moment, the rotational quantum number of 
the line transition upper level, and the rigid rotor rotational constant,
respectively \citep{mangum2015}.
The excitation temperature $T_\mathrm{ex}$ is equal to $T_\mathrm{g}$ in the LTE analysis.
The gas surface density is then calculated by 
$\Sigma_\mathrm{g} = m_\mathrm{H_2} N_\mathrm{tot} / \chi$,
where $m_\mathrm{H_2}$ is the molecular mass of H$_2$.

Figure \ref{fig:sigma}(b) shows the derived results of $\Sigma_\mathrm{g}$.
At $\mathrm{P.A.} = 180^\circ$ and $\mathrm{P.A.} = 240^\circ$,
the surface density derived from the C$^{18}$O $J=1-0$ and C$^{18}$O $J=3-2$ lines
differs by less than $10\%$, indicating a smooth variation of 
$\Sigma_\mathrm{g}$ derived from the two lines.
Along the ridge of $\Sigma_\mathrm{g}$, 
the maximum is located at $(1\farcs0, 3^\circ)$ where
$\Sigma_\mathrm{g} = 9.14\times10^{-1}~\mathrm{g~cm^{-2}}$,
while the minimum
is located at $(1\farcs0, 189^\circ)$
where $\Sigma_\mathrm{g} = 1.75\times 10^{-1}~\mathrm{g~cm^{-2}}$.
The contrast between the two directions
is approximately $5$, similar to the contrast of the C$^{18}$O $J=1-0$ 0th-moment ridge.
\citet{boehler2017} derived the peak gas surface density to be 
$1.125~\mathrm{g~cm^{-2}}$ and $0.3~\mathrm{g~cm^{-2}}$ in the north and south $\mathrm{P.A.}$ sectors, respectively. 
Similar to the beam convolution performed to the model dust surface density in discussed in Section \ref{sec:analyses_dust},
after convolving the model gas surface density derived by \citet{boehler2017} with our observations beam size
the peaks are derived to be $\approx 1.0~\mathrm{g~cm^{-2}}$
and $\approx 0.2~\mathrm{g~cm^{-2}}$.
Our results are thus consistent with those derived by \citet{boehler2017}.

Along the ridge of $^{13}$CO $J=1-0$ 0th-moment 
in $\mathrm{P.A.} = 200^\circ - 240^{\circ}$,
$\tau_\mathrm{g}$ can also be solved from Equation (\ref{eq:gas_rte}).
Within this $\mathrm{P.A.}$ and the radial range $r=0\farcs6-1\farcs6$,
the column density derived from the $^{13}$CO $J=1-0$ and C$^{18}$O $J=3-2$ line emission are  
$N_\mathrm{tot} (^{13}\mathrm{CO}) = (1.98\pm0.21) \times10^{19}~\mathrm{cm^{-2}}$ and
$N_\mathrm{tot} (\mathrm{C^{18}O}) = (1.98\pm0.12) \times10^{18}~\mathrm{cm^{-2}}$, respectively;
the uncertainties only include the propagation of $1\sigma$ noise level.
The ratio
$N_\mathrm{tot} (^{13}\mathrm{CO})/N_\mathrm{tot} (\mathrm{C^{18}O})$ is derived to be $ 9.99 \pm 1.67$,
and it agrees with the value of the local interstellar medium ($8.1\pm1.1$) within the uncertainty \citep{wilson1999}.

\section{Discussions} \label{sec:discussions}
\subsection{G/D and correlation between gas and dust surface densities} \label{sec:gdratio}
\begin{figure*}\centering
\begin{tabular}{cc}
\includegraphics[width=0.45\textwidth]{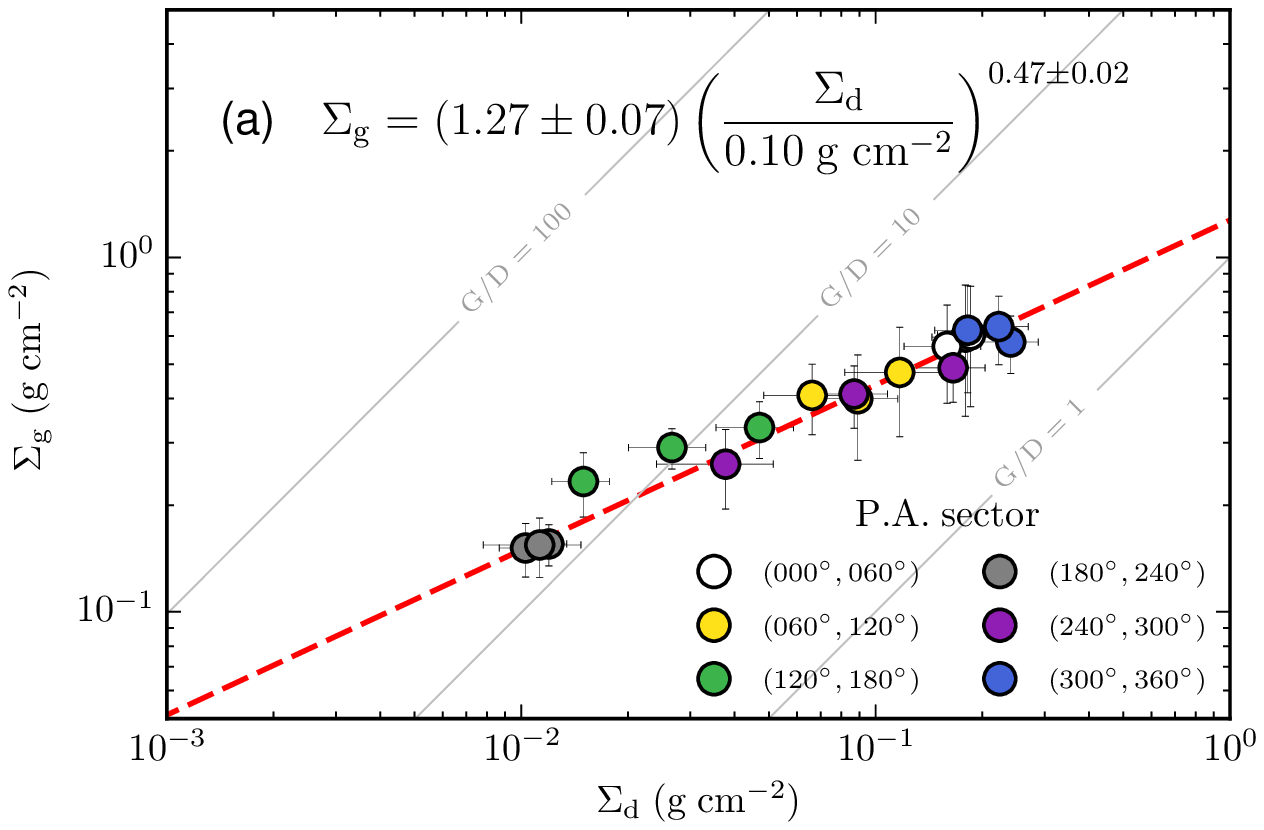} &
\includegraphics[width=0.45\textwidth]{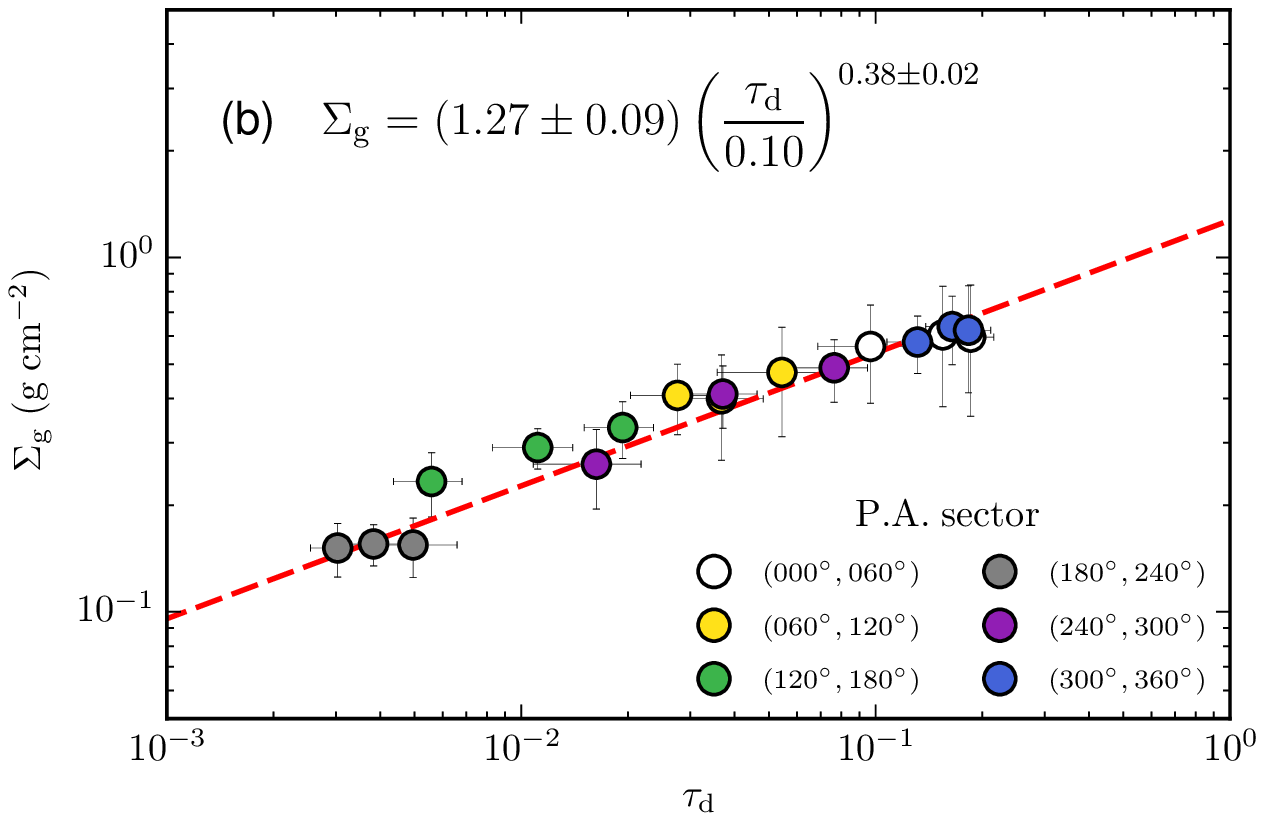} 
\end{tabular}
\caption{
Panel (a) shows the correlation between the gas surface density $\Sigma_\mathrm{g}$
and the dust surface density $\Sigma_\mathrm{d}$;
the gray lines indicate the gas-to-dust ratio of $\mathrm{G/D} = 1,~10,~100$.
Panel (b) shows the correlation between $\Sigma_\mathrm{g}$ and the dust optical depth at $98.5~\mathrm{GHz}$, $\tau_\mathrm{d}$.
The best-fit power law of exponent $p$ is drawn as red dashed line.
In both panels, the data points are averaged values in a bin of angular size $20^{\circ}$ and radial size $0\farcs5$,
and the error bars indicate the standard deviation of the values in the bin.
The colors of the data points denote the $\mathrm{P.A.}$ of the points.
}\label{fig:correlation}\end{figure*}

Figure \ref{fig:sigma}(c) shows the G/D, defined as the ratio of $\Sigma_\mathrm{g}$ to $\Sigma_\mathrm{d}$;
figure \ref{fig:polar_plots}(c) shows the same results in polar coordinates.
The estimated $\mathrm{G/D}$ is $\sim 3$ and $\sim 20$
in the northern and southern regions, respectively,
and $\mathrm{G/D}$ gradually varies along the azimuthal direction.
Our results are 
consistent with the modeling results by \citet{muto2015} and \citet{boehler2017},
but they only focused on two sectors centered at
$\mathrm{P.A.} = 21^\circ$ and $\mathrm{P.A.} = 221^\circ$.
We successfully derived the projected $\mathrm{G/D}$ distribution across the disk
and found that $\mathrm{G/D}$ in the outer disk varies along the azimuthal direction.
The low $\mathrm{G/D}$ in the northern region may be important 
in the formation of planetesimals \citep{lambrechts2012,raettig2015}.
Two-dimensional simulations predicted that a vortex would be destroyed by the dust back reaction
when $\mathrm{G/D}$ got low \citep{fu2014},
but succeeding three-dimensional simulations have shown that this effect disappears \citep{lyra2018}.

Figure \ref{fig:correlation}(a) shows the relation between $\Sigma_\mathrm{g}$
and $\Sigma_\mathrm{d}$ on the dust ridge,
where the peak $T_\mathrm{B}$ of $^{13}$CO $J=3-2$ is highest,
and the estimate of the temperature is most reliable.
The plot can be fitted by a power law for which the exponent $p$ (indicated in the top left corner of the plots)
is $p = 0.47$.
The relation $\Sigma_\mathrm{g} \propto \Sigma_\mathrm{d}^p$
corresponds to $\mathrm{G/D} \propto \Sigma_\mathrm{d}^{p-1}$,
and therefore figure \ref{fig:correlation}(a) suggests that $\mathrm{G/D}$ varies approximately
with $\propto \Sigma_\mathrm{d}^{-0.53}$.
This relation may be a critical test for
future theoretical studies to understand the trapping efficiency of dust grains
and the origin of the asymmetric disk structure around HD~142527.

Figure \ref{fig:correlation}(b) is similar
to figure \ref{fig:correlation}(a),
but with the horizontal axis replaced by the dust optical depth at $98.5~\mathrm{GHz}$.
Again, we use a power law and obtain the best fit exponent $p=0.38$.
The exponent in this case is smaller than that of the relation
shown in figure \ref{fig:correlation}(a),
since there is an azimuthal variation of $\beta$ (or $\kappa_\mathrm{d}$) as
shown in figure \ref{fig:continuum_tau}(c).

\subsection{Relative spatial distribution of gas and dust surface densities} \label{sec:spatial_gas_dust}
\begin{figure*}\centering
\includegraphics[width=0.8\textwidth]{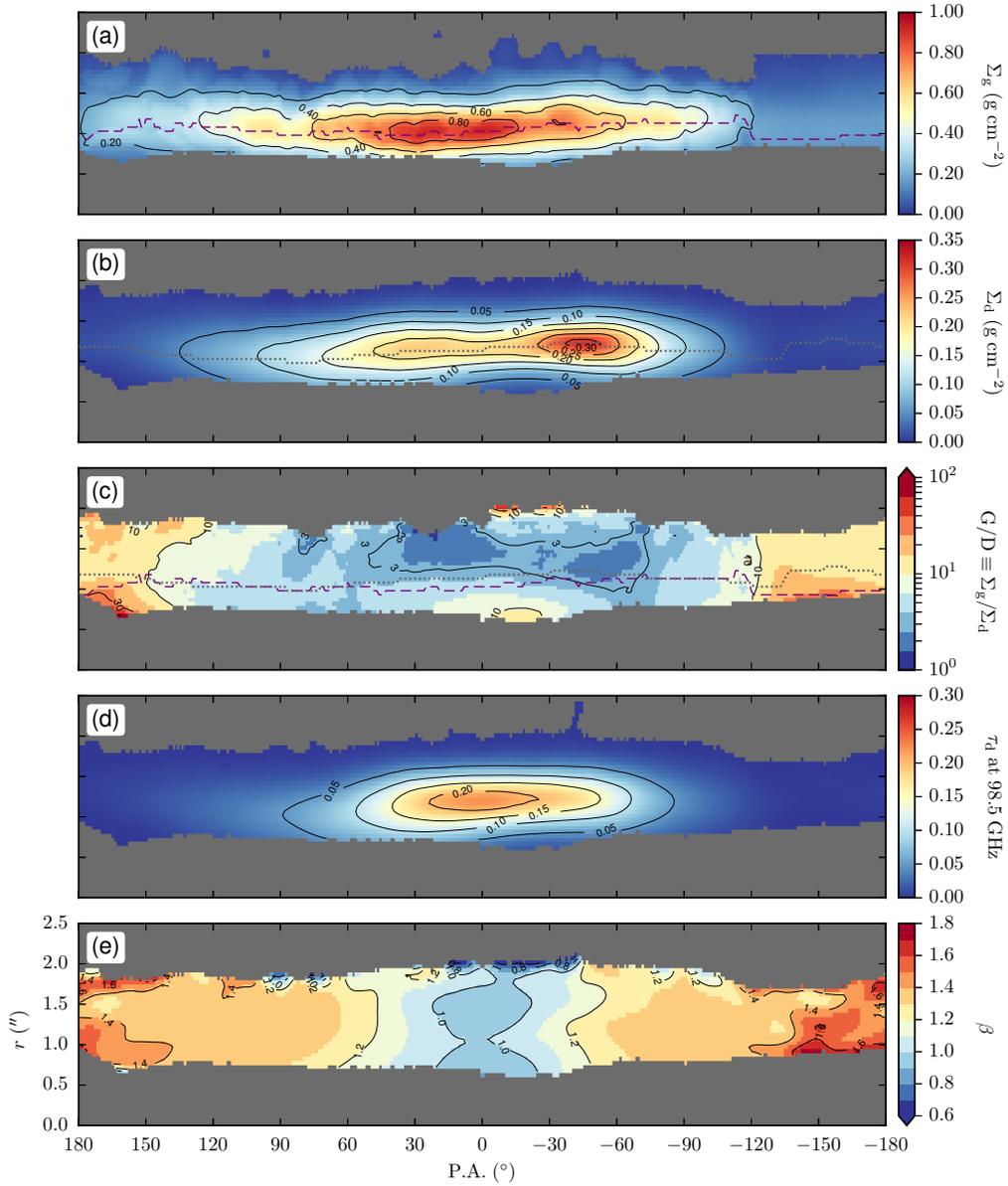}
\caption{
Derived results of (a) gas surface density $\Sigma_\mathrm{g}$,
(b) dust surface density $\Sigma_\mathrm{d}$,
(c) gas-to-dust ratio $\mathrm{G/D}$,
(d) dust optical depth at $98.5~\mathrm{GHz}$,
and (e) dust opacity spectral index $\beta$
plotted in polar coordinates.
The purple dashed line and the gray dotted line in panels (a) and (b)
denote the $\Sigma_\mathrm{g}$ and $\Sigma_\mathrm{d}$ ridges,
which are also superimposed on the $\mathrm{G/D}$ map in panel (c).
}\label{fig:polar_plots}\end{figure*}

Figure \ref{fig:polar_plots} compares the spatial distributions of $\Sigma_\mathrm{g}$
and $\Sigma_\mathrm{d}$ in the polar coordinates.
From figures \ref{fig:moment_13C16O_030kms}(c), \ref{fig:moment_12C18O_030kms}(c),
and the infrared images \citep{fujiwara2006}, the disk is thought to be  rotating in a clockwise rotation.
Comparing figures \ref{fig:polar_plots}(a) and \ref{fig:polar_plots}(b),
we see that the peak $\Sigma_\mathrm{d}$ is located at $\mathrm{P.A.} \approx 315^\circ$,
which is downstream of the peak $\Sigma_\mathrm{g}$
located at $\mathrm{P.A.} \approx 3^\circ$.
If the northern region of high $\Sigma_\mathrm{g}$ corresponds to a vortex with higher pressure,
then the dust will accumulate 
at a region shifted ahead of the vortex.
This picture is consistent with 
the theoretical prediction by \citet{baruteau2016}
if the Stokes number, $\mathrm{St}$, of the dust particles is $\gtrsim 1$;
we estimate the physical radius of the dust particle $s$ by using
equation (1) in the paper by the same authors,
i.e.,
\begin{equation}
\label{eq:stokes}
s \approx 4.67 \times \left( \frac{\mathrm{St}}{1}\right)  \ \left( \frac{\rho_\mathrm{pc}}{1~\mathrm{g~cm^{-3}}} \right)^{-1} \left( \frac{\Sigma_\mathrm{g}}{0.7~\mathrm{g~cm^{-2}}} \right)~\mathrm{mm},
\end{equation}
where $\rho_\mathrm{pc}$ is the internal mass density of the particles.
The gas surface density is $\Sigma_\mathrm{g} \approx 0.7~\mathrm{g~cm^{-2}}$ where $\Sigma_\mathrm{d}$ peaks,
and by assuming that $\rho_\mathrm{pc} = 1~\mathrm{g~cm^{-3}}$,
$s \gtrsim 5~\mathrm{mm}$ if $\mathrm{St} \gtrsim 1$.

If the lower $\beta$ in the north indicates a particle growth,
$\beta \approx 1$ implies a particle size of $\approx 1~\mathrm{mm} - 10~\mathrm{mm}$,
and it agrees with the above size estimation.
In addition, the estimated size is also consistent with 
the continuum observations at $34~\mathrm{GHz}$ which should be dominated by thermal emission 
from particles of size $\sim 1~\mathrm{mm}$;
at $34~\mathrm{GHz}$ one of two clumps of concentrated dust is observed at $\mathrm{P.A.} \approx 330^\circ$
(the other is at $\mathrm{P.A.} \approx 15^\circ$, \cite{casassus2015}).
On the other hand,
the dust size estimated here is larger than the estimated size of $150~\mu\mathrm{m}$ based on polarization modeling \citep{ohashi2018}.
One possible explanation to the discrepancy is a segregation of dust size in the disk vertical direction,
such that smaller particles (efficient scatterers) float at the upper atmosphere
while larger particles (efficient emitters) sediment at the midplane.

Figure \ref{fig:polar_plots}(d) shows
the dust optical depth at $98.5~\mathrm{GHz}$, $\tau_\mathrm{d}$, at polar coordinates.
Comparing figures \ref{fig:polar_plots}(a) and \ref{fig:polar_plots}(d),
we see that the peaks of $\tau_\mathrm{d}$ and $\Sigma_\mathrm{g}$ share a similar spatial location in the north,
which is contrary to the relative distributions of the peaks of $\Sigma_\mathrm{d}$ and $\Sigma_\mathrm{g}$.
We caution that the derivation of $\Sigma_\mathrm{d}$ is highly dependent on $\kappa_\mathrm{d}$,
and in this paper we derived $\Sigma_\mathrm{d}$ using a $\kappa_\mathrm{d}$ whose spatial variation 
is determined by the $\beta$ distribution.
As shown in Appendix \ref{sec:continuum_700},
our derived parameters of the dust disk can also reproduce the intensity distributions of the continuum emission at 
$700~\mathrm{GHz}$ reported by \citet{casassus2015}.
To better disentangle $\kappa_\mathrm{d}$ and $\Sigma_\mathrm{d}$, however,
radiative transfer modeling of a wider frequency range may be required,
and this is beyond the scope of this study.

The observations of asymmetric protoplanetary disk around HD~135444B
reveal a frequency dependent azimuthal shift in the peak continuum position,
which can be interpreted as a consequence of size segregation of dust grains trapped by a vortex \citep{cazzoletti2018}.
Comparing figure \ref{fig:continuum}(a) and (b) (see also Table \ref{tab:continuum_ridge}),
there is also an azimuthal offset between the $98.5~\mathrm{GHz}$ and $336~\mathrm{GHz}$ peaks in the north.
The peak shift in the case of HD~142527, however,
may not be due to the size 
segregation since the $336~\mathrm{GHz}$ emission is
fairly optically thick.
In fact, the $\beta$ distribution in the outer disk shown in 
figure \ref{fig:polar_plots}(e) is nearly mirror symmetric with respect to $\mathrm{P.A.=0^\circ}$,
and there is no gradient
along the azimuthal direction from upstream to downstream indicative of dust size segregation. 
The dynamic range of the observed frequencies in this study may be insufficient to discuss size segregation in the disk
(cf. \cite{casassus2015}).

\subsection{Effects of uncertainties in CO abundance} \label{sec:validity_results}
In a protoplanetary disk, the abundance of CO as relative to H$_2$
can be lower than that in the interstellar medium if CO is depleted due to freeze-out onto dust grains.
In addition, CO can also be depleted by photodissociation in the disk upper layer
by UV radiation
\citep{miotello2014,miotello2016,miotello2017}.
In fact, the gas masses of several T Tauri disks derived
from the HD $J=1-0$ transition, which are considered to be a reliable mass tracer,
are found to be a factor of five greater than
the masses derived independently from the CO observations without accounting for
these CO depleting processes \citep{bergin2013,mcclure2016}.

Here, we discuss the effects of CO freeze-out and photodissociation
in the outer disk HD~142527.
CO freeze-out occurs in the disk midplane where the disk temperature is lower than the CO freeze-out temperature, 
i.e., approximately $20~\mathrm{K}$ \citep{qi2011}.
The disk models used by \citet{muto2015} 
show that the temperature in the outer disk of HD~142527 varies radially,
and in the disk midplane the temperature is mostly higher than 
$20~\mathrm{K}$ (see Figure 8 of their paper), implying that the CO freeze-out may be insignificant.
On the other hand, the photodissociation is subject to self-shielding
and therefore the process is isotope-selective.
The less abundant isotopologue will have a thicker
photodissociation layer,
and the relative abundances among CO, $^{13}$CO, and C$^{18}$O
are thus expected to be varying spatially \citep{shimajiri2014}.
In Section \ref{sec:analyses_gas} we derived
$N_\mathrm{tot}(\mathrm{^{13}CO})/N_\mathrm{tot}(\mathrm{C^{18}O})$
to be similar to the $^{13}$CO to C$^{18}$O ratio in the local interstellar medium;
these results might suggest that isotope-selective photodissociation is insignificant in the case of HD~142527.
The abundances of these isotopologues relative to CO, however, are still currently unknown
and therefore further investigation is required to confirm the photodissociation of CO.

Even if the effects of freeze-out and isotope-selective photodissociation are small,
the abundance of CO relative to H$_2$ still
remains uncertain if carbon is locked up in other form of molecules.
The detail thermochemical models show that 
the mass conversion from CO to H$_2$ would be underestimated
by a factor of about three to eight \citep{yu2016,yu2017a,molyarova2017}.
The depletion of more than a factor of ten, however, is unlikely
because the gas disk would be gravitationally unstable \citep{fukagawa2013}.
In short,
though $\Sigma_\mathrm{g}$, $\mathrm{G/D}$,
and the dust size estimated from equation (\ref{eq:stokes})
may be underestimated,
the exponent $p \approx 0.5$
found in the power law relation between
$\Sigma_\mathrm{g}$ and $\Sigma_\mathrm{d}$ remains valid
because $p$ does not depend on the 
absolute values of $\Sigma_\mathrm{g}$ and $\Sigma_\mathrm{d}$.

\section{Summary} \label{sec:summary}
We present the ALMA Band 3 observations of the $98.5~\mathrm{GHz}$
dust continuum and the $^{13}$CO $J=1-0$ and C$^{18}$O $J=1-0$ lines of
the protoplanetary disk associated with HD~142527 at a spatial resolution of $\sim 0\farcs5$,
and compare the results to the ALMA observations at Band 7.
The $98.5~\mathrm{GHz}$ continuum and C$^{18}$O $J=1-0$
is optically thin, and we derived the gas-to-dust ratio, $\mathrm{G/D}$, of the outer disk.
The main conclusions are as follows.
\begin{enumerate}
\item The $98.5~\mathrm{GHz}$ dust continuum shows a similar distribution to that at $336~\mathrm{GHz}$,
         where the northern region is brighter than the southern region.
         The contrast of the $98.5~\mathrm{GHz}$ along its ridge is approximately $58$.
         The spectral index $\alpha$ is $\approx 2.8$ and $\approx 3.4$ in the northern and southern regions, respectively.
\item The integrated intensity of the $^{13}$CO $J=1-0$ line emission is more axisymmetric compared to the dust continuum emission,
	where the northern region is brighter than the south by a factor of $\sim1.4$.
	The C$^{18}$O $J=1-0$ emission is confined in a narrower radial extent,
	where its peak emission is located near to the $98.5~\mathrm{GHz}$ dust continuum peak;
	the integrated intensity of the line in the northern region is brighter than the south by a factor of $~4$.
\item The dust opacity spectral index $\beta$ is derived to be
	$\approx 1$ and $\approx 1.7$ in the northern and southern regions of the disk, respectively;
	the difference in $\beta$ between the two regions indicate the difference in dust properties.
	We use the $J=1-0$ and $J=3-2$ lines of C$^{18}$O
	and the $98.5~\mathrm{GHz}$ and $336~\mathrm{GHz}$ continuum emission
	to derive the disk gas and dust surface densities, $\Sigma_\mathrm{g}$ and $\Sigma_\mathrm{d}$.
	We assume the local thermodynamic equilibrium,
	the interstellar abundance $\chi(\mathrm{C^{18}O/H_2}) = 1.79\times10^{-7}$,
	and the canonical dust opacity described by \citet{beckwith1990} by
	varying $\beta$ spatially.
	The derived surface densities are
	$\Sigma_\mathrm{g} \sim 0.9~\mathrm{g~cm^{-2}}$ and $\Sigma_\mathrm{d} \sim 0.3~\mathrm{g~cm^{-2}}$
	in the northern regions, with results of $\Sigma_\mathrm{g} \sim 0.2~\mathrm{g~cm^{-2}}$ and
	$\Sigma_\mathrm{d} \sim 0.01~\mathrm{g~cm^{-2}}$ in the southern regions.
	The contrast along the $\Sigma_\mathrm{g}$ and $\Sigma_\mathrm{d}$ ridges are
	$5$ and $33$, respectively. 
	The gas-to-dust ratio, $\mathrm{G/D} \equiv \Sigma_\mathrm{g}/\Sigma_\mathrm{d}$, is derived 
	to vary smoothly in the azimuthal direction of the disk, where it is $\sim 3$ and $\sim 20$ in the northern and southern regions, respectively.
\item	By using the results of $\Sigma_\mathrm{g}$ and $\Sigma_\mathrm{d}$
	derived at the $\Sigma_\mathrm{d}$ ridge, we found that $\Sigma_\mathrm{g}$ varies approximately as $\Sigma_\mathrm{d}^{0.47}$, or 
	equivalently $\mathrm{G/D} \propto \Sigma_\mathrm{d}^{-0.53}$.
	This relation will be a critical test for future theoretical studies to understand the azimuthal-asymmetric disk structure.
\item Our results show that the $\Sigma_\mathrm{d}$ peak is slightly shifted ahead of the $\Sigma_\mathrm{g}$,
	which is predicted by theoretical studies of the trapping of dust by vortices of high gaseous pressures. The estimated dust size is $\gtrsim 5~\mathrm{mm}$
	if $\chi(\mathrm{C^{18}O/H_2})$ in the disk is similar to the interstellar value.
\item The $^{13}$CO $J=1-0$ line emission in $\mathrm{P.A.} = 200^\circ - 240^\circ$ is marginally optically thin,
	where we derive
	$N_\mathrm{tot}(\mathrm{^{13}CO})/N_\mathrm{tot}(\mathrm{C^{18}O})=9.99\pm1.67$; the value agrees with the interstellar abundance ratio
	$\chi(\mathrm{^{13}CO/C^{18}O}) = 8.11\pm1.1$ within uncertainties.
\end{enumerate}

\begin{ack}
This work was supported by JSPS KAKENHI Grant Numbers 17H01103 and 18H05441.
This paper makes use of the following ALMA data:
ADS/JAO.ALMA\#2011.0.00318S and ADS/JAO.ALMA\#2013.1.00670S. 
ALMA is a partnership of ESO (representing its member states), NSF (USA) and NINS (Japan), together with NRC (Canada), MOST and ASIAA (Taiwan), and KASI (Republic of Korea), in cooperation with the Republic of Chile. The Joint ALMA Observatory is operated by ESO, AUI/NRAO and NAOJ.
\end{ack}

\appendix
\section{Channel maps of $^{13}$CO $J=1-0$ and C$^{18}$O $J=1-0$}\label{sec:channel_map}
\begin{figure*}[ht!]
\centering
\begin{tabular}{cc}
\includegraphics[width=0.4\textwidth,height=\textheight,keepaspectratio]{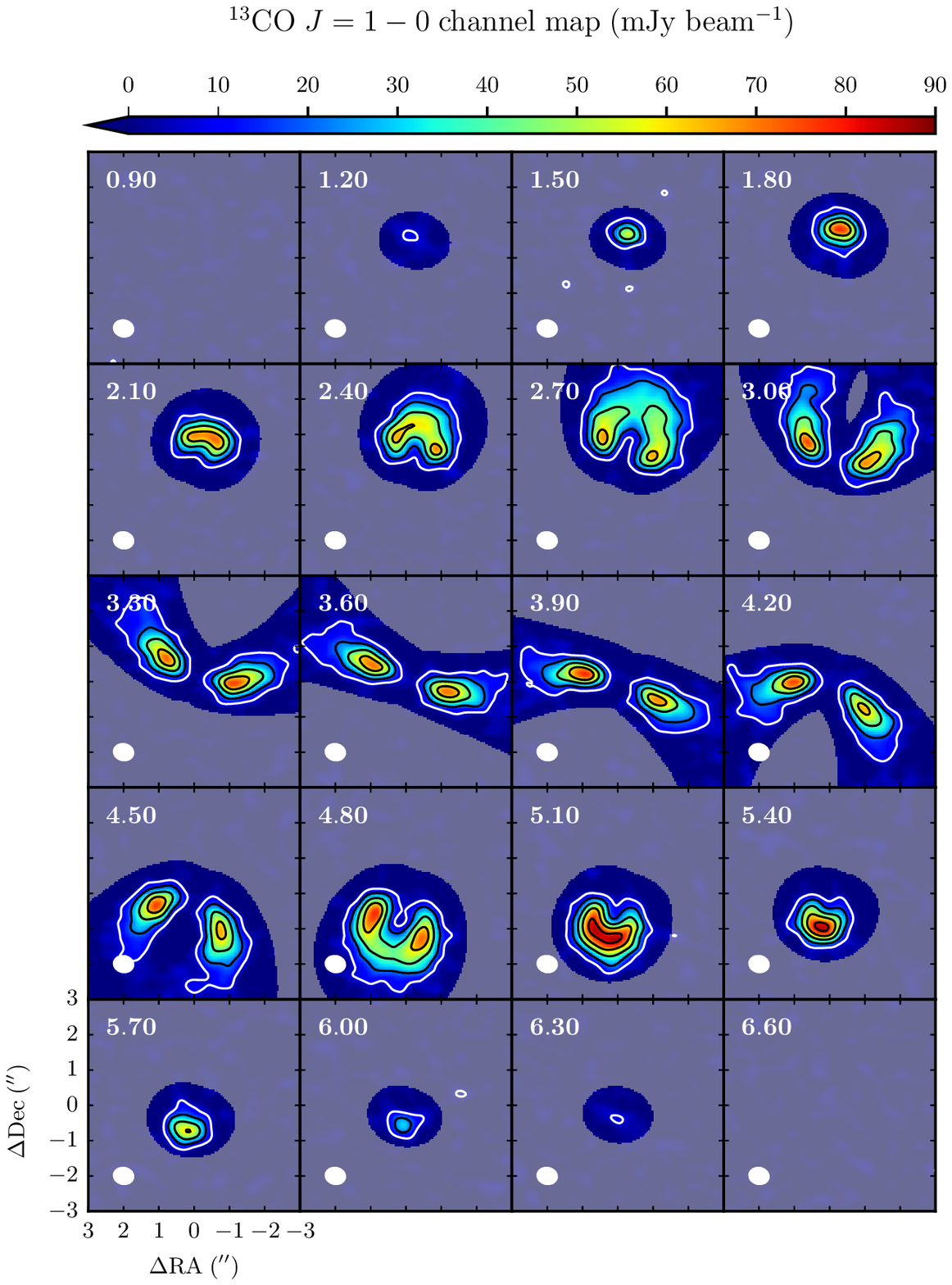} &
\includegraphics[width=0.4\textwidth,height=\textheight,keepaspectratio]{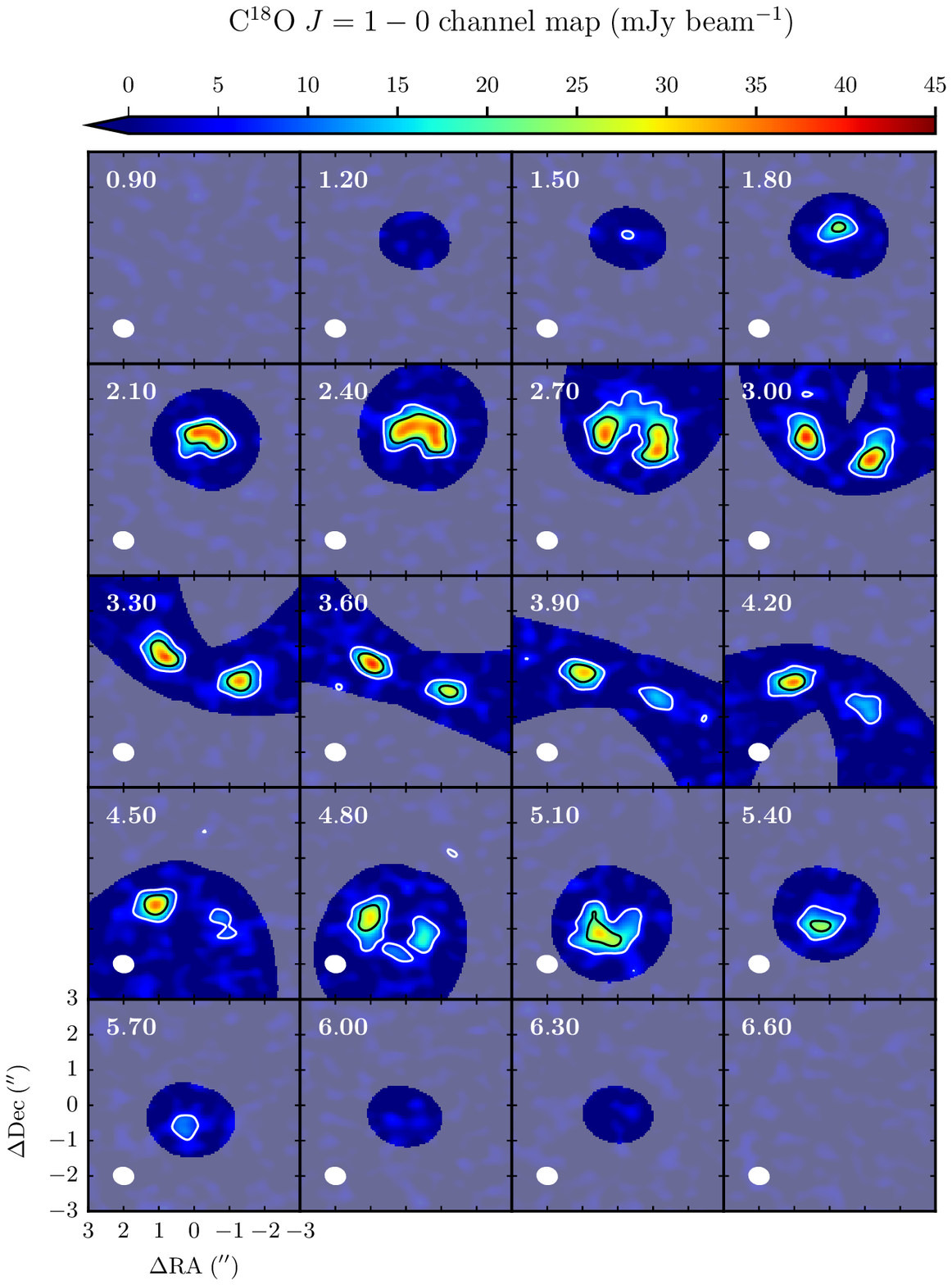}
\end{tabular}
\caption{The velocity channel maps of the $^{13}$CO $J=1-0$ (left panel) and 
C$^{18}$O $J=1-0$ (right panel) line emission.
The velocity (in units of $\mathrm{km~s^{-1}}$) is written in the top left corner in each channel map.
The white ellipse in the bottom left indicates the synthesized beam,
which is identical to that of the $98.5~\mathrm{GHz}$
continuum image shown in figure \ref{fig:continuum}
($0\farcs54 \times 0\farcs44$, $\mathrm{P.A.}=  78\fdg1$).
The white contour denotes the $3.5\sigma = 7~\mathrm{mJy~beam^{-1}}$
and the black contours are drawn at $(20,40,60,80)~\mathrm{mJy~beam^{-1}}$.
The white shaded regions denote the Keplerian masks used to create
the moment maps in figures \ref{fig:moment_13C16O_030kms} and \ref{fig:moment_12C18O_030kms}.
}\label{fig:channel_map}\end{figure*}

Figure \ref{fig:channel_map} shows the channel maps of $^{13}$CO and C$^{18}$O line emission
from which the moment maps shown in 
figures \ref{fig:moment_13C16O_030kms} and \ref{fig:moment_12C18O_030kms}
are created.
When creating the moment maps,
we first mask out the regions where Keplerian motion of the disk is not expected in the velocity channels
\citep{salinas2017,ansdell2018}, 
and the emission below $3.5\sigma$ in the unmasked region is then clipped.
We adopt a stellar mass of $M_* = 2.4~M_\odot$ and a disk inclination of 
$i = 27^\circ$ when creating the Keplerian mask.

\section{Two-layer disk model}\label{sec:two_layer_disk}
\begin{figure*}\centering
\begin{tabular}{ccc}
\includegraphics[width=0.3\textwidth]{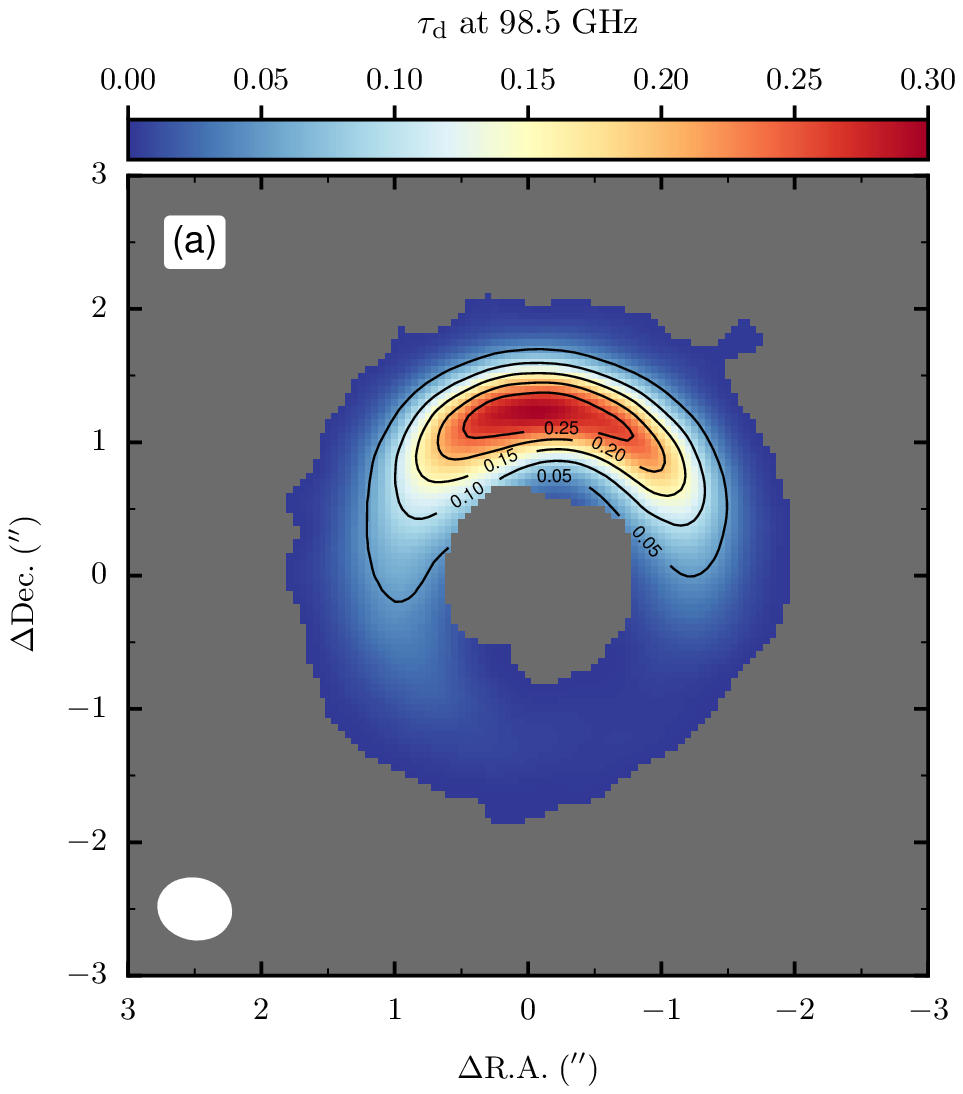} &
\includegraphics[width=0.3\textwidth]{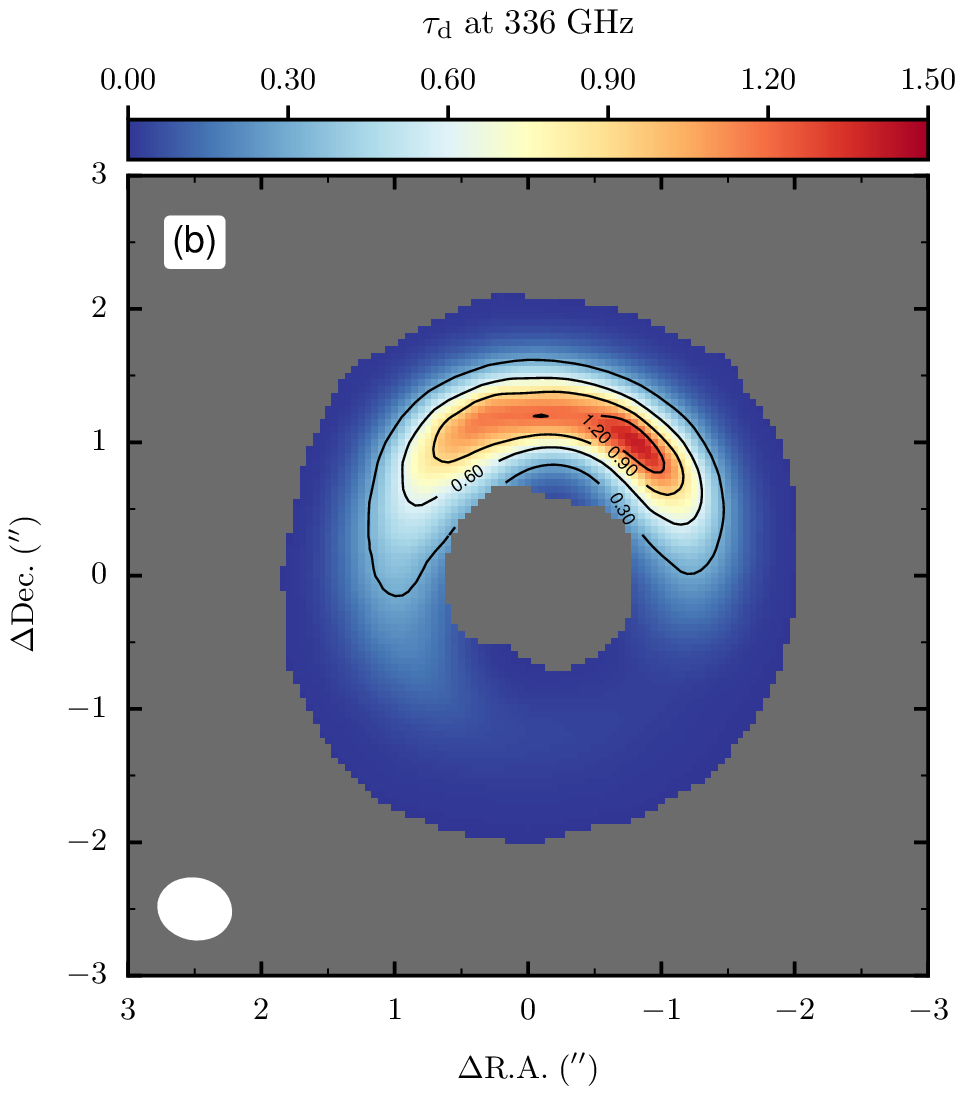} &
\includegraphics[width=0.3\textwidth]{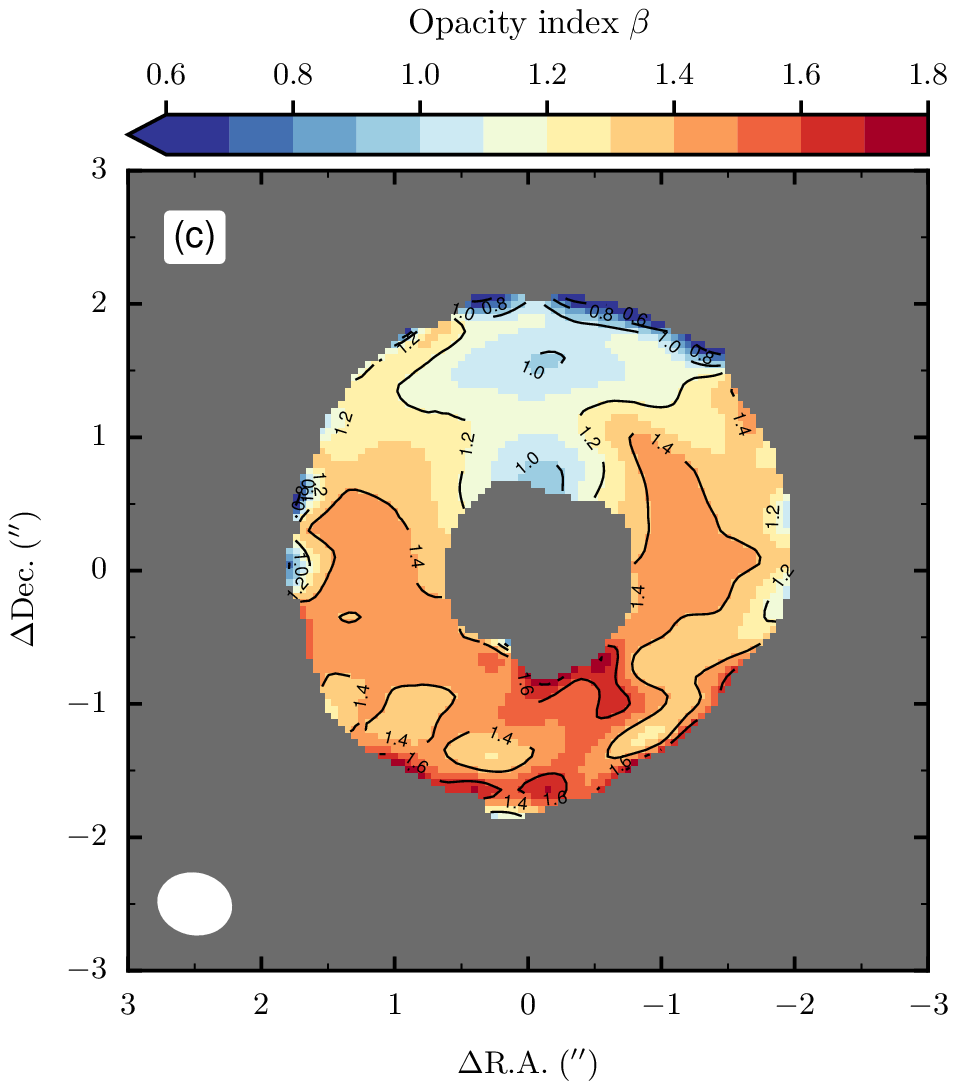}
\end{tabular}
\caption{
Similar to figure \ref{fig:continuum_tau}, 
for the results derived from the two-layer disk model.
}\label{fig:continuum_tau_lowtex}\end{figure*}
\begin{figure*}\centering
\includegraphics[width=0.8\textwidth]{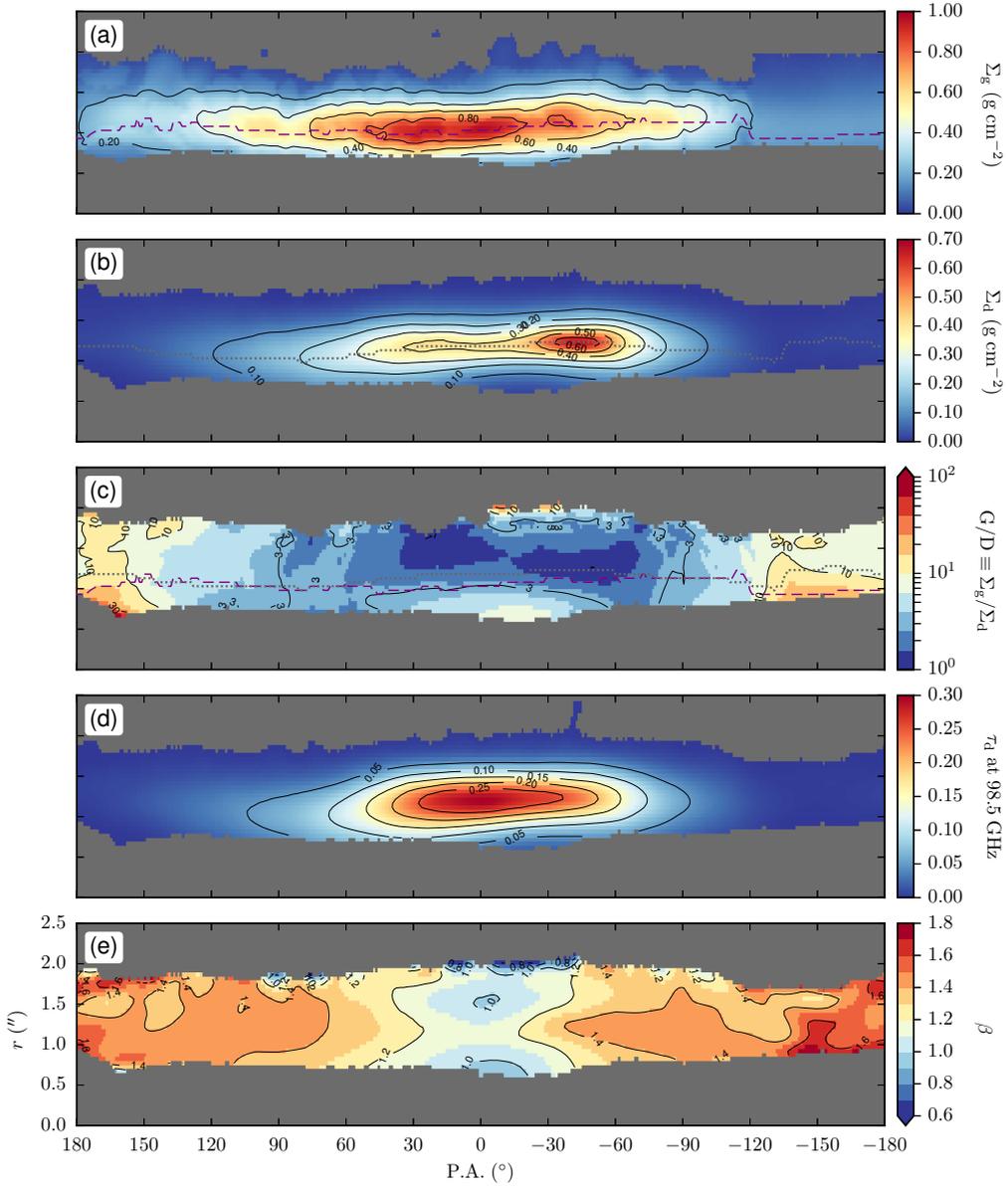}
\caption{Similar to figure \ref{fig:polar_plots},
but the results are derived from the two-layer disk model.
Note that only the color scale range of (b) is twice wider than that in figure \ref{fig:polar_plots}(b),
while the others have the same scaling range.
}\label{fig:polar_plots_lowtex}\end{figure*}
\begin{figure*}\centering
\begin{tabular}{cc}
\includegraphics[width=0.45\textwidth]{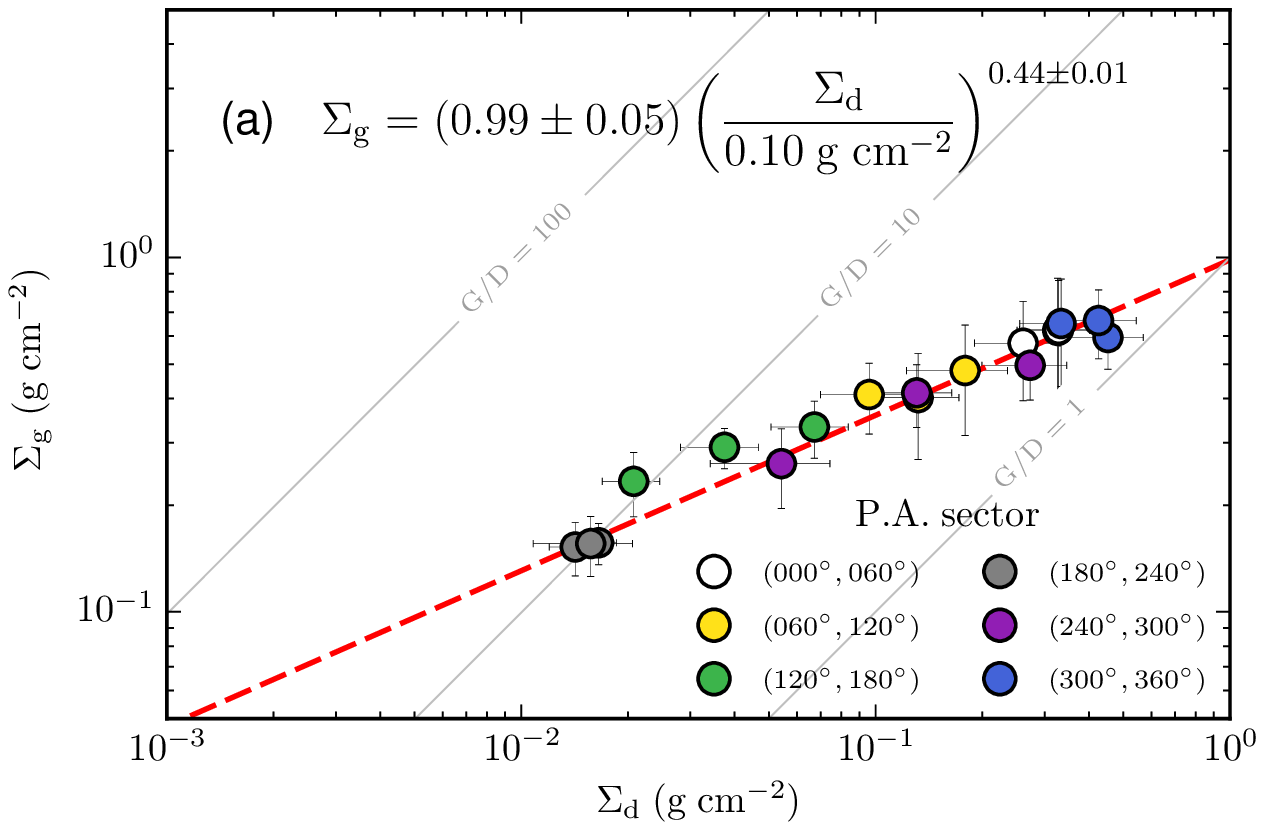} &
\includegraphics[width=0.45\textwidth]{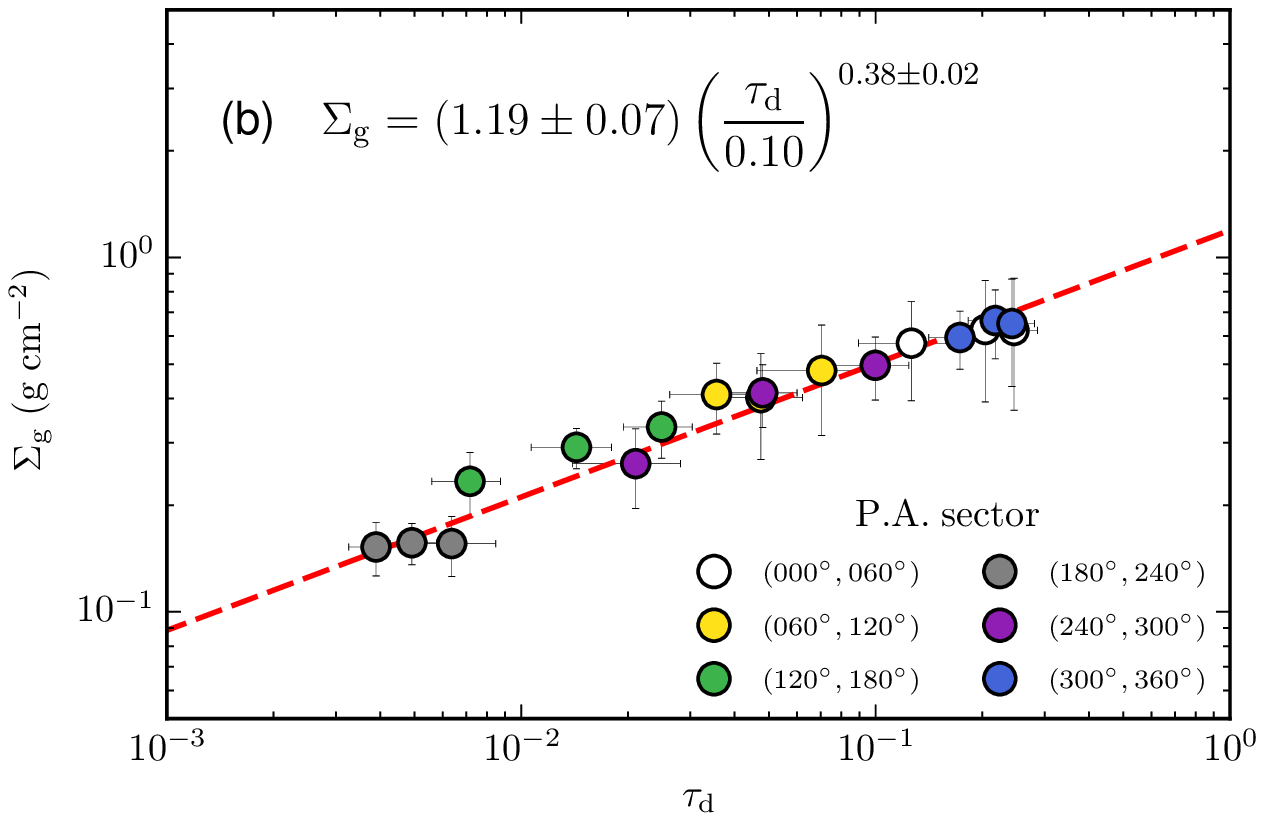} 
\end{tabular}
\caption{
Similar to figure \ref{fig:correlation},
for the results derived from the two-layer disk model.
}\label{fig:correlation_lowtex}\end{figure*}

To evaluate the uncertainties of $\mathrm{G/D}$ and the exponent $p$,
we here adopt a two-layer disk model when deriving the surface densities,
where the temperature of gas and dust particles are different.
The gas temperature $T_\mathrm{g}$ is assumed to be the same as the brightness temperature at the peak 
$T_\mathrm{B}$ of $^{13}$CO $J=3-2$ including the continuum emission, as shown in figure \ref{fig:13co_peakTB}(d).
The dust temperature $T_\mathrm{d}$, on the other hand,
is simply taken to be $80\%$ of the gas temperature across the map;
this physical condition mimics the effect of dust sedimentation.
The $20\%$ temperature drop results in a peak 
$T_\mathrm{d} \approx 29~\mathrm{K}$
in the dust temperature field. Indeed, this agrees with the peak dust temperature
derived by Casassus et al. (2015) using the continuum observations at Bands 7 and 9, 
and therefore it is a reasonable estimate for the dust temperature.

Figure \ref{fig:continuum_tau_lowtex} shows the optical depth of
the continuum emission at $98.5~\mathrm{GHz}$ and $336~\mathrm{GHz}$,
and the dust opacity spectral index $\beta$ derived from the two-layer disk model;
they are larger than those derived from the one-layer disk model due to the lower temperature.
In addition, the larger $\beta$ results in a smaller dust opacity $\kappa_\mathrm{d}$,
where it is smaller than that of the one-layer disk model by $\approx 10\%$ in the disk southern region
and by $\approx 30\% - 50\%$ in the dust concentrated northern region.

The radiative transfer 
for the gas molecular line in the two-layer disk model reads 
\begin{eqnarray}
\label{eq:gas_rte_lowtex}
\lefteqn{I_\mathrm{g+d} = \left[  B_\nu(T_\mathrm{g}) - B_\nu(T_\mathrm{bg}) \right] \left[  1 - \exp(-\tau_\mathrm{g}/2)    \right] }\nonumber \\ 
                          &\ &+\left[  B_\nu(T_\mathrm{d}) - B_\nu(T_\mathrm{bg}) \right] \left[  1 - \exp(-\tau_\mathrm{g})    \right] \exp(-\tau_\mathrm{g}/2)  \nonumber \\ 
                          &\ &+\left[  B_\nu(T_\mathrm{g}) - B_\nu(T_\mathrm{bg}) \right] \left[  1 - \exp(-\tau_\mathrm{g}/2)    \right] \exp(-\tau_\mathrm{g}/2 - \tau_\mathrm{d}), \nonumber \\
\end{eqnarray}
where $I_\mathrm{g+d}$ denotes the line emission including the continuum emission \citep{nomura2016}. 
In the line of sight,
the first term accounts for the line emission from the disk atmosphere at the front side,
the second term the dust emission in the disk midplane,
while the last term the line emission from the back side that will propagate through the disk midplane and the front atmosphere.

Figure \ref{fig:polar_plots_lowtex} shows the results derived from this two-layer disk model.
We mask out the disk inner region in which the disk temperature
is lower than $24~\mathrm{K}$, i.e., $80\%$ of the temperature criteria used
to mask out the disk inner region in the one-layer disk model (Section \ref{sec:analyses}).
The overall distributions of $\Sigma_\mathrm{g}$, $\Sigma_\mathrm{d}$, and $\mathrm{G/D}$
are similar to the one-layer disk model;
due to the lower $\kappa_\mathrm{d}$, however, in the northern region
$\Sigma_\mathrm{d}$ is derived to be twice as high as that in the one-layer disk model,
and therefore the $\mathrm{G/D}$ distribution is derived to be lower.
Figure \ref{fig:correlation_lowtex}(a) shows the correlation between $\Sigma_\mathrm{g}$ and $\Sigma_\mathrm{d}$.
Similar to figure \ref{fig:correlation}(a) we find a power law with an exponent $p = 0.44$
to be a good fit to the results.
The derived value of exponent $p$ is consistent
with that of the one-layer disk model
despite the different temperature assumption between these two disk models.
Figure \ref{fig:correlation_lowtex}(b) is analogous to figure \ref{fig:correlation}(b),
where the derived values of $p$ are also consistent between the two models.

\section{Mock-up dust continuum image at $700~\mathrm{GHz}$}\label{sec:continuum_700}
\begin{figure*}\centering
\begin{tabular}{cc}
\includegraphics[width=0.3\textwidth]{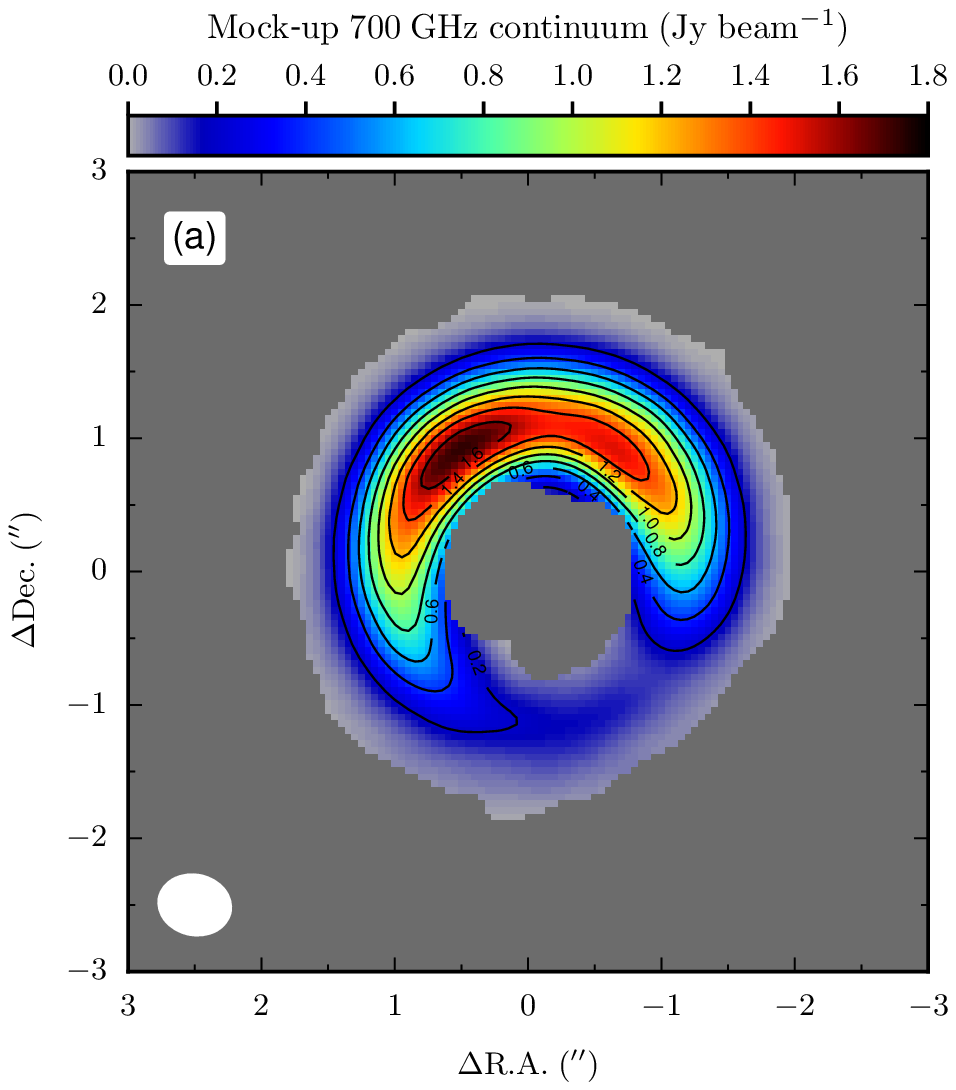} &
\includegraphics[width=0.3\textwidth]{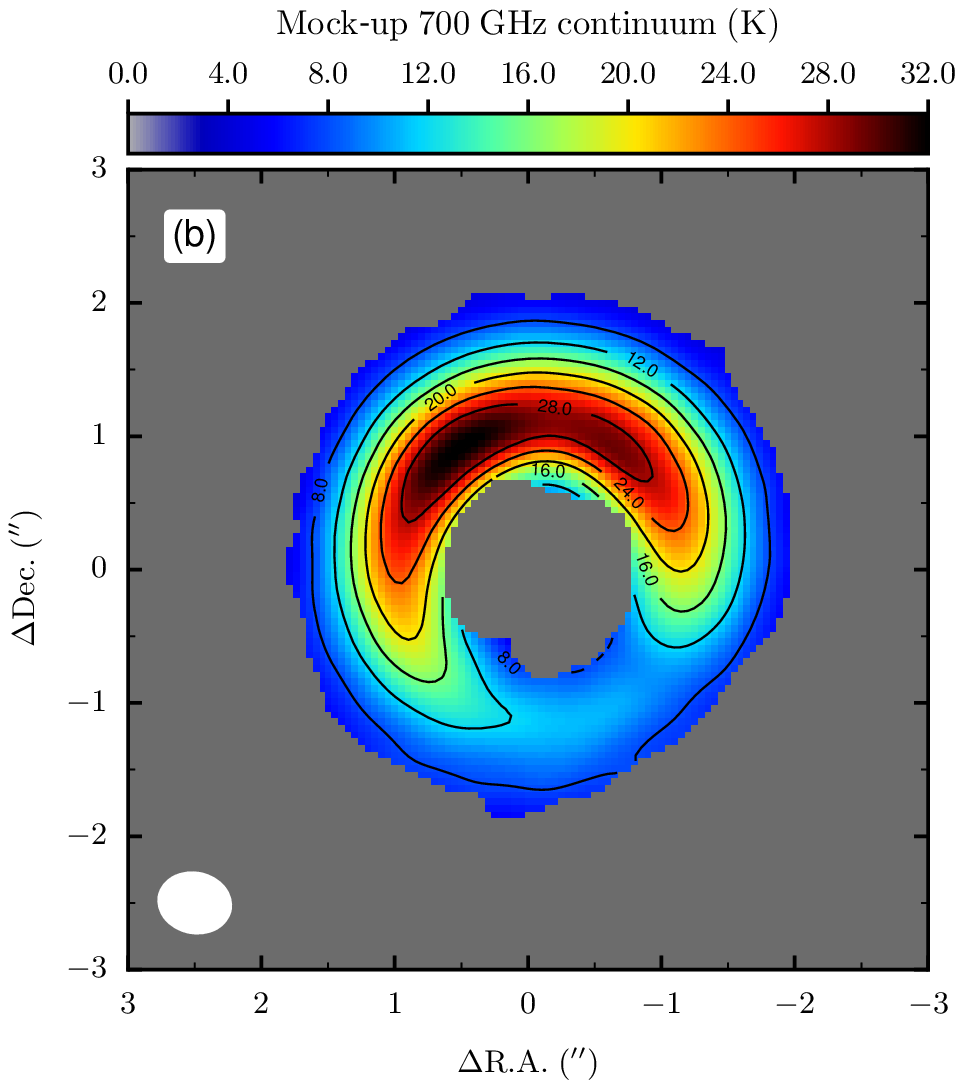} \\
\end{tabular}
\caption{
The mock-up dust continuum image at $700~\mathrm{GHz}$.
Panels (a) and (b) present the continuum emission in units of 
$\mathrm{Jy\ beam^{-1}}$ and $\mathrm{K}$, respectively.
The ellipse in the lower left corner indicates
the beam size ($0\farcs54 \times 0\farcs44$, $\mathrm{P.A,} = 78\fdg1 $)
of the Band 3 and Band 7 observations,
which are used to the derive the $\Sigma_\mathrm{d}$ distribution.
}\label{fig:continuum_700}\end{figure*}

By using the dust surface density $\Sigma_\mathrm{d}$ distribution derived from the one layer-disk model (figure \ref{fig:sigma}a),
we create a mock-up image at $700~\mathrm{GHz}$ to compare with the continuum observations at the same frequency reported by \citet{casassus2015}.
We use the same temperature distribution, i.e., the peak $T_\mathrm{B}$ of $^{13}$CO $J=3-2$ including the continuum (figure \ref{fig:13co_peakTB}d)
as the disk temperature
and the $\beta$ distribution (figure \ref{fig:continuum_tau}c) to calculate the dust opacity $\kappa_\mathrm{d}$ at $700~\mathrm{GHz}$ (Equation \ref{eq:dust_opacity}).
The radiative transfer follows Equation (\ref{eq:dust_rte}).
Figure \ref{fig:continuum_700} shows the mock-up image.
Though the beam size is larger than that of the image obtained by \citet{casassus2015},
our mock up image successfully reproduces the observed intensity distribution, in which there
are two emission peaks and the northwestern one is brighter (because the temperature in this region is higher).
This comparison shows that the disk parameters estimated from Bands 3 and 7 are reasonable.


\end{document}